\documentclass[prd,nofootinbib,reprint,twocolumn,showpacs,superscriptaddress,floatfix]{revtex4-1}
\pdfoutput=1 
\usepackage{bm, color} 
\usepackage{amssymb,amsfonts,slashed,amsthm,amsmath, graphicx, soul, empheq, cancel}
\usepackage{hyperref}
\usepackage[caption=false]{subfig}

\begin{document}

\newcommand{\vev}[1]{ \left\langle {#1} \right\rangle }
\newcommand{\bra}[1]{ \langle {#1} | }
\newcommand{\ket}[1]{ | {#1} \rangle }
\newcommand{\eV}{ \ {\rm eV} }
\newcommand{\KeV}{ \ {\rm keV} }
\newcommand{\MeV}{\  {\rm MeV} }
\newcommand{\GeV}{\  {\rm GeV} }
\newcommand{\TeV}{\  {\rm TeV} }
\newcommand{\1}{\mbox{1}\hspace{-0.25em}\mbox{l}}
\newcommand{\Red}[1]{{\color{red} {#1}}}

\newcommand{\lmk}{\left(}  
\newcommand{\rmk}{\right)}
\newcommand{\lkk}{\left[}  
\newcommand{\rkk}{\right]}
\newcommand{\lhk}{\left \{ }  
\newcommand{\rhk}{\right \} }
\newcommand{\del}{\partial}  
\newcommand{\la}{\left\langle} 
\newcommand{\ra}{\right\rangle}
\newcommand{\half}{\frac{1}{2}}

\newcommand{\bea}{\begin{array}}
\newcommand{\eea}{\end{array}}
\newcommand{\beq}{\begin{eqnarray}}
\newcommand{\eeq}{\end{eqnarray}}
\newcommand{\eq}[1]{Eq.~(\ref{#1})}

\newcommand{\dd}{\mathrm{d}}
\newcommand{\Mpl}{M_{\rm Pl}}
\newcommand{\mg}{m_{3/2}}
\newcommand{\abs}[1]{\left\vert {#1} \right\vert}
\newcommand{\mphi}{m_{\phi}}
\newcommand{\Hz}{\ {\rm Hz}}
\newcommand{\for}{\quad \text{for }}
\newcommand{\Min}{\text{Min}}
\newcommand{\Max}{\text{Max}}
\newcommand{\Kahler}{K\"{a}hler }
\newcommand{\cphi}{\varphi}
\newcommand{\Tr}{\text{Tr}}
\newcommand{\diag}{{\rm diag}}

\newcommand{\SUf}{SU(3)_{\rm f}}
\newcommand{\Upq}{U(1)_{\rm PQ}}
\newcommand{\Zpq}{Z^{\rm PQ}_3}
\newcommand{\Cpq}{C_{\rm PQ}}
\newcommand{\ubar}{u^c}
\newcommand{\dbar}{d^c}
\newcommand{\ebar}{e^c}
\newcommand{\nubar}{\nu^c}
\newcommand{\Ndw}{N_{\rm DW}}
\newcommand{\Fpq}{F_{\rm PQ}}
\newcommand{\fpq}{v_{\rm PQ}}
\newcommand{\Br}{{\rm Br}}
\newcommand{\Lag}{\mathcal{L}}
\newcommand{\Lqcd}{\Lambda_{\rm QCD}}

\newcommand{\ji}{j_{\rm inf}} 
\newcommand{\jb}{j_{B-L}} 
\newcommand{\M}{M} 
\newcommand{\im}{{\rm Im} }
\newcommand{\re}{{\rm Re} }

\def\lrf#1#2{ \left(\frac{#1}{#2}\right)}
\def\lrfp#1#2#3{ \left(\frac{#1}{#2} \right)^{#3}}
\def\lrp#1#2{\left( #1 \right)^{#2}}
\def\REF#1{Ref.~\cite{#1}}
\def\SEC#1{Sec.~\ref{#1}}
\def\FIG#1{FIG.~\ref{#1}}
\def\EQ#1{Eq.~(\ref{#1})}
\def\EQS#1{Eqs.~(\ref{#1})}
\def\TEV#1{10^{#1}{\rm\,TeV}}
\def\GEV#1{10^{#1}{\rm\,GeV}}
\def\MEV#1{10^{#1}{\rm\,MeV}}
\def\KEV#1{10^{#1}{\rm\,keV}}
\def\blue#1{\textcolor{blue}{#1}}
\def\red#1{\textcolor{blue}{#1}}

\newcommand{\eff}{\Delta N_{\rm eff}}
\newcommand{\neff}{\Delta N_{\rm eff}}
\newcommand{\cc}{\Omega_\Lambda}
\newcommand{\Mpc}{\ {\rm Mpc}}
\newcommand{\Msolar}{M_\odot}

\def\sn#1{\textcolor{red}{#1}}
\def\SN#1{\textcolor{red}{[{\bf SN:} #1]}}

%%%%%%%%%%%%%%%%%%%%%%%%%%%%%%%%%%%%%%%%%%%%%%%%%%%%%%%%%%%%%%%

\title{
Conformal Phase Transition in Supersymmetric QCD
}

\author{Kohei Fujikura}
\affiliation{Graduate School of Arts and Sciences, University of Tokyo, Komaba, \\ 
Meguro-ku, Tokyo 153-8902, Japan}

\author{Shota Nakagawa}
\affiliation{Tsung-Dao Lee Institute, Shanghai Jiao Tong University, \\
No.~1 Lisuo Road, Pudong New Area, Shanghai 201210, China}
\affiliation{School of Physics and Astronomy, Shanghai Jiao Tong University, \\
800 Dongchuan Road, Shanghai 200240, China}

\author{Yuichiro Nakai}
\affiliation{Tsung-Dao Lee Institute, Shanghai Jiao Tong University, \\
No.~1 Lisuo Road, Pudong New Area, Shanghai 201210, China}
\affiliation{School of Physics and Astronomy, Shanghai Jiao Tong University, \\
800 Dongchuan Road, Shanghai 200240, China}

\author{Peng Sun}
\affiliation{Tsung-Dao Lee Institute, Shanghai Jiao Tong University, \\
No.~1 Lisuo Road, Pudong New Area, Shanghai 201210, China}
\affiliation{School of Physics and Astronomy, Shanghai Jiao Tong University, \\
800 Dongchuan Road, Shanghai 200240, China}

\author{Yufei Zhang}
\affiliation{Tsung-Dao Lee Institute, Shanghai Jiao Tong University, \\
No.~1 Lisuo Road, Pudong New Area, Shanghai 201210, China}
\affiliation{School of Physics and Astronomy, Shanghai Jiao Tong University, \\
800 Dongchuan Road, Shanghai 200240, China}

\begin{abstract}
We construct a four-dimensional supersymmetric QCD in conformal window with a marginally relevant deformation
which triggers the spontaneous breaking of (approximate) scale invariance and the subsequent confinement, generating a mass gap,
at an energy scale hierarchically smaller than the Planck scale without fine-tuning.
We analyze the finite temperature system and show that the phase transition associated with the breaking of conformal invariance is of the strong first order. 
When such a phase transition takes place at a temperature of the Universe around the electroweak scale,
it generates a stochastic gravitational wave (GW) background probed by future space-based interferometers,
while a conformal phase transition in a dark sector at $\mathcal{O}(1)$ GeV
generates GWs to explain the reported pulsar timing array signal.
\end{abstract}

\maketitle

%%%%%%%%%%%%%%%%%%%%%%%%%%%%%%%%%%%%%%%%%%%%%%%%%%
%%%%%%%%%%%%%%%%%%%%%%%%%%%%%%%%%%%%%%%%%%%%%%%%%%
\section{Introduction
\label{introduction}}

Higher symmetries might be realized at the beginning of the hot Big Bang Universe. 
As the temperature decreases by the cosmic expansion,
our Universe may have experienced phase transitions associated with the spontaneous breaking of symmetries.
When a symmetric and a symmetry broken phases coexist and the symmetry broken phase is energetically favored,
the phase transition is of the first order, which proceeds
via nucleation of true vacuum bubbles and their subsequent expansions.
With a sufficiently large discontinuity in the order parameter, it is called a {\it strong first-order phase transition}.
Although the Standard Model (SM) experiences smooth crossover behavior~\cite{Kajantie:1995kf,Kajantie:1996qd,Rummukainen:1998as,Aoki:2006we} rather than a first-order phase transition in the early universe,
physics beyond the SM may provide its candidates. 
Since a strong first-order phase transition is an out-of-equilibrium process,
its dynamics has been extensively discussed in the context of the origin of baryon asymmetry
(see e.g. refs.~\cite{Turok:1990in,Krauss:1999ng,Garcia-Bellido:1999xos,Konstandin:2011ds,Fujikura:2021abj,Dasgupta:2022isg,Ellis:2022lft,Chun:2023ezg,Cataldi:2024pgt,Fujikura:2024jto,Agashe:2024uvp}) as well as the production of dark matter~\cite{Witten:1984rs,Chway:2019kft,Hong:2020est,Azatov:2021ifm,Kawana:2021tde,Giudice:2024tcp,Zhang:2024dgv,Allahverdi:2024ofe}.
Moreover, the violent nature of the phase transition often leads to the production of a stochastic gravitational wave (GW) background through bubble collisions~\cite{Turner:1990rc,Turner:1992tz,Kosowsky:1992rz,Kosowsky:1992vn},
sound waves~\cite{Hindmarsh:2013xza,Giblin:2014qia,Hindmarsh:2015qta,Hindmarsh:2017gnf}, and turbulence~\cite{Kamionkowski:1993fg,Kosowsky:2001xp,Caprini:2006jb,Caprini:2009yp,Gogoberidze:2007an,Niksa:2018ofa} of the background fluid.
Its unique signature carries a valuable information about the early Universe and
is detectable by ongoing and near-future GW observatories.

Various theories to show a strong first order phase transition have been explored, but
a widely shared characteristic is the presence of an approximate scale invariance.
For instance, the model of radiative symmetry breaking with a classical scale invariance (via Coleman-Weinberg mechanism~\cite{Coleman:1973jx}) always admits a local minimum induced by the finite-temperature effect, no matter how low the temperature is~\cite{Witten:1980ez}.
In fact, this model leads to a strong first order phase transition under a mean field analysis
(see e.g. refs.~\cite{Iso:2017uuu,Jinno:2016knw}) although a significant amount of fine-tuning on a renormalized mass parameter is required once additional heavy fields are included.\footnote{In reality, we have to include the effect of thermal environment to precisely determine the order of a phase transition (see e.g. ref.~\cite{Hiramatsu:2014uta}).}

Conformal field theory (CFT) offers a more natural way to implement an approximate scale invariance at the quantum level.
We can consider a CFT whose marginally relevant deformation drives the spontaneous breaking of (approximate) scale invariance
at a certain energy scale.
The theory must contain a pseudo-Nambu-Goldstone boson (pNGB)
associated with the spontaneously broken scale invariance called {\it dilaton}.
While the phase transition dynamics may be studied in terms of the dilaton effective theory
\cite{Coradeschi:2013gda,Chacko:2012sy},
this possibility has been usually discussed without an explicit model of CFT,
partly due to its non-perturbative nature.
To have an explicit model relies on the AdS/CFT correspondence
\cite{Maldacena:1997re,Gubser:1998bc,Witten:1998qj} (see also refs.~\cite{Arkani-Hamed:2000ijo,Rattazzi:2000hs}).
In the Randall–Sundrum (RS) model
\cite{Randall:1999ee},
the 5D Universe is bounded by two 4D branes called UV and IR branes,
whose separation distance is parametrized by the expectation value of the {\it radion},
and its stabilization has been studied in refs.~\cite{Goldberger:1999uk,Garriga:2000jb,Goldberger:2000dv,Hofmann:2000cj,Brevik:2000vt,Flachi:2001pq,Nojiri:2001ai,Garriga:2002vf,Haba:2019zjc,Fujikura:2019oyi,Girmohanta:2024kyx,Girmohanta:2024ywz}.
The radion can be identified as the dilaton in the dual 4D picture.
The authors of ref.~\cite{Creminelli:2001th} have initiated the study of the phase transition in this 5D model at finite temperature.
At high temperatures, the Universe is described by the AdS-Schwarzschild (AdS-S) geometry with an event horizon replacing the IR brane, while at a low temperature, it shows a phase transition to the RS spacetime,
which provides a dual picture of the {\it conformal phase transition}
associated with the spontaneous breaking of (approximate) conformal invariance. 
Although the study in terms of the 5D model claims that the phase transition is of the strong first order,\footnote{
The Goldberger-Wise model for radion stabilization
\cite{Goldberger:1999uk} even shows a problematically long supercooling phase.
}
a discussion based on a concrete renormalizable model of 4D CFT is desirable
to put the conformal phase transition on a firm basis.\footnote{An another approach based on the perturbative walking model revealed a long period of a supercooled phase transition \cite{Miura:2018dsy,Azatov:2020nbe}.
}

In the present paper, we provide a concrete UV complete 4D model to realize a strong first order phase transition.
Our model is based on a supersymmetric QCD in conformal window
whose (s)quarks marginally couple to a singlet chiral superfield.
The theory flows into an IR fixed point.
A supersymmetric mass term of the singlet field gives a marginally-relevant perturbation to the CFT,
and leads to the spontaneous breaking of (approximate) scale invariance and the subsequent confinement of the gauge theory.
The phase transition dynamics of the model at finite temperature is analyzed, and it is found that
the phase transition can be ubiquitously of the strong first-order with a mean field approximation.
We estimate the GW spectrum generated from the strong first order phase transition
by numerically calculating the probability of nucleation of critical bubbles.
It turns out that when the phase transition takes place at a temperature 
around the electroweak scale, it generates GWs probed by future space-based interferometers,
while the phase transition in a dark sector at $\mathcal{O}(1)$
GeV generates GWs to explain the reported pulsar timing array signal
\cite{NANOGrav:2023gor,EPTA:2023fyk,Xu:2023wog,Reardon:2023gzh}.

The rest of the paper is organized as follows.
In section~\ref{sec:Model}, we present our 4D model that shows the spontaneous breaking of approximate scale invariance
triggered by a marginally relevant operator.
Section~\ref{sec:PT} then discusses the conformal phase transition in the finite temperature system.
Then, the GW generation is explored in section~\ref{sec:cosmo}.
Section~\ref{sec:discussion} is devoted to conclusions and discussions.

%%%%%%
%%%%%%%%%%%%%%%%%%%%%%%%%%%%%%%%%%%%%%%%%%%%%%%%%%
\section{The Model
\label{sec:Model}}

Let us consider a supersymmetric $SU(N)$ gauge theory with $N_F$ vector-like chiral superfields, $Q_a, \bar{Q}_a$
$(a = 1,2,\dots,N_F)$, which transform as the (anti-)fundamental representations.
The number of color $N~(\geq2)$ and that of flavor $N_F$ satisfy $3N/2 < N_F < 3N$, so that the theory is in conformal window and has a nontrivial IR fixed point \cite{Intriligator:2007cp}.
The flavor number is then parameterized as $N_F = (2 + \epsilon)N$, 
where $\epsilon$ is a nonzero constant, $-1/2<\epsilon<1$.
Since a larger value of $\epsilon$ better describes the perturbative IR fixed point, we focus on $0<\epsilon<1$
in the following discussion,
\beq
\epsilon = \frac{k}{N} \ ,
\eeq
with $k=1,2,\cdots, N-1$.

We now introduce a gauge-singlet chiral superfield $\Phi$, which has a superpotential,\footnote{
With a global $U(1)$ charge for the gauge singlet field $\Phi$, the theory can be endowed with the Peccei-Quinn (PQ) mechanism
to address the strong CP problem
\cite{Peccei:1977hh}.
Refs.~\cite{Nakai:2021nyf,Nakagawa:2023shi,Nakagawa:2024kcb} have presented high-quality axion models
where Planck-suppressed operators
explicitly breaking the PQ symmetry are suppressed by a large anomalous dimension of $\Phi$.}
\beq
W_Q = \lambda \Phi Q_a \bar{Q}_a  \, , 
\label{WQ}
\eeq
where $\lambda$ is a dimensionless coupling which also flows into a nontrivial IR fixed point.
The superpotential respects the $U(1)_R$ symmetry whose charge assignments are given by 
$R(Q)=R(\bar{Q})=(N_F-N)/N_F$ and $R(\Phi)=2N/N_F$.

When $\Phi$ takes a nonzero field value $ \Phi  \neq 0$,
the supermultiplets $Q_a, \bar{Q}_a$ acquire masses of
$\lambda \Phi $. 
At the energy scale around their masses, these supermultiplets are decoupled.
Since the effective theory becomes a pure super-Yang-Mills theory, it shows gaugino condensation.
This generates the effective superpotential,
\beq
 W_{\rm gaugino} = N \Lambda_{\rm new}^3 \, , \label{gaugino}
\eeq
with the matching relation,
\beq
 \Lambda_{\rm new}^3 = \lmk \lambda  \Phi  \rmk^{2+\epsilon} 
 \Lambda'^{1-\epsilon} \, . \label{gaugino2}
\eeq
Here, $\Lambda'$ is defined as the holomorphic dynamical scale of the original theory with $Q_a, \bar{Q}_a$,
and $\Lambda_{\rm new}$ is that of the pure super-Yang-Mills theory.

To develop a vacuum with spontaneously broken scale invariance,
we require a marginally relevant deformation.
Let us introduce a supersymmetric mass term for $\Phi$,\footnote{Since $\Phi$ is a gauge singlet, other renormalizable operators are not forbidden, but
for simplicity, we focus on only the mass term with some discrete symmetry, e.g. $Z_2$, and it is extendable but will be studied elsewhere.}
\beq
W_\Phi = - M_\Phi \Phi^2 \, , 
\label{Rbreak} 
\eeq
where $M_\Phi$ is a mass parameter.
This mass term explicitly breaks $U(1)_R$ symmetry which is restored in the limit of $\epsilon\to 0$.
Note that the scaling dimension of the operator is only a small marginally relevant deformation for $|\epsilon| \ll 1$.
Through the anomalous dimension of $\Phi$ given below, one can calculate the scaling dimension of the operator $\Phi^2$ at the fixed point as $3-3\epsilon/2$.
The theory is still scale invariant for $\epsilon=0$, while $\epsilon\neq 0$ gives the relevant deformation of the scale invariance.

The dynamically generated superpotential \eqref{gaugino} with the relation~\eqref{gaugino2} and the mass term \eqref{Rbreak} gives rise to the $\mathcal{F}$-term potential,
\beq
V_{\mathcal{F}} (\Phi,Q_a,\bar{Q}_a) &=& 
Z_\Phi^{-1}\left| 2M_\Phi \Phi - \lmk2+\epsilon\rmk N \lambda \Lambda'^{1-\epsilon} \lmk \lambda \Phi \rmk^{1+\epsilon} \right|^2\nonumber\\
&+& Z_Q^{-1}|\lambda\Phi\bar{Q}_a|^2+Z_{\bar{Q}}^{-1}|\lambda\Phi Q_a|^2.
\label{VF}
\eeq
Here, we have used the same characters as the chiral superfields to denote the scalar components for $\Phi, Q_a, \bar{Q}_a$,
and $Z_\Phi, Z_Q, Z_{\bar{Q}}$ represent the wavefunction renormalization factors for $\Phi, Q_a,\bar{Q}_a$, respectively.
When $\epsilon \neq 0$ in the supersymmetric limit, the $\mathcal{F}$-term potential has two degenerate minima with $\langle Q_a \rangle = \langle \bar{Q}_a \rangle = 0$ and 
\beq
 \la \Phi \ra =0 \,  ,
\eeq
or
\beq
 \la \Phi \ra &=& \frac{\Lambda'}{\lambda } \lmk \frac{2}{2+\epsilon}\frac{M_\Phi}{N \lambda^2 \Lambda'} \rmk^{1/\epsilon}.
\label{flatdir}
\eeq
While the former vacuum corresponds to the symmetric phase, the latter does the broken phase.

K\"{a}hler potential terms are subject to the wavefunction renormalization. 
We find
\beq
Z_\Phi &=& \left(\frac{M_c}{\Lambda}\right)^{-\gamma_\Phi},\\[1ex]
Z_Q &=& \left(\frac{M_c}{\Lambda}\right)^{-\gamma_Q},
\label{wfrenorm}
\eeq
where $\gamma_\Phi \equiv (2-2\epsilon)/(2+\epsilon)$ is the anomalous dimension of $\Phi$ and $\gamma_Q \equiv (\epsilon-1)/(2+\epsilon)$ is that of $Q_a, \bar{Q}_a$ at the IR fixed point.
The theory enters into the conformal regime, where the beta functions of the gauge and Yukawa couplings almost vanish,
at a scale $\Lambda$, and exits at a scale $M_c$.
With initial conditions at a UV scale, the scale $\Lambda$ can be obtained by solving renormalization group (RG) equations of the $SU(N)$ gauge coupling $g$ and the Yukawa coupling $\lambda$. 
Taking the UV scale as the (reduced) Planck mass scale $\Mpl \equiv1/\sqrt{8\pi G}$ with the Newton constant $G$,
one typically obtains $\Lambda = \mathcal{O}(10^{-3}-10^{-1})\Mpl$ \cite{Nakai:2021nyf}.
We define the canonically normalized chiral superfields as
\beq
&& \hat{\Phi} = 
 \sqrt{Z_\Phi}\Phi \, ,
 \label{eq:hatPhi}
 \\[1ex]
&& \hat{Q}_a =  
 \sqrt{Z_Q} 
 Q_a \, ,
\quad
 \hat{\bar{Q}}_a =  
 \sqrt{Z_Q} 
 \bar{Q}_a \, .
\label{eq:hatQ}
\eeq
In our model, the theory exits from the conformal regime at the energy scale of $M_c \sim \lambda \langle \hat{\Phi} \rangle$.
Therefore, the vacuum condensate can be rewritten as
\beq
\langle\hat{\Phi}\rangle &\sim& \frac{M_c}{\lambda}\nonumber\\
&=& \lmk\frac{\Lambda}{\Lambda'}\rmk^{\frac{(2+\epsilon)(1-\epsilon)}{3\epsilon}} \frac{\Lambda}{\lambda} \lmk \frac{2}{2+\epsilon}\frac{M_\Phi}{N\lambda^2\Lambda} \rmk^{\frac{2+\epsilon}{3\epsilon}}.~~
\label{PTscale}
\eeq
Note again that while $\Lambda$ is defined as the conformal entering scale, $\Lambda'$ is the holomorphic dynamical scale.
The hierarchy between these two scales depends on the initial condition on the gauge coupling at a UV scale. Without fine-tuning, we can typically take $\Lambda/\Lambda' = \mathcal{O}(1-100)$. 
For the theory to enter into the conformal regime, we require $\lambda \langle \hat{\Phi} \rangle < \Lambda$, or
\beq
\frac{\Lambda}{\Lambda'} < \lmk\frac{2+\epsilon}{2}\frac{N\lambda^2\Lambda}{M_\Phi}\rmk^{1/(1-\epsilon)}.
\label{condition}
\eeq
The ratio $\Lambda/\Lambda' = \mathcal{O}(1-100)$ means $M_\Phi\lesssim\Lambda$, and thus 
Eq.~\eqref{PTscale} tells us that the large hierarchy $\langle\hat{\Phi}\rangle\ll \Lambda$ can be achieved.

We parameterize $\hat{\Phi}$ in terms of the radial mode $\phi$ and the phase mode $\sigma$ as
\beq
\hat{\Phi} = \frac{\phi}{\sqrt{2}} \exp\lmk\frac{i\sigma}{v}\rmk \, , 
\label{parameterize}
\eeq
with $v \equiv \langle \hat{\Phi} \rangle$.
The vacuum expectation values (VEVs) in the broken phase are then
$\langle\phi\rangle=\sqrt{2}v, \langle\sigma\rangle=0$.
Using this parameterization, we can expand the $\mathcal{F}$-term potential as
\beq
&& V_{\mathcal{F}}(\phi, \sigma) \nonumber   \\
&=& \lambda^2 \phi^4 \left|\lmk\frac{\lambda\phi}{\sqrt{2}\Lambda}\rmk^{-\frac{3\epsilon}{2+\epsilon}}\frac{M_\Phi}{\Lambda} - \frac{2+\epsilon}{2}N\lambda^2 
\lmk\frac{\Lambda'}{\Lambda}\rmk^{1-\epsilon} 
e^{\frac{i\epsilon\sigma}{v}}\right|^2\nonumber\\
&&+ \,\, |\lambda\hat{\Phi}\hat{\bar{Q}}_a|^2 + |\lambda\hat{\Phi} \hat{Q}_a|^2.
\label{VFexpand}
\eeq
Note that the wavefunction renormalization for $Q_a,\bar{Q}_a$ is canceled in the last two terms.
Since $\phi$ is canonically normalized, we obtain its mass at $\langle \phi \rangle =\sqrt{2}v$ as
\beq
m_\phi &\equiv& \left.\sqrt{\frac{\del^2 V_{\mathcal{F}}}{\del\phi^2}}\right|_{\langle\phi\rangle=\sqrt{2}v,\langle\sigma\rangle=0}\nonumber\\[1ex]
&=& \frac{3}{\sqrt{2}}
\lmk\frac{\Lambda'}{\Lambda}\rmk^{1-\epsilon}
\epsilon N\lambda^3 v \, .
\label{phimass}
\eeq
The mass for $\sigma$ is of the same order with $\phi$.
Note that these masses are more significantly suppressed for smaller $\epsilon$, which means that the potential is shallower around the broken phase.

Although we have found the degenerate vacua in the supersymmetric limit, the degeneracy is lifted by supersymmetry breaking effects.
Let us introduce a soft supersymmetry breaking term for our (canonically normalized) scalar field,\footnote{Additional soft terms can be implemented so that the phase mode $\sigma$ is stabilized at a nonzero value, e.g. $b\hat{\Phi}^2+{\rm h.c.}$ with $b$ a complex constant. In the present paper, we focus on the simplest case.} 
\beq
\mathcal{L}_{\rm soft} =  -V_{\rm soft}
= -m^2_{\Phi}|\hat{\Phi}|^2 \, .
\label{soft}
\eeq
To make the vacuum energy in the broken phase lower than that in the symmetric phase,
the sign of $m_\Phi^2$ must be negative.
This soft term also explicitly violates the scale invariance.
The total zero-temperature potential is given by $V_0 = V_{\mathcal{F}}+V_{\rm soft}$.

At $\phi/\sqrt{2}v\ll1$, the zero temperature potential is approximately given by
\beq
V_0(\phi) 
&=& \lambda^2 \phi^4 \left|\lmk\frac{\lambda\phi}{\sqrt{2}\Lambda}\rmk^{-\frac{3\epsilon}{2+\epsilon}}\frac{M_\Phi}{\Lambda} - \frac{2+\epsilon}{2}N\lambda^2 
\lmk\frac{\Lambda'}{\Lambda}\rmk^{1-\epsilon} \right|^2\nonumber\\
&+&\frac{1}{2}m_\Phi^2\phi^2\nonumber\\
&\simeq&  \lmk\frac{2+\epsilon}{2}\rmk^2 N^2\lambda^6 
\lmk\frac{\Lambda'}{\Lambda}\rmk^{2-2\epsilon}\phi^4 + \frac{1}{2}m_\Phi^2\phi^2.
\eeq
Solving $V'_0(\phi)=0$, we obtain the VEV of the false vacuum,
\beq
\frac{\phi_f}{\sqrt{2}v} \simeq \frac{3}{4}\frac{2}{2+\epsilon}\epsilon\frac{\sqrt{-m_\Phi^2}}{m_\phi} \, .
\label{phif}
\eeq

\begin{figure*}[t!]
\begin{minipage}[t]{16.5cm}
\includegraphics[width=7.7cm]{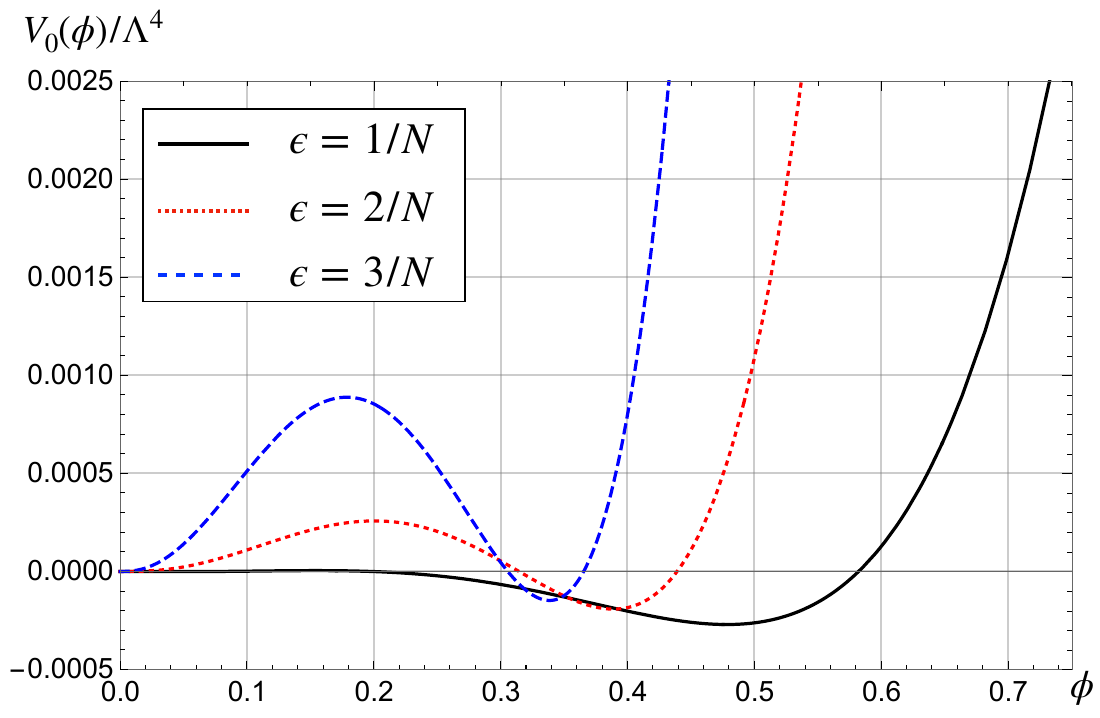}
\hspace{5mm}
\includegraphics[width=7.9cm]{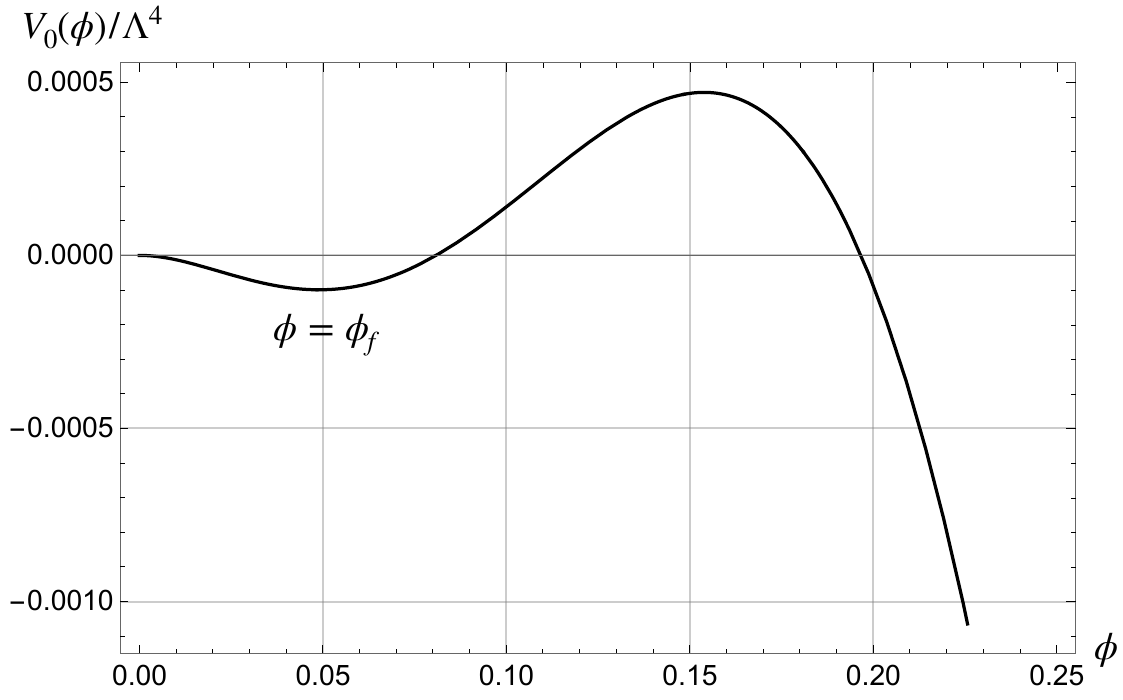}
\end{minipage}
\caption{
The shape of the potential $V_0=V_{\mathcal{F}}+V_{\rm soft}$ at $T=0$, normalized by $\Lambda^4$ as a function of $\phi$ in unit of GeV.
We take $N=5$, $\lambda=1$, $M_\Phi/\Lambda=1$, $\Lambda/\Lambda'=5$, $\Lambda/\sqrt{|m_\Phi^2|}=14$, and $m_\Phi^2=-(0.1\GeV)^2$ for better visibility.
In the left panel, the black solid, red dotted, and blue dashed lines denote the cases of $\epsilon = 1/N, 2/N, 3/N$, respectively.
In the right panel, we show the shape for $\epsilon=1/N$, zooming up the part around the origin, where a nonzero VEV $(\phi_f\neq0)$ in the symmetric phase is induced.
}
\label{fig:potential}
\end{figure*}

\FIG{fig:potential} shows the shape of the $V_0$ for $\phi$ at zero temperature.
Here, we take $N=5$, $\lambda=1$, $M_\Phi/\Lambda=1$, $\Lambda/\Lambda'=5$, $\Lambda/\sqrt{|m_\Phi^2|}=14$, and $m_\Phi^2=-(0.1\GeV)^2$.
This choice is only for better visibility, e.g. $\lambda$ would depend on $N$ and $\epsilon$, and $\Lambda/\sqrt{|m_\Phi^2|}$ would be much larger.
In the left panel, the black solid, red dotted, and blue dashed lines correspond to the cases of $\epsilon = 1/N, 2/N, 3/N$, respectively. 
Since the suppression by anomalous dimension is more significant for smaller $\epsilon$, the potential barrier height is given hierarchically and the true vacuum moves farther away, i.e. the potential is shallower for smaller $\epsilon$ as explained around (\ref{phimass}).
In the right panel, we show the potential shape for $\epsilon=1/N$ with the same parameter choice,
zooming up the part around the origin.
One can see that the VEV of the false vacuum $\phi_f$ has a finite value for a very shallow potential due to the soft mass term, which would move to the origin in the supersymmetric limit.

Let us compare the potential for $\phi$ of Eq.~\eqref{VFexpand} in our model with
that of the Goldberger-Wise mechanism for radion stabilization.
In the latter case, the potential for the radion (or dilaton in the dual CFT picture) $\mu$ is restricted to the following form,
\beq
V_{\rm GW}(\mu)=\mu^4P\left[\lmk\frac{\mu}{\mu_0}\rmk^\varepsilon\right].\label{eq:dilaton eft}
\eeq
Here, $\mu_0$ is a typical scale, $\varepsilon$ represents the violation of the scale invariance,
and $P(x)$ is a slowly-varying function, $P'(x)/P(x)\ll 1$. 
One can see that our potential \eqref{VFexpand} has the same form as Eq.~\eqref{eq:dilaton eft}
due to the marginally relevant deformation.
Nevertheless, when we discuss the phase transition with finite temperature effects,
we will find a notable difference between these two models.

%%%%%%
%%%%%%%%%%%%%%%%%%%%%%%%%%%%%%%%%%%%%%%%%%%%%%%%%%
\section{Phase transition
\label{sec:PT}}

We now consider the finite temperature system to discuss the transition from the symmetric phase to the broken phase.

\subsection{Thermal effects}

There have been attempts to understand the dynamics of the conformal phase transition in terms of 
the dilaton effective field theory whose potential is of the form given in Eq.~\eqref{eq:dilaton eft}.
However, in the high-temperature limit, one cannot use such an effective field theory.
Instead, thermal effects can be captured by the holographic approach~\cite{Creminelli:2001th}
where the phase transition dynamics is believed to be understood as the transition from the AdS black-brane solution to the RS spacetime.
Since we have an explicit UV complete model,
the phase transition dynamics can be analyzed without relying on holography in principle.
One fascinating point of our model is that there exists a candidate of the well-defined local order parameter $\langle \hat{\Phi} \rangle$ which distinguishes two phases.

Before going to a detailed discussion, we give a crude argument in the following.
For $\langle \hat{\Phi} \rangle =0$, the theory possesses an approximate scale invariance, so that the mass gap is almost closed.
Hence, there are many excitations that contribute to the free energy at finite temperature.
This simple consideration implies that the free energy in the limit of large $N\sim N_F \gg 1$ behaves as
\begin{align}
    F(\langle \hat{\Phi} \rangle = 0) = -c_1N^2T^4 \, , \label{eq:free energy gapless}
\end{align}
where $c_1$ is a positive numerical constant.
The $N^2$ dependence comes from the fact that we consider the $SU(N)$ gauge theory, and contributions from other fields that are singlet under $SU(N)$ are neglected.
In the holographic approach, $c_1=\pi^2/8$ by computing the partition function of the AdS black brane solution
with saddle-point approximation~\cite{Creminelli:2001th}.

In the opposite limit of $\langle \hat{\Phi} \rangle \gg T$, the scalar condensate generates masses for (s)quarks.
Furthermore, because of confinement of the pure super-Yang-Mills theory, we do not expect any light states other than the dilaton which comes from the condensate of the singlet $\Phi$.
Hence, we can approximate the free energy dependence on $\langle \hat{\Phi} \rangle$ as the zero-temperature one,
\begin{align}
    F(\langle \hat{\Phi}\rangle)\simeq V_{\mathcal{F}}+V_{\rm soft}  \, , \label{eq: potential zero temperature limit}
\end{align}
for $\langle \hat{\Phi} \rangle/T\gg 1$.

To clarify the free energy dependence on $\langle \hat{\Phi} \rangle$ for the intermediate region $0<\langle\hat{\Phi} \rangle/T\lesssim 1 $ may require a non-perturbative study because there could be contributions from bound states rather than elementary excitations.
In the 5D dual picture, this can be understood as thermal excitations of Kaluza-Klein modes.
In fact, a perturbative computation of the free energy around this region is not under control in our setup and suffers from higher-order effects (See refs.~\cite{Weinberg:1974hy,Linde:1980ts,Arnold:1992rz,Glioti:2018roy} for the condition of the perturbativity).
Although there exist theoretical uncertainties in the free energy dependence on $\langle\hat{\Phi} \rangle$ for this region,
we perform the mean-field approximation for $\Phi$ and compute the thermal effective potential at one-loop order using the standard imaginary time formulation~\cite{Dolan:1973qd,Quiros:1999jp} as a first attempt.
Therefore our computation of the free energy should be interpreted as the extrapolation from the condensed region, $\langle\hat{\Phi}\rangle/T\gg 1$.

As discussed in the previous section, the tree-level potential at zero temperature is given by the $\mathcal{F}$-term potential and the soft mass term,
\beq
V_0 (\phi) = V_{\mathcal{F}}+V_{\rm soft} \, .
\eeq
The broken phase is energetically favored by the soft mass term, which shows that the global minimum is realized
at a nonzero expectation value of $\phi$.

One can compute the thermal effective potential at the one-loop order~\cite{Dolan:1973qd,Quiros:1999jp},
\beq
&&V_{T} (\phi, T)\nonumber\\[1ex]
&&=\frac{T^4}{2\pi^2}\sum_s (-1)^{2s} 4NN_F \nonumber\\
&&\times \int_0^{+\infty}dq q^2 \mathrm{ln} \left[1 + (-1)^{2s+1} \mathrm{exp} \lmk -\sqrt{q^2 + \frac{\lambda^2 \phi^2}{2T^2}}\rmk\right],\nonumber\\
\label{VT}
\eeq
with $s=0,1/2$ for bosonic and fermionic fields, respectively.
For a small condensate $\langle \phi\rangle\simeq 0$, one obtains the free energy dependence of
Eq.~\eqref{eq:free energy gapless} while here
the precise numerical value of $c_1$ suffers from non-perturbative effects.
In the limit of $\langle \hat{\Phi} \rangle/T\gg 1$, we recover the expression~\eqref{eq: potential zero temperature limit}
because of the Boltzmann factor.

\begin{figure}[t!]
\centering
\includegraphics[width=8.5cm]{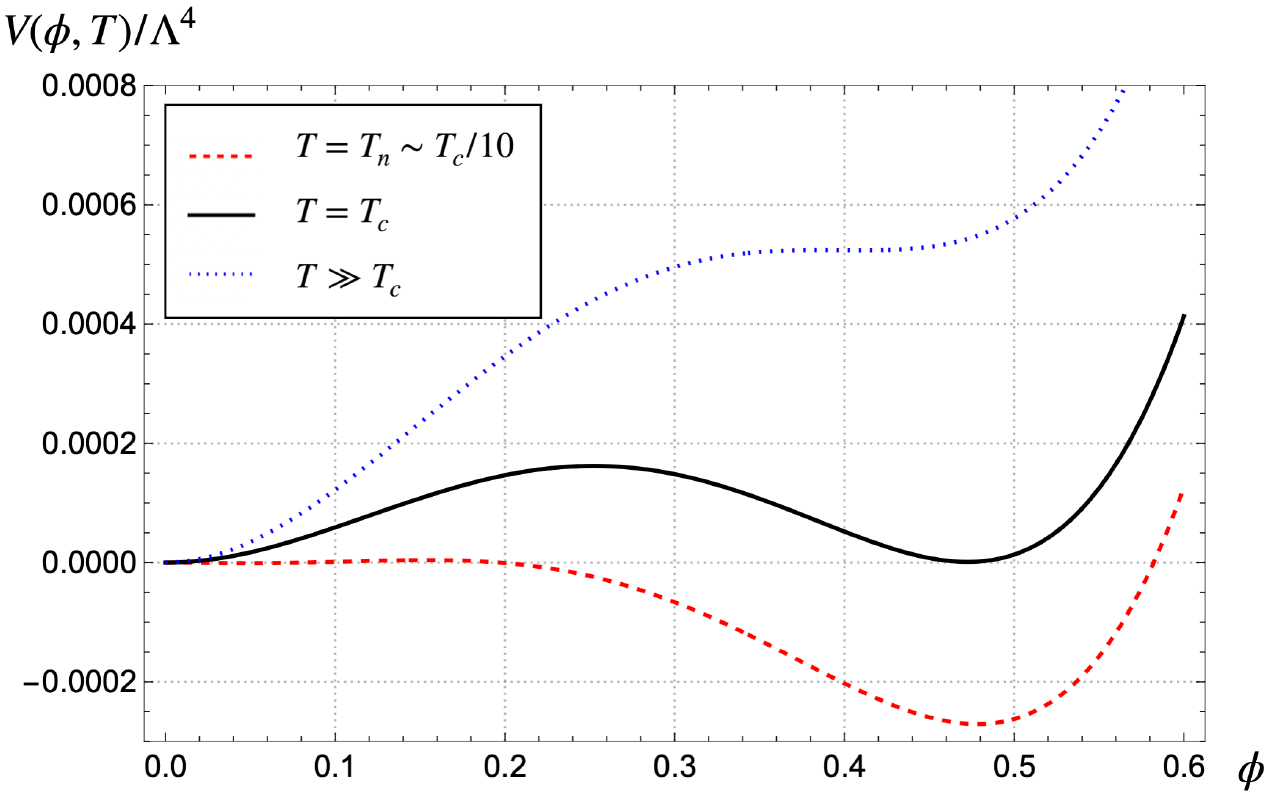}
\caption{The shape of the total potential $V(\phi,T) = V_0 + V_{T}$ normalized by $\Lambda^4$ as a function of $\phi$ in unit of GeV with the same parameter set as \FIG{fig:potential} for $\epsilon=1/N$, and different temperatures are chosen to illustrate the behavior of thermal corrections, where we have normalized the potential value at the origin. The black solid line denotes the potential at $T=T_c$, while the red dashed and blue dotted lines show the potential shapes at the approximate nucleation temperature $T_n\sim T_c/10$ and a high temperature $T\gg T_c$, respectively.
}
\label{fig:thermalpotential}
\end{figure}

We now obtain the total effective potential as
\beq
V (\phi, T) = V_0 (\phi) + V_T (\phi, T) \, .
\eeq
\FIG{fig:thermalpotential} shows the temperature dependence of the total effective potential.
Here we set the vacuum energy density by the condition $V_T(\phi=0,T)=0$.
One can see from the figure that $\langle \hat{\Phi}\rangle\simeq 0$ is favored in the high-temperature phase, while there exists the metastable local minimum at $\langle\hat{\Phi}\rangle\sim v$.
As the temperature cools, two phases coexist at a certain temperature $T_c$, which is defined as the critical temperature (corresponding to the black solid curve in Fig.~\ref{fig:thermalpotential}).
For $T<T_c$, the broken phase $\langle \hat{\Phi} \rangle\gg T$ is energetically favorable.
Since there exists the potential barrier between two minima, the phase transition is of the first order.

\subsection{Critical temperature}

Let us evaluate the critical temperature $T_c$ of the phase transition.
At $T=T_c$, our system satisfies the following approximate relation:
\beq
V_{\rm soft}(\phi_{\rm min}) + V_T(\phi_{\rm min},T_c)\simeq 0 \, ,
\label{Tccond}
\eeq
with the temporal global minimum, $\phi_{\rm min} \sim \sqrt{2} v$.
We assume $-m_{\Phi}^2 < v^2$, so that there exists a potential barrier between two minima at zero temperature,
and the scale of temperature is roughly estimated as $T_c^4\sim -m_\Phi^2(\lambda v)^2$.
{Hence around $\phi\simeq v$ and $T\simeq T_c$, one can perform the low-temperature approximation $\lambda \phi\gg T$ of the integrated function defined in Eq.~\eqref{VT}.
The thermal effective potential is then approximated as}
\begin{equation}
\begin{split}
V_T(\phi,T) \simeq \frac{4(2+\epsilon)N^2}{\pi^2}T^4 &\left[2-\frac{\lambda^2\phi^2}{2T^2}K_2 \lmk\frac{\lambda\phi}{\sqrt{2}T}\rmk\right] \\[1ex]
&\qquad \qquad \quad (\lambda\phi \gg T) \, ,
\end{split}
\end{equation}
where $K_2$ is the second modified Bessel function and a factor $2$ in the parenthesis is a normalization factor so that $V_T(\phi=0,T)=0$.
Using this approximate formula and the relation (\ref{Tccond}),
we find
\beq
T_c \simeq \frac{\sqrt{\pi}}{2}\lmk\frac{2}{2+\epsilon}\rmk^{1/4}\frac{1}{\sqrt{N}}(-m_\Phi^2)^{1/4}\sqrt{v}\label{analyTc} \, .
\eeq
Here, we have approximately used the fact that the Bessel function $X^2K_2(X)$ asymptotes to $0$ at a large $X$ or $T_c<\lambda v$.

\begin{figure}[t!]
\centering
\includegraphics[width=7.8cm]{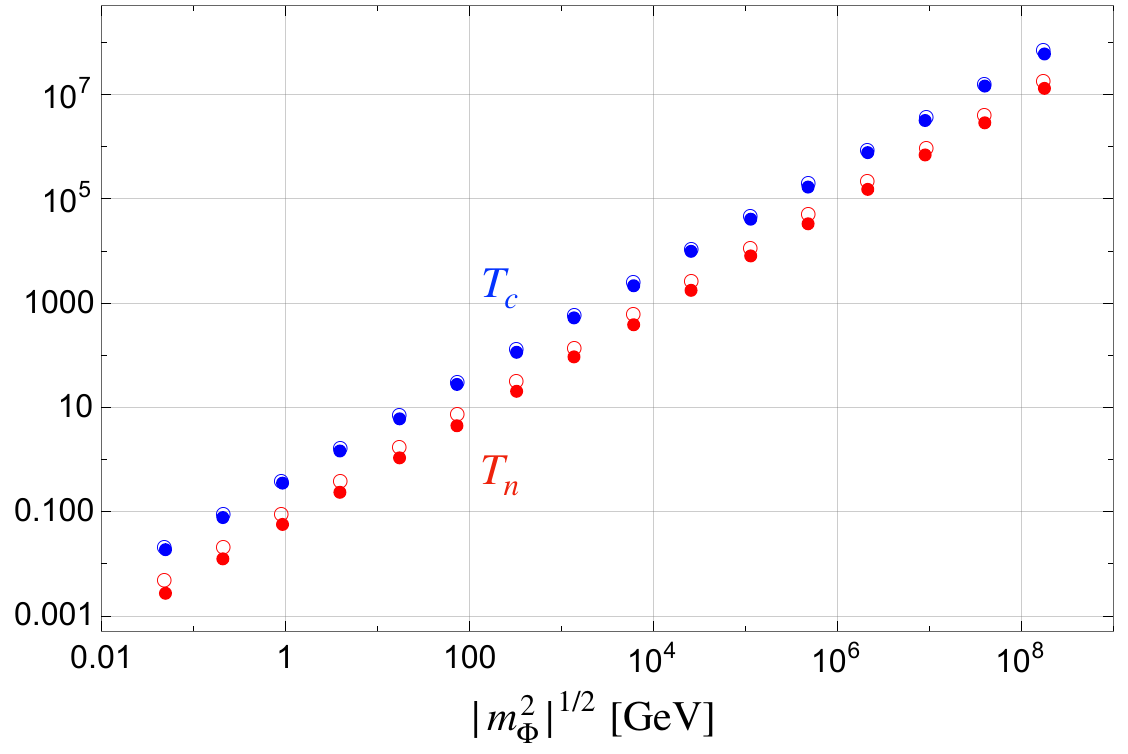}
\hspace{5mm}
\caption{
The critical temperature $T_c$ and the nucleation temperature $T_n$ as a function of $|m_\Phi^2|^{1/2}$ in unit of GeV.
Here we take $N=9, \epsilon=1/N, \lambda=\lambda_{*(1)}, \Lambda/\Lambda'=1$, $\Lambda=10^{17}\GeV$, and
$M_\Phi/\Lambda$ and $m_\Phi^2$ are fixed so that $S_4^{\rm th} - S_{4,0} \simeq 10$.
The blue (red) circles and bullets represent the numerical and analytical results for $T_c ~(T_n)$, respectively.
}
\label{fig:TcTn}
\end{figure}

\FIG{fig:TcTn} shows the critical temperature $T_c$ as a function of $|m_\Phi^2|^{1/2}$ in unit of GeV.
The blue circle and bullet represent the numerical result and the approximate analytical estimate (\ref{analyTc}),
respectively.
We set $N=9, \epsilon=1/N, \Lambda/\Lambda'=1$, and $\Lambda=10^{17}\GeV$, and use the fixed point value at the one-loop level for $\lambda$,\footnote{Although the two-loop contribution gives a correction of $\mathcal{O}(0.1)$ \cite{Nakai:2021nyf}, the one-loop calculation would be sufficient for our purposes.}
\beq
\lambda_{*(1)}^{2} \simeq 16\pi^2\frac{\gamma_\Phi}{NN_F} \, .
\eeq
The value of $M_\Phi/\Lambda$ and $m_\Phi^2$ are numerically determined by the zero temperature bounce action $S_4^{\rm th}-S_{4,0}\simeq10$,\footnote{We have seven parameters, $N, \epsilon, \lambda$, $\Lambda/\Lambda', M_\Phi/\Lambda$, $\Lambda$, and $m_\Phi^2$.
Since $\lambda$ is determined at the fixed point, one can find that the number of independent parameters is five.}
where $S_4^{\rm th}$ is the threshold value of the bounce action and will be given in \EQ{nuclcond}.
The values of $M_\Phi/\Lambda$ are in the range of $10^{-2} - 10^{-1/2}$.
We can see that the analytical results are consistent with the numerical results with high accuracy.

\subsection{Bounce action}

\begin{figure*}[t!]
\begin{minipage}[t]{16.5cm}
\includegraphics[width=7.9cm]{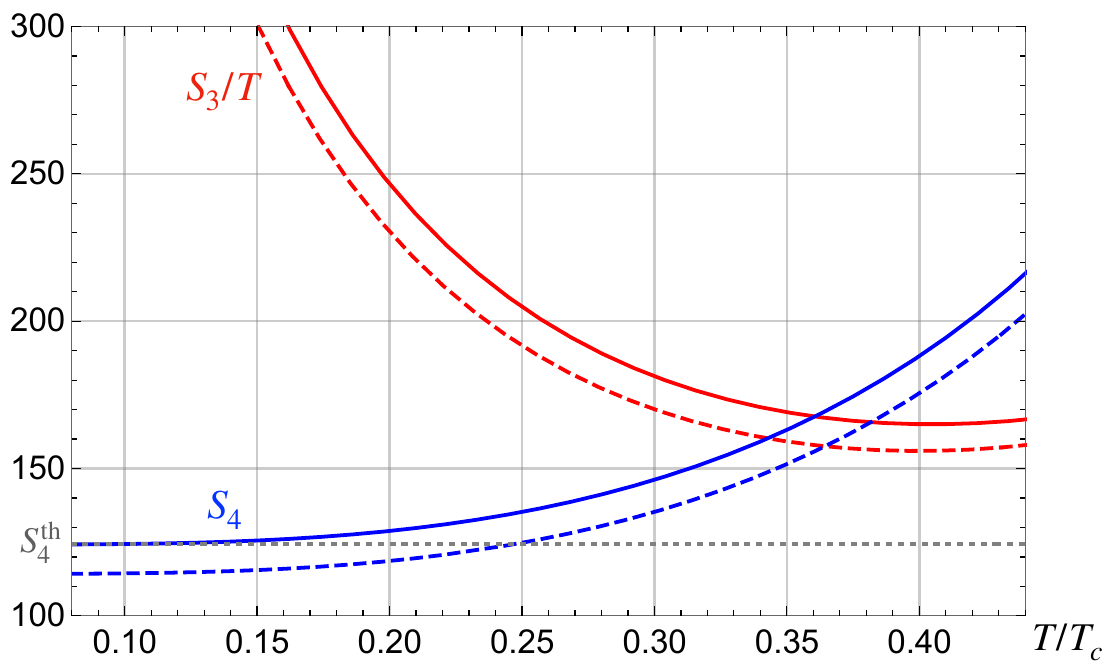}
\hspace{5mm}
\includegraphics[width=7.9cm]{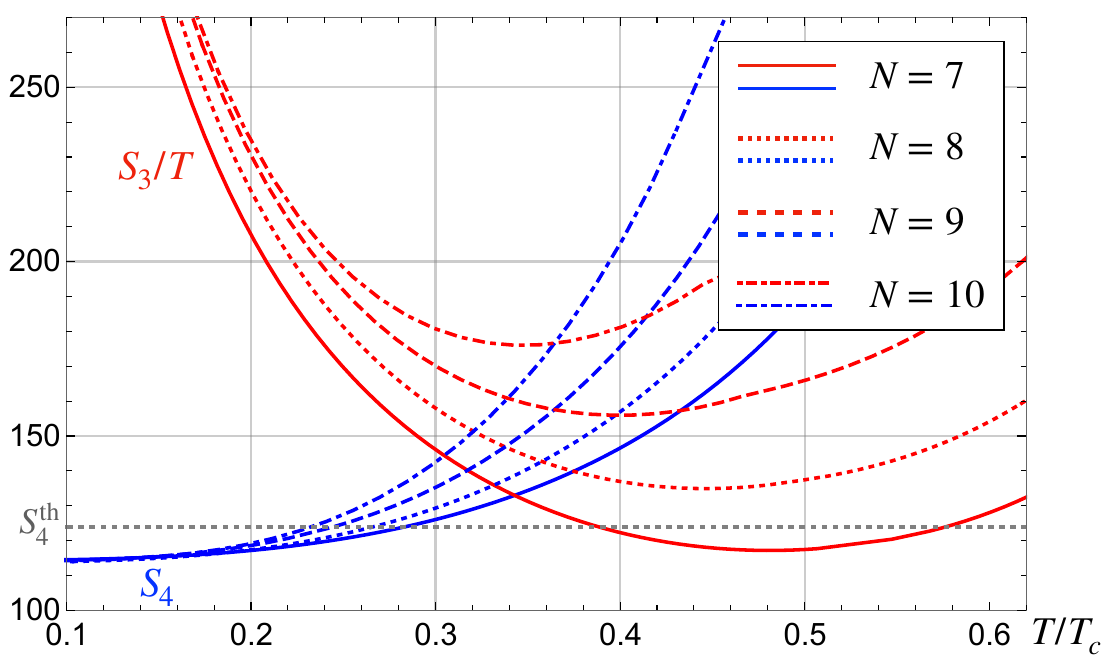}
\end{minipage}
\caption{
The bounce actions as a function of $T/T_c$.
Left: The red (blue) solid and dashed lines denote the $O(3)~(O(4))$ symmetric solutions for $N=9$ and $S_4^{\rm th}-S_{4,0}=0.1, 10$, respectively. 
Right: The red (blue) solid, dotted, dashed, dot-dashed lines respectively corresponds to the $O(3)~(O(4))$ symmetric solutions for $N=7,8,9,10$ and $S_4^{\rm th}-S_{4,0}=10$.
The other parameters in both panels are set to $\epsilon=1/N$, $\lambda=\lambda_{*(1)}$, $\Lambda=10^{17}\GeV$, $\Lambda=\Lambda'$, and the values of $M_\Phi/\Lambda$ are chosen so that $T_c\sim10^4\GeV$.
The horizontal gray dotted line represents the threshold value $S_4^{\rm th}$.
}
\label{fig:S3}
\end{figure*}

The first-order phase transition proceeds via the nucleation of bubbles which can be interpreted as the decay of the metastable state.
The decay rate of metastable state can be evaluated using a semi-classical approximation~\cite{Coleman:1985rnk}.
With an analytic continuation $t=-i\tau$ where $\tau$ is the Euclidean time, the Euclidean action is given by
\beq
S = \int d^4x \left[\frac{1}{2}(\del_\mu\phi)^2 +V(\phi, T)\right] .
\eeq
Here, $\langle\sigma\rangle=0$ has been taken.
Since the system is in thermal equilibrium, the Euclidean time is periodic with a period $T^{-1}$.
Regarding this effect, the Euclidean action is roughly given by~\cite{Linde:1981zj}
\beq
S = {\rm min}\left\{ S_4(T),~\frac{S_3(T)}{T} \right\},
\eeq
where $S_{4(3)}$ is the bounce action for the ${O}(4)~({O}(3))$ symmetric solution.
In \FIG{fig:S3}, we show the numerical estimates of the time dependence of the $O(3)$ and $O(4)$ bounce actions. 
In the left panel, we show the result for $N=9$ and $S_4^{\rm th}-S_{4,0}=0.1,10$ denoted by the solid and dashed lines.
Here we set $\epsilon=1/N$, $\lambda=\lambda_{*(1)}$, $\Lambda=10^{17}\GeV$, $\Lambda=\Lambda'$, and the values of $M_\Phi/\Lambda$ are chosen so that $T_c\sim10^4\GeV$.
Although there is a mild dependence on $S_4^{\rm th}-S_{4,0}$, the $O(4)$ bounce action dominates the nucleation for $N=9$, as long as $S_4^{\rm th}-S_{4,0}\lesssim\mathcal{O}(10)$.
In addition, we show the results for $N=7,8,9,10$ and $S_4^{\rm th}-S_{4,0}=10$ in the right panel, with the same other parameters as the left panel.
A smaller $N$ makes the $O(3)$ action more dominant.
This is because, when the phase transition strength is so strong, the decay of metastable state is almost induced by quantum fluctuation, and hence $O(4)$ symmetric Euclidean action is favored. 
We focus on this case which can be conservatively achieved for $N\geq9$.\footnote{It is possible to obtain mildly weak supercooling for $N\lesssim8$.
For example, as can be seen from the right panel of \FIG{fig:S3}, $T_n/T_c\sim 0.6$ is realized for $N=7$.}

Since the solution we focus on is $O(4)$ symmetric~\cite{Coleman:1977th}, we use the radial coordinate $r\equiv \sqrt{x_\mu^2}$
to find the spherically symmetric solution,
\beq
S_4 = \int dr \, 2\pi^2r^3\left[\frac{1}{2}\lmk\frac{d\phi(r)}{dr}\rmk^2+V(\phi(r), T)\right].
\eeq
The classical saddle-point solution of the above action is called a bounce which satisfies the following equation of motion,
\beq
\frac{d^2}{dr^2}\phi(r) +\frac{3}{r}\frac{d}{dr}\phi(r) = V'(\phi, T) \, .
\eeq
The boundary condition is given by $\phi(r=\infty)=0$ and $d\phi/dr|_{r=0}=0$.

\begin{figure}[t!]
\centering
\includegraphics[width=8cm]{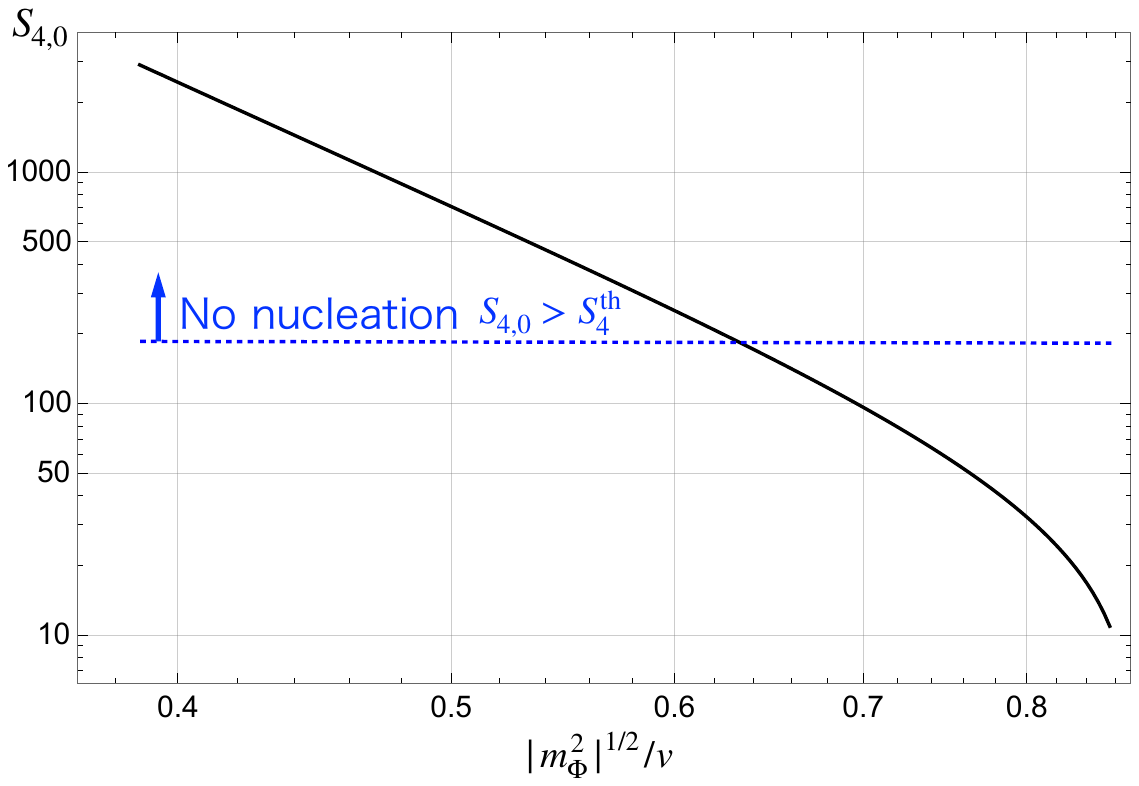}
\caption{The ${O}(4)$ symmetric bounce solution at $T=0$ as a function of $|m_\Phi^2|^{1/2}/v$. 
We set $N=9$, $\lambda=\lambda_{*(1)}$, $M_\Phi/\Lambda=0.01$, $\Lambda/\Lambda'=1$, and $\Lambda=10^{17}\GeV$ (or $v\simeq 77\MeV$). 
The black solid curve represents the numerical estimate of the  bounce action, while the blue dotted line denotes the threshold value $S_4^{\rm th}$. 
}
\label{fig:bounce}
\end{figure}

Using the standard over/under shooting algorithm, we compute the bounce action at $T=0$, denoted as $S_{4,0}$, which is plotted as a function of $|m_\Phi^2|^{1/2}/v$
in \FIG{fig:bounce}. 
The bounce action is represented by the black solid curve,
while the blue dashed line denotes the threshold value at which bubble nucleation can occur,
as will be explained in the next subsection.
In the region of $|m_\Phi^2|^{1/2}/v\gtrsim0.86$, the potential barrier disappears, which implies that there is no bounce configuration.
One can see from this figure that the soft mass scale should be close to the VEV otherwise the bubble nucleation cannot take place.
This implies that conformal invariance should be sufficiently violated such that the completion of the phase transition is ensured.

In addition to numerical calculations, an approximate analytical method can be used to estimate the bounce action.
It turns out that the bounce configuration is thick enough to use a thick wall approximation. 
In the thick wall approximation, the bounce action is given by  \cite{Nardini:2007me}
\beq
S_4 \simeq \frac{\pi^2}{2}\frac{|\phi_f-\phi_t|^4}{V(\phi_f,T)-V(\phi_t,T)} \, ,
\label{analy}
\eeq
where $\phi_f$ and $\phi_t$ are respectively defined as the field values at the false state (see also the right panel of \FIG{fig:potential}) and the tunneling point.
The tunneling point is determined by the saddle point condition, $\del S_4/\del\phi_t=0$. 
For a thicker wall, the tunneling point is closer to the false state.

\subsection{Nucleation temperature}

Using the bounce action, let us now evaluate the transition rate per unit volume and time as $\Gamma\sim A e^{-S_4}$,
where the prefactor is estimated by the dimensional analysis $A\simeq m_\phi^4$ for simplicity.
Since the vacuum energy density dominates over the radiation energy density below the critical temperature,
the Universe undergoes short period of inflation, where the Hubble parameter is given by $H^2\simeq (V_0(\phi_f)-V_0(\phi_{\rm min}))/3\Mpl^2$ by the completion of the phase transition.
A single bubble can be nucleated in the Hubble volume $H^{-3}$ at the time when $\Gamma H^{-4}\gtrsim 1$ is satisfied.
This condition can be translated into
\beq
S_4 &\lesssim& 4\log\lmk\frac{\Mpl m_\phi}{\sqrt{V_0(\phi_f)-V_0(\phi_{\rm min})}}\rmk
\label{nuclcond}\\[1ex]
&\simeq& 4\log\left[\lmk\frac{\Lambda}{\Lambda'}\rmk^{1-\epsilon}\frac{\Mpl}{\sqrt{-m_\Phi^2}}\right].
\eeq
The nucleation temperature $T_n$ is defined as the temperature when the inequality is saturated.
The bounce action at that threshold is represented by $S_{4}^{\rm th}$.
Here, the vacuum energy is approximated for $\phi_{\rm min}\simeq\sqrt{2}v$.
For example, we find $S_{4}^{\rm th}\simeq 175$ for $m_\Phi^2=-(0.1\GeV)^2$ and $S_{4}^{\rm th}\simeq 140$ for $m_\Phi^2=-(1\TeV)^2$.

In the parameter region where the thermal potential is subdominant compared to the zero-temperature one, an analytic computation of $T_n$ is possible under the thick wall approximation.
Using the thick-wall approximation, one obtains
\beq
S_4 &\simeq& \frac{\pi^2}{2}\frac{(\phi_f-\phi_t)^4}{V_0(\phi_f)-V_0(\phi_t)}\left[1+\frac{V_T(\phi_f, T)-V_T(\phi_t, T)}{V_0(\phi_f)-V_0(\phi_t)}\right]^{-1}\nonumber\\[1ex]
&\simeq& S_{4,0}\left[1-\frac{V_T(\phi_f, T)-V_T(\phi_t, T)}{V_0(\phi_f)-V_0(\phi_t)}\right].
\eeq
Here, the zero-temperature bounce action $S_{4,0}$ 
must be less than $S_4^{\rm th}$ to complete the nucleation.
In the second line, we assume that the finite temperature effect around the false state and the tunneling point is sufficiently suppressed so that the tunneling point $\phi_t$ can be safely approximated by the effective zero temperature potential.

Consequently, at $T=T_n$, the bounce action satisfies the following relation,
\beq
-\frac{V_T(\phi_f,T_n)-V_T(\phi_t,T_n)}{V_0(\phi_f)-V_0(\phi_t)}=\frac{S_4^{\rm th}-S_{4,0}}{S_{4,0}} \, .
\eeq
As long as the bubble wall is sufficiently thick, we have $\lambda\phi_{f(t)}/\sqrt{2}T_n\gg1$
($\phi_f$ and $\phi_t$ are typically $\mathcal{O}(0.1)\cdot\sqrt{2}v$),
which justifies the low temperature approximation.
Recall that $\phi_f\neq 0$ otherwise the bubble nucleation cannot take place (See FIG.~\ref{fig:bounce}).
Using $\lambda\phi_{f(t)}\gg T_n$, the difference in thermal potential can be approximated as
\beq
&&V_T(\phi_f,T_n)-V_T(\phi_t,T_n)\nonumber\\[1ex]
&\simeq& \frac{4(2+\epsilon)N^2}{\pi^2}T^4\nonumber\\
&\times& \left[-\frac{\lambda^2\phi_f^2}{2T_n^2}K_2\lmk\frac{\lambda \phi_f}{\sqrt{2}T_n}\rmk+\frac{\lambda^2\phi_t^2}{2T_n^2}K_2\lmk\frac{\lambda \phi_t}{\sqrt{2}T_n}\rmk\right].~~
\label{BesselApp}
\eeq
Since $\phi_t\sim\phi_f$ in most of the parameter region where the bubble nucleation takes place, we can approximate the finite difference by a derivative,\footnote{Even if $\phi_t\sim\phi_f$, we find that $\lambda(\phi_t-\phi_f)/\sqrt{2}T_n$ cannot be so small. 
Although this approximation may lose some quantitative information, it is a better prescription compared to the difficulty of calculating $\phi_t$ analytically which reduces the computational cost significantly.}
\beq
&&-\frac{\lambda^2\phi_f^2}{2T_n^2}K_2\lmk\frac{\lambda \phi_f}{\sqrt{2}T_n}\rmk+\frac{\lambda^2\phi_t^2}{2T_n^2}K_2\lmk\frac{\lambda \phi_t}{\sqrt{2}T_n}\rmk\nonumber\\[1ex]
&\simeq& -\lmk\frac{\lambda \phi_f}{\sqrt{2}T_n}\rmk^2 K_1\lmk\frac{\lambda \phi_f}{\sqrt{2}T_n}\rmk\cdot\lmk\frac{\lambda \phi_t}{\sqrt{2}T_n}-\frac{\lambda \phi_f}{\sqrt{2}T_n}\rmk.\nonumber\\
\eeq
Here, we have used $(X^\nu K_\nu(X))' = -X^\nu K_{\nu-1}(X)$ with $\nu$ an integer.
Consequently nucleation temperature can be expressed by the following simple form,
\beq
T_n &\simeq& \frac{\pi}{2}\sqrt{\frac{S_4^{\rm th}-S_{4,0}}{S_{4,0}}}\sqrt{\frac{2}{2+\epsilon}}\frac{\sqrt{-m_\Phi^2}}{N\lambda} \nonumber\\
&\times&\left[\frac{\lambda \phi_f}{\sqrt{2}T_n} K_1\lmk\frac{\lambda \phi_f}{\sqrt{2}T_n}\rmk\right]^{-1/2}.
\label{Tn}
\eeq

Using \EQ{Tn} and \EQ{phif}, the semi-analytic result of the nucleation temperature $T_n$ is plotted as the red bullet with the corresponding numerical one (the red circle) in \FIG{fig:TcTn}.
We can see that the overall behavior is consistent between the semi-analytical and numerical results.
The slight deviation comes from the approximation (\ref{BesselApp}) and the uncertainty in the calculation of $\phi_f$.

We now estimate the ratio of $T_n$ and $T_c$.
Using \EQ{analyTc} and \EQ{Tn}, it is given by
\beq
\frac{T_n}{T_c} &\simeq& \frac{\sqrt{\pi}}{\sqrt{N}\lambda} \sqrt{\frac{S_4^{\rm th}-S_{4,0}}{S_{4,0}}}\lmk\frac{2}{2+\epsilon}\rmk^{1/4} \sqrt{\frac{\sqrt{-m_\Phi^2}}{v}}\nonumber\\
&\times&\left[\frac{\lambda\phi_f}{\sqrt{2}T_n}K_1\lmk\frac{\lambda\phi_f}{\sqrt{2}T_n}\rmk\right]^{-1/2}.
\label{TcTndependence}
\eeq
Note that $T_n/T_c\to 0$ in the limit of $m_\Phi \to 0$ or $\epsilon\to 0$.

%%%%%%
%%%%%%%%%%%%%%%%%%%%%%%%%%%%%%%%%%%%%%%%%%%%%%%%%%
\section{Cosmological implications}
\label{sec:cosmo}

In this section, we study the implications of our strong first-order phase transition, in particular the estimation
of the GW spectrum generated by the phase transition.

\subsection{Entropy production}

For $T\lesssim T_c$,
the vacuum energy density dominates over the radiation energy density,
and the universe experiences a short period of the inflation.
Equating $\rho_{\rm vac}\simeq \rho_{\rm rad}$, we obtain the temperature of the onset of supercooling (mini-inflation) as
\beq
T_{\rm inf} \simeq \lmk\frac{-m_\Phi^2v^2}{-m_\Phi^2v^2+\rho_{\rm rad}(T_c)}\rmk^{1/4}T_c \, ,
\eeq
where $\rho_{\rm rad}(T)=(\pi^2/30)g_*T^4$ with $g_*$ the number of relativistic degrees of freedom for energy density,
and we have assumed that the vacuum energy density $\rho_{\rm vac}$ is approximated by the soft mass and the thermal potential. 
Since $\rho_{\rm rad}(T_c)\sim -m_\Phi^2v^2$ up to an $\mathcal{O}(1)$ factor, we find $T_{\rm inf}\sim T_c$.
The duration of supercooling is parameterized by the number of e-folds,
\beq
N_e \simeq \log\lmk\frac{T_c}{T_n}\rmk.
\eeq
The ratio $T_c/T_n$ is typically of $\mathcal{O} (10)$ in our model,
and the inflation lasts in a short time, $N_e\lesssim 4$.

Assuming that the whole vacuum energy is converted to the radiation instantaneously after the phase transition,
the reheating temperature is given by $T_{\rm reh}\simeq T_c$.
The dilution factor by entropy production is then estimated as 
\beq
\frac{s(T_n)}{s(T_{\rm reh})} \simeq \frac{g_{*s}(T_n)}{g_{*s}(T_c)}\lmk\frac{T_n}{T_c}\rmk^3,
\eeq
where $g_{*s}$ denotes the number of relativistic degrees of freedom for entropy.
With $T_c/T_n$ of $\mathcal{O} (10)$ in our model, any comoving quantities existing before the phase transition,
such as dark matter density and baryon asymmetry, are diluted by
$\lesssim\mathcal{O}(10^{-3})$.

\subsection{Characteristic parameters}

\begin{figure}[t!]
\centering
\includegraphics[width=8cm]{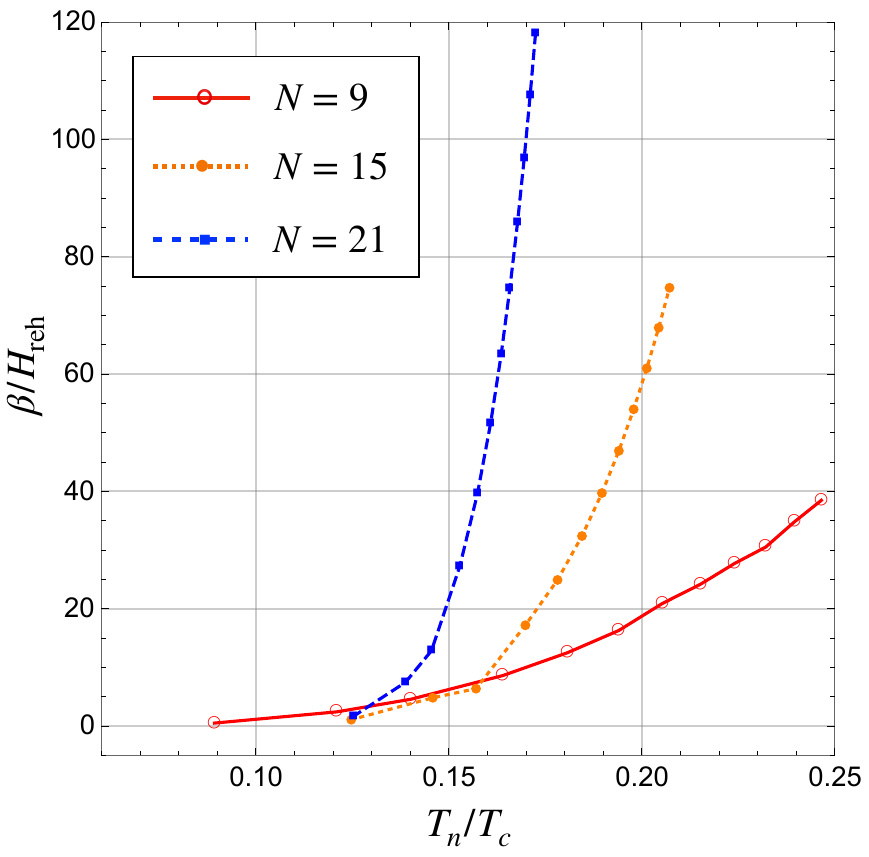}
\caption{
The relation between $T_n/T_c$ and $\beta/H_{\rm reh}$.
Each dot corresponds to $S_4^{\rm th}-S_{4,0}=0.1,0.5,1,2,\cdots, 10$ from the left.
We choose $\epsilon=1/N,  \lambda=\lambda_{*(1)}, \Lambda/\Lambda'=1$, and $\Lambda=10^{17}\GeV$.
The red circles, orange bullets, and blue squares represent the numerical results for $(N,M_\Phi/\Lambda)=(9,0.1),(15,0.3),(21,0.5)$, respectively. 
}
\label{fig:contour}
\end{figure}

The bubbles of a true vacuum nucleate at $T_n$ and expand. 
When the collision of bubbles take place or later, GWs are produced by several sources.
The amplitude and peak frequency of GWs typically depend on two parameters: the released energy during the phase transition $\alpha$, and the inverse duration of the phase transition $\beta$.
The latent heat released inside the bubble is defined as \cite{Caprini:2015zlo}
\beq
\alpha = \frac{\rho_{\rm vac}}{\rho_{\rm rad}(T_n)} \, ,
\eeq
where $\rho_{\rm rad}(T_n)$ is estimated in a false vacuum.
Since the Universe is dominated by the vacuum energy at $T=T_n$, we obtain $\alpha\gg1$
for a strong first-order phase transition.
The second key parameter, the inverse time duration of phase transition, is defined as \cite{Caprini:2015zlo}
\beq
\frac{\beta}{H(T_{\rm reh})} &\equiv& -\frac{1}{H(T_{\rm reh})}\left.\frac{dS_4}{dt}\right|_{T=T_n}\nonumber\\[1ex] &\simeq& \frac{H(T_n)}{H(T_{\rm reh})} T_n\left.\frac{dS_4}{dT}\right|_{T=T_n}\nonumber\\[1ex]
&\simeq& (S_4^{\rm th}-S_{4,0}) \left[2+\frac{X_fK_0(X_f)}{K_1(X_f)}\right],
\label{beta}
\eeq
with $X_f\equiv \lambda\phi_f/\sqrt{2}T_n$.
In the second equality, we have used the entropy conservation law by ignoring the temperature dependence of $g_{*s}$, and in the third equality, we used $(X^\nu K_\nu(X))' = -X^\nu K_{\nu-1}(X)$ under an assumption of $\phi_f\sim\phi_t$.
For a smaller $\beta/H(T_{\rm reh})$, a phase transition becomes stronger and produces more GWs.
We can also calculate $\beta/H(T_{\rm reh})$ numerically. 
Although some inaccuracies in both the analytical estimate of $X_f$ and the numerical derivative
with respect to temperature exist, we have confirmed that the analytical result (\ref{beta}) is consistent with the numerical one
up to an $\mathcal{O}(1)$ factor.
Noting that the inverse duration is proportional to $S_4^{\rm th}-S_{4,0}$, we obtain $\beta/H(T_{\rm reh})=\mathcal{O}(1-10)$ for $S_{4}^{\rm th}-S_{4,0}\lesssim 10$.

\FIG{fig:contour} shows the relation between $T_n/T_c$ and $\beta/H_{\rm reh}$
for $S_4^{\rm th}-S_{4,0}=0.1,0.5,1,2,\cdots, 10$ from the left.
Here we choose $\epsilon=1/N$,  $\lambda=\lambda_{*(1)}$, $\Lambda/\Lambda'=1$, and $\Lambda=10^{17}\GeV$, and
the red circles, orange bullets, and blue squares represent the numerical results for $(N,M_\Phi/\Lambda)=(9,0.1),(15,0.3),(21,0.5)$, respectively, where
the values of $M_\Phi/\Lambda$ are chosen so that the critical temperature is $T_c\sim10^4\GeV$.
We have approximated the numerical derivatives by finite size variation $\delta S_{4}/\delta T$ with $\delta T/T_n \sim\mathcal{O}(0.1)\%$, for which the convergence has been confirmed for some parameter sets.  
Since $\beta/H\gtrsim1$ is required to complete the phase transition, there is a bound, $S_4^{\rm th}-S_{4,0}\gtrsim0.1$ or $T_n/T_c \gtrsim 0.05$.
Although our analytical result shows the quadratic behavior, $\beta/H_{\rm reh}\propto (T_n/T_c)^2$, other factors in \EQ{TcTndependence} and \EQ{beta}, such as $\phi_f$ and the modified Bessel functions, can enhance the exponent.
One can see from the figure that the exponent appears to be larger for a larger $N$,
but the detailed analysis is beyond our scope.

Let us briefly compare our result with the case of the Goldberger-Wise model,
where a very strong supercooling or $T_n/T_c\ll1$ is typically predicted for a small $\beta$,
leading to a large entropy dilution,
while we have found that $T_n/T_c$ is not so small in our model.
In the Goldberger-Wise case, the high-temperature system in the AdS-S is described in terms of the Hawking temperature as the order parameter,
while in the RS spacetime the radion field corresponds to the IR order parameter
\cite{Creminelli:2001th}.
The brane position changes with the ambient temperature, whereas the effective potential around the true vacuum is approximated by the Goldberger-Wise mechanism.
In addition to such a difference in temperature dependence, we naively find that the barrier of the Goldberger-Wise potential is wider and the tunneling rate is more suppressed, leading to a hierarchically smaller $T_n/T_c$
compared to our four-dimensional model.

\subsection{Gravitational waves}

Parameterized by the characteristic parameters, $\alpha$ and $\beta$, the spectrum of GWs in terms of frequency $f$ is approximately expanded as
\beq
\Omega_{\rm GW} (f) \simeq \Omega_{\phi}(f) + \Omega_{\rm sw}(f) \, ,
\eeq
where the first contribution is associated with collisions of the bubble walls \cite{Turner:1990rc,Kosowsky:1991ua,Kosowsky:1992vn,Turner:1992tz,Jinno:2016vai,Jinno:2017fby}, and the second one is the sound wave in the plasma after the collisions \cite{Hindmarsh:2013xza,Giblin:2014qia,Hindmarsh:2015qta,Hindmarsh:2017gnf}.
Although not displayed here, there is also the third contribution from the magnetohydrodynamic turbulence \cite{Kamionkowski:1993fg,Caprini:2006jb,Caprini:2009yp,Kosowsky:2001xp,Gogoberidze:2007an,Niksa:2018ofa}. However, this is subdominant compared to the other two contributions \cite{Caprini:2015zlo}, and we omit it throughout the paper.

By numerical simulations, we can estimate the parameters, $\alpha$ and $\beta$, to obtain the amount of GWs from both contributions, the scalar kinetic energy and sound wave in the bulk fluid.
The spectrum functions of GWs are respectively given by
\beq
\Omega_\phi (f) \simeq \mathcal{R}\tilde{\Omega}_\phi \lmk\frac{\alpha}{1+\alpha}\rmk^2\lmk\frac{H_{\rm reh}}{\beta}\rmk^2 \mathcal{S}\lmk\frac{f}{f_\phi}\rmk,
\eeq
\beq
\Omega_{\rm sw}(f) &\simeq& \mathcal{R}\tilde{\Omega}_{\rm sw} v_w \lmk\frac{\kappa_{\rm sw}\alpha}{1+\alpha}\rmk^2\lmk\frac{H_{\rm reh}}{\beta}\rmk\nonumber\\ &\times& \lmk1-\frac{1}{\sqrt{1+2\tau_{\rm sh}H_{\rm reh}}}\rmk \mathcal{S}\lmk\frac{f}{f_{\rm sw}}\rmk,
\eeq
where the spectral shape is generally parameterized in terms of the following function,
\beq
\mathcal{S}(x) = \frac{1}{\mathcal{N}}\frac{(a+b)^c}{[bx^{-a/c} + ax^{b/c}]^c} \, ,
\label{spectral}
\eeq
with normalization factor,
\beq
\mathcal{N} = \lmk\frac{b}{a}\rmk^{a/n}\lmk\frac{nc}{b}\rmk^c \frac{\Gamma(a/n)\Gamma(b/n)}{n\Gamma(c)}, ~~~n=\frac{a+b}{c} \, .
\eeq
Here the peak amplitudes scale with $\tilde{\Omega}_\phi\simeq 0.042$ \cite{Jinno:2017fby} and $\tilde{\Omega}_{\rm sw}\simeq 0.051$ \cite{Hindmarsh:2013xza,Hindmarsh:2015qta,Hindmarsh:2017gnf} with assumption of $v_w\sim1$,
and $\mathcal{R}$ is used to reproduce the present value of amplitude,
\beq
\mathcal{R} = \frac{\pi^2}{90}\frac{T_0^4}{\Mpl^2H_0^2}g_*(T_{\rm reh})\lmk\frac{g_{*s,0}}{g_{*s}(T_{\rm reh})}\rmk^{4/3},
\eeq
where $T_0$ denotes the current temperature of cosmic microwave background (CMB) photon.
The peak frequencies are given by 
\beq
f_{\phi, {\rm sw}} &=& 1.65\times10^{-5}{\rm Hz} \lmk\frac{\tilde{f}_{\phi, {\rm sw}}}{\beta}\rmk \lmk\frac{\beta}{H_{\rm reh}}\rmk\nonumber\\
&\times& \lmk\frac{T_{\rm reh}}{100\GeV}\rmk\lmk\frac{g_*(T_{\rm reh})}{100}\rmk^{1/6},
\label{peakf}
\eeq
where $\tilde{f}_{\phi ({\rm sw})}/\beta\simeq0.20~(0.54)$ represents a peak frequency at $T=T_{\rm reh}$.
For the contribution from sound wave, we define the fraction of vacuum energy converted into the bulk motion as \cite{Espinosa:2010hh}
\beq
\kappa_{\rm sw} = \frac{\alpha}{0.73+0.083\sqrt{\alpha}+\alpha} \, .
\eeq
In addition, we define $\tau_{\rm sh}$ as the timescale of shock formation,
\beq
\tau_{\rm sh} = (8\pi)^{1/3}\frac{v_w}{\beta\sqrt{3\kappa_{\rm sw}\alpha/(4(1+\alpha))}} \, ,
\eeq
and after this period, the production from sound wave is suppressed.

In addition to the key parameters, $\alpha$ and $\beta$, the bubble wall velocity $v_w$ is also an important factor.
In principle, $v_w$ can be determined by the balance between pressure and friction on the bubble wall \cite{Bodeker:2009qy,Bodeker:2017cim,Dorsch:2018pat,BarrosoMancha:2020fay,Laurent:2020gpg,Azatov:2020ufh,Wang:2020zlf,Gouttenoire:2021kjv}.
Depending on the strength of interaction with the thermal plasma, the bubble expansion mode can be classified into two classes: terminal velocity and runaway.
For a stronger interaction with the thermal plasma, the bubble wall velocity reaches the terminal velocity more easily, and most of the kinetic energy of bubble walls is transferred to the bulk motion of fluid, in which the sound wave gives the most dominant contribution, $\Omega_{\rm GW}\simeq \Omega_{\rm sw}$.
On the other hand, if the frictional force by the thermal plasma is negligible, the bubble wall is accelerated largely and
GWs are dominantly induced from the scalar field itself, $\Omega_{\rm GW}\simeq \Omega_\phi$.
However, the precise estimate of $v_w$ is still an unresolved issue and highly model-dependent with  large theoretical uncertainties.
Thus we assume that the bubble wall velocity is comparable to the speed of light, $v_w\sim1$,\footnote{As clarified in ref.~\cite{Bodeker:2017cim}, even if $\alpha\gg1$ corresponding to the present case, the bubble wall can reach the terminal velocity by the emission of transition radiations, which is very close to the speed of light. See also refs.~\cite{Ai:2023see,Krajewski:2023clt,Li:2023xto} for recent developments.}
and consider both two extreme cases explained above.

\begin{figure*}[t!]
\centering
\includegraphics[width=17cm]{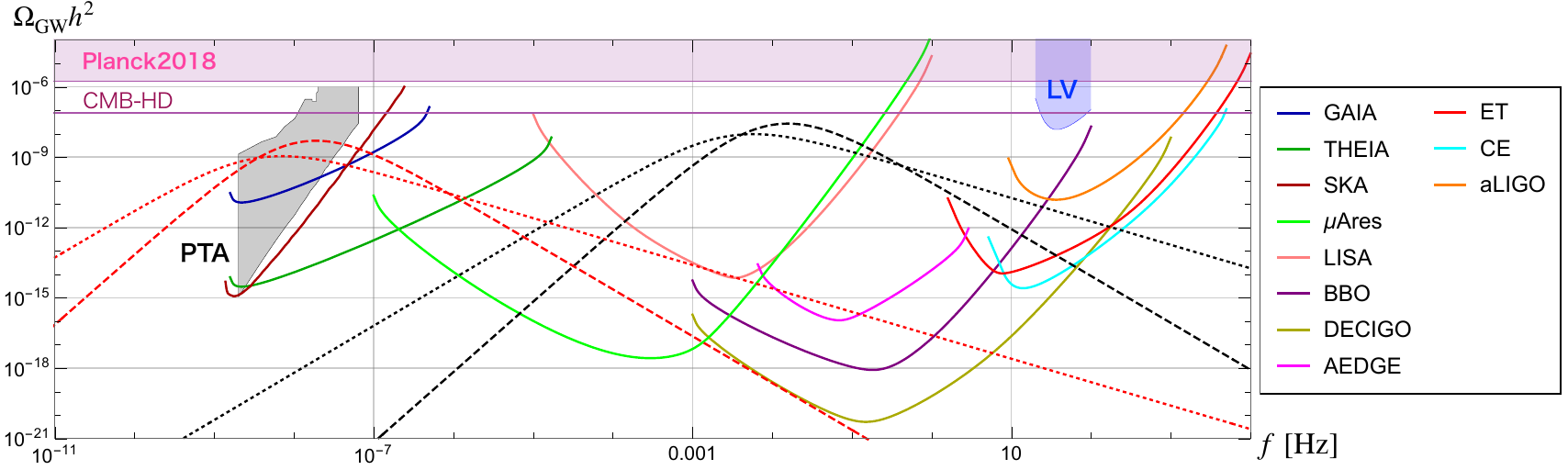}
\caption{
The GW spectrum generated from the first order phase transition in our model.
The black (red) dotted and dashed lines denote the spectral shapes from the scalar and sound wave contributions in the range of mHz (nHz), respectively. See the main text for each parameter set.
The gray shaded region is the signal from PTA collaborations \cite{NANOGrav:2023gor,EPTA:2023fyk,Xu:2023wog,Reardon:2023gzh} (only NANOGrav 15-year data depicted in the figure), and the blue shaded region is the constraint from the LIGO-VIRGO \cite{KAGRA:2021kbb}.
The colored curved lines represent the power-law integrated sensitivities of the future GW detectors.
The light purple shaded region is excluded by the bound on dark radiation abundance from the Planck2018 \cite{Planck:2018vyg}, and the region above the purple horizontal line will be probed by the projected CMB search, CMB-HD \cite{Sehgal:2019ewc,CMB-HD:2022bsz}. 
}
\label{fig:full_range}
\end{figure*}

The spectral shape of GW energy density $\Omega_{\rm GW}h^2$ from the first order phase transition are shown by black and red, dotted and dashed lines in \FIG{fig:full_range}.
Here the Hubble constant is given by $H_0=100h {\rm km/s/Mpc}$ with $h=0.7$. 
We demonstrate two cases in different frequency ranges, milli-Hz and nano-Hz.
For the case of the mHz range (black dotted and dashed), we choose $N=9$, $\epsilon=1/N$, $\lambda=\lambda_{*(1)}$, $m_\Phi^2=-(100\TeV)^2$, $\Lambda=10^{17}\GeV$, $\Lambda/\Lambda'=1$, and $M_\Phi/\Lambda\simeq 0.0982$, which corresponds to $S_4^{\rm th}-S_{4,0}=1$.
The numerical calculation of the bounce action gives us cosmological parameters, $\alpha\simeq8.64\times 10^3$, $\beta/H_{\rm reh}\simeq4.60$, $T_n/T_c\simeq0.140$, and $T_{\rm reh}\simeq 37.9\TeV$.
On the other hand, for the case of the nHz range (red dotted and dashed), we choose a parameter set as follows: $N=9$, $\epsilon=1/N$, $\lambda=\lambda_{*(1)}$, $m_\Phi^2=-(40\MeV)^2$, $\Lambda=10^{17}\GeV$, $\Lambda/\Lambda'=1$, $M_\Phi/\Lambda\simeq 0.00968$, which corresponds to $S_4^{\rm th}-S_{4,0}=5$,
and $g_*=10.75$.
We numerically obtain cosmological parameters, $\alpha\simeq2.44\times 10^4$, $\beta/H_{\rm reh}\simeq 19.9$, $T_n/T_c\simeq0.192$, and $T_{\rm reh}\simeq 15.5\MeV$.
The spectral shape parameters are taken as $(a,b,c)=(2.01,1,2.93)$ for the sound wave contribution
and $(a,b,c)=(3,2,5)$ for the scalar contribution.
The gray shaded region denotes the observed GW signal by NANOGrav 15-year data \cite{NANOGrav:2023gor},
which may be explained by the latter case (for the interpretation of the pulsar timing array (PTA) signal in terms of a dark phase transition,
see e.g. refs.~\cite{Nakai:2020oit,Fujikura:2023lkn}).\footnote{
If a dark sector responsible for the phase transition is completely secluded from the visible sector,
it contradicts with cosmological measurements
(for example, it may give a dangerously large abundance of dark radiation components).
Therefore, the dark sector should interact with the visible sector particles,
so that it can decay into the visible sector after the phase transition. 
See \cite{Bringmann:2023opz} for the detailed analysis.
}

In addition to the PTA signal region, \FIG{fig:full_range} describes 
power-law integrated sensitivities of the future GW detectors,
GAIA/THEIA \cite{Moore:2017ity,Theia:2017xtk,Garcia-Bellido:2021zgu}, SKA \cite{Breitbach:2018ddu}, $\mu$Ares \cite{Sesana:2019vho}, LISA \cite{Cornish:2017vip,Robson:2018ifk}, BBO \cite{Yagi:2011wg}, DECIGO \cite{Kawamura:2020pcg}, AEDGE \cite{AEDGE:2019nxb}, Einstein Telescope (ET) \cite{Hild:2010id}, Cosmic Explorer (CE) \cite{Reitze:2019iox}, and the advanced LIGO (aLIGO) \cite{LIGOScientific:2014pky},
following ref.~\cite{Thrane:2013oya} for the estimation (see also refs.~\cite{Allen:1997ad,Kudoh:2005as,Caprini:2019pxz}).
The blue shaded region is excluded by the LIGO and VIRGO (denoted as LV) \cite{KAGRA:2021kbb}.

Since GWs behave like radiation fluid, the abundance is bounded from the above so as not to spoil the successful prediction of Big Bang Nucleosynthesis (BBN) and CMB anisotropy.
We parameterize the dark radiation density by the so-called effective number of neutrinos, $\Delta N_{\rm eff}$.
If we assume no other dark radiation components, the density parameter is given by
\beq
\int {\rm dln}f~\Omega_{\rm GW}(f) h^2 \simeq 5.6\times10^{-6}\Delta N_{\rm eff} \, . 
\eeq
In \FIG{fig:full_range}, we use the most severe bound on $\Delta N_{\rm eff}$ from the CMB anisotropy by the Planck2018 \cite{Planck:2018vyg} to approximately evaluate the upper bound at the first order.
In addition, the CMB-S4 \cite{CMB-S4:2022ght} and CMB-HD collaboration \cite{Sehgal:2019ewc,CMB-HD:2022bsz} will extend the upper bound down to $10^{-6}$.

%%%%%%
%%%%%%%%%%%%%%%%%%%%%%%%%%%%%%%%%%%%%%%%%%%%%%%%%%
\section{Conclusion
\label{sec:discussion}}

We have explored a model to realize a strong first order phase transition
based on a supersymmetric QCD in conformal window.
With a marginally relevant deformation,
the spontaneous breaking of conformal invariance is driven by the balance with a non-perturbatively generated superpotential,
making the breaking scale hierarchically smaller than a typical mass scale of the theory such as the Planck scale
without fine-tuning.
The conformal breaking makes all (s)quarks massive and the remaining pure super-Yang-Mills theory confines.
We analyzed the (super)conformal phase transition
by considering perturbative finite temperature corrections, and estimated the cosmological parameters,
such as $T_c, T_n, \beta$, both analytically and numerically.
It was found that the resulting GW amplitude can be large enough to be detected by projected observations.
A conformal phase transition at around the electroweak scale generates GWs probed by future space-based interferometers,
while a dark phase transition at $\mathcal{O}(0.01-1)$ GeV can explain the reported PTA signal.

Although we do not need a fine-tuning to realize the spontaneous breaking of conformal invariance, the soft supersymmetry breaking scalar mass parameter must be close to the conformal breaking scale
to ensure the completion of the phase transition, which raises a coincidence issue.
If the same strong gauge dynamics can break supersymmetry dynamically, the issue may be addressed.
Model-building in this direction is left for a future study.

%%%%%%
%%%%%%%%%%%%%%%%%%%%%%%%%%%%%%%%%%%%%%%%%%%%%%%%%%
\section*{Acknowledgments}
We thank Sudhakantha Girmohanta for fruitful discussions.
YN is supported by Natural Science Foundation of Shanghai.
KF is supported
by JSPS Grant-in-Aid for Research Fellows Grant No.22J00345 and JST CREST Grant Number JPMJCR24I3.

\appendix

\bibliography{reference}

\begin{thebibliography}{125}%
\makeatletter
\providecommand \@ifxundefined [1]{%
 \@ifx{#1\undefined}
}%
\providecommand \@ifnum [1]{%
 \ifnum #1\expandafter \@firstoftwo
 \else \expandafter \@secondoftwo
 \fi
}%
\providecommand \@ifx [1]{%
 \ifx #1\expandafter \@firstoftwo
 \else \expandafter \@secondoftwo
 \fi
}%
\providecommand \natexlab [1]{#1}%
\providecommand \enquote  [1]{``#1''}%
\providecommand \bibnamefont  [1]{#1}%
\providecommand \bibfnamefont [1]{#1}%
\providecommand \citenamefont [1]{#1}%
\providecommand \href@noop [0]{\@secondoftwo}%
\providecommand \href [0]{\begingroup \@sanitize@url \@href}%
\providecommand \@href[1]{\@@startlink{#1}\@@href}%
\providecommand \@@href[1]{\endgroup#1\@@endlink}%
\providecommand \@sanitize@url [0]{\catcode `\\12\catcode `\$12\catcode
  `\&12\catcode `\#12\catcode `\^12\catcode `\_12\catcode `\%12\relax}%
\providecommand \@@startlink[1]{}%
\providecommand \@@endlink[0]{}%
\providecommand \url  [0]{\begingroup\@sanitize@url \@url }%
\providecommand \@url [1]{\endgroup\@href {#1}{\urlprefix }}%
\providecommand \urlprefix  [0]{URL }%
\providecommand \Eprint [0]{\href }%
\providecommand \doibase [0]{http://dx.doi.org/}%
\providecommand \selectlanguage [0]{\@gobble}%
\providecommand \bibinfo  [0]{\@secondoftwo}%
\providecommand \bibfield  [0]{\@secondoftwo}%
\providecommand \translation [1]{[#1]}%
\providecommand \BibitemOpen [0]{}%
\providecommand \bibitemStop [0]{}%
\providecommand \bibitemNoStop [0]{.\EOS\space}%
\providecommand \EOS [0]{\spacefactor3000\relax}%
\providecommand \BibitemShut  [1]{\csname bibitem#1\endcsname}%
\let\auto@bib@innerbib\@empty
%</preamble>
\bibitem [{\citenamefont {Kajantie}\ \emph {et~al.}(1996)\citenamefont
  {Kajantie}, \citenamefont {Laine}, \citenamefont {Rummukainen},\ and\
  \citenamefont {Shaposhnikov}}]{Kajantie:1995kf}%
  \BibitemOpen
  \bibfield  {author} {\bibinfo {author} {\bibfnamefont {K.}~\bibnamefont
  {Kajantie}}, \bibinfo {author} {\bibfnamefont {M.}~\bibnamefont {Laine}},
  \bibinfo {author} {\bibfnamefont {K.}~\bibnamefont {Rummukainen}}, \ and\
  \bibinfo {author} {\bibfnamefont {M.~E.}\ \bibnamefont {Shaposhnikov}},\
  }\href {\doibase 10.1016/0550-3213(96)00052-1} {\bibfield  {journal}
  {\bibinfo  {journal} {Nucl. Phys. B}\ }\textbf {\bibinfo {volume} {466}},\
  \bibinfo {pages} {189} (\bibinfo {year} {1996})},\ \Eprint
  {http://arxiv.org/abs/hep-lat/9510020} {arXiv:hep-lat/9510020} \BibitemShut
  {NoStop}%
\bibitem [{\citenamefont {Kajantie}\ \emph {et~al.}(1997)\citenamefont
  {Kajantie}, \citenamefont {Laine}, \citenamefont {Rummukainen},\ and\
  \citenamefont {Shaposhnikov}}]{Kajantie:1996qd}%
  \BibitemOpen
  \bibfield  {author} {\bibinfo {author} {\bibfnamefont {K.}~\bibnamefont
  {Kajantie}}, \bibinfo {author} {\bibfnamefont {M.}~\bibnamefont {Laine}},
  \bibinfo {author} {\bibfnamefont {K.}~\bibnamefont {Rummukainen}}, \ and\
  \bibinfo {author} {\bibfnamefont {M.~E.}\ \bibnamefont {Shaposhnikov}},\
  }\href {\doibase 10.1016/S0550-3213(97)00164-8} {\bibfield  {journal}
  {\bibinfo  {journal} {Nucl. Phys. B}\ }\textbf {\bibinfo {volume} {493}},\
  \bibinfo {pages} {413} (\bibinfo {year} {1997})},\ \Eprint
  {http://arxiv.org/abs/hep-lat/9612006} {arXiv:hep-lat/9612006} \BibitemShut
  {NoStop}%
\bibitem [{\citenamefont {Rummukainen}\ \emph {et~al.}(1998)\citenamefont
  {Rummukainen}, \citenamefont {Tsypin}, \citenamefont {Kajantie},
  \citenamefont {Laine},\ and\ \citenamefont
  {Shaposhnikov}}]{Rummukainen:1998as}%
  \BibitemOpen
  \bibfield  {author} {\bibinfo {author} {\bibfnamefont {K.}~\bibnamefont
  {Rummukainen}}, \bibinfo {author} {\bibfnamefont {M.}~\bibnamefont {Tsypin}},
  \bibinfo {author} {\bibfnamefont {K.}~\bibnamefont {Kajantie}}, \bibinfo
  {author} {\bibfnamefont {M.}~\bibnamefont {Laine}}, \ and\ \bibinfo {author}
  {\bibfnamefont {M.~E.}\ \bibnamefont {Shaposhnikov}},\ }\href {\doibase
  10.1016/S0550-3213(98)00494-5} {\bibfield  {journal} {\bibinfo  {journal}
  {Nucl. Phys. B}\ }\textbf {\bibinfo {volume} {532}},\ \bibinfo {pages} {283}
  (\bibinfo {year} {1998})},\ \Eprint {http://arxiv.org/abs/hep-lat/9805013}
  {arXiv:hep-lat/9805013} \BibitemShut {NoStop}%
\bibitem [{\citenamefont {Aoki}\ \emph {et~al.}(2006)\citenamefont {Aoki},
  \citenamefont {Endrodi}, \citenamefont {Fodor}, \citenamefont {Katz},\ and\
  \citenamefont {Szabo}}]{Aoki:2006we}%
  \BibitemOpen
  \bibfield  {author} {\bibinfo {author} {\bibfnamefont {Y.}~\bibnamefont
  {Aoki}}, \bibinfo {author} {\bibfnamefont {G.}~\bibnamefont {Endrodi}},
  \bibinfo {author} {\bibfnamefont {Z.}~\bibnamefont {Fodor}}, \bibinfo
  {author} {\bibfnamefont {S.~D.}\ \bibnamefont {Katz}}, \ and\ \bibinfo
  {author} {\bibfnamefont {K.~K.}\ \bibnamefont {Szabo}},\ }\href {\doibase
  10.1038/nature05120} {\bibfield  {journal} {\bibinfo  {journal} {Nature}\
  }\textbf {\bibinfo {volume} {443}},\ \bibinfo {pages} {675} (\bibinfo {year}
  {2006})},\ \Eprint {http://arxiv.org/abs/hep-lat/0611014}
  {arXiv:hep-lat/0611014} \BibitemShut {NoStop}%
\bibitem [{\citenamefont {Turok}\ and\ \citenamefont
  {Zadrozny}(1990)}]{Turok:1990in}%
  \BibitemOpen
  \bibfield  {author} {\bibinfo {author} {\bibfnamefont {N.}~\bibnamefont
  {Turok}}\ and\ \bibinfo {author} {\bibfnamefont {J.}~\bibnamefont
  {Zadrozny}},\ }\href {\doibase 10.1103/PhysRevLett.65.2331} {\bibfield
  {journal} {\bibinfo  {journal} {Phys. Rev. Lett.}\ }\textbf {\bibinfo
  {volume} {65}},\ \bibinfo {pages} {2331} (\bibinfo {year}
  {1990})}\BibitemShut {NoStop}%
\bibitem [{\citenamefont {Krauss}\ and\ \citenamefont
  {Trodden}(1999)}]{Krauss:1999ng}%
  \BibitemOpen
  \bibfield  {author} {\bibinfo {author} {\bibfnamefont {L.~M.}\ \bibnamefont
  {Krauss}}\ and\ \bibinfo {author} {\bibfnamefont {M.}~\bibnamefont
  {Trodden}},\ }\href {\doibase 10.1103/PhysRevLett.83.1502} {\bibfield
  {journal} {\bibinfo  {journal} {Phys. Rev. Lett.}\ }\textbf {\bibinfo
  {volume} {83}},\ \bibinfo {pages} {1502} (\bibinfo {year} {1999})},\ \Eprint
  {http://arxiv.org/abs/hep-ph/9902420} {arXiv:hep-ph/9902420} \BibitemShut
  {NoStop}%
\bibitem [{\citenamefont {Garcia-Bellido}\ \emph {et~al.}(1999)\citenamefont
  {Garcia-Bellido}, \citenamefont {Grigoriev}, \citenamefont {Kusenko},\ and\
  \citenamefont {Shaposhnikov}}]{Garcia-Bellido:1999xos}%
  \BibitemOpen
  \bibfield  {author} {\bibinfo {author} {\bibfnamefont {J.}~\bibnamefont
  {Garcia-Bellido}}, \bibinfo {author} {\bibfnamefont {D.~Y.}\ \bibnamefont
  {Grigoriev}}, \bibinfo {author} {\bibfnamefont {A.}~\bibnamefont {Kusenko}},
  \ and\ \bibinfo {author} {\bibfnamefont {M.~E.}\ \bibnamefont
  {Shaposhnikov}},\ }\href {\doibase 10.1103/PhysRevD.60.123504} {\bibfield
  {journal} {\bibinfo  {journal} {Phys. Rev. D}\ }\textbf {\bibinfo {volume}
  {60}},\ \bibinfo {pages} {123504} (\bibinfo {year} {1999})},\ \Eprint
  {http://arxiv.org/abs/hep-ph/9902449} {arXiv:hep-ph/9902449} \BibitemShut
  {NoStop}%
\bibitem [{\citenamefont {Konstandin}\ and\ \citenamefont
  {Servant}(2011)}]{Konstandin:2011ds}%
  \BibitemOpen
  \bibfield  {author} {\bibinfo {author} {\bibfnamefont {T.}~\bibnamefont
  {Konstandin}}\ and\ \bibinfo {author} {\bibfnamefont {G.}~\bibnamefont
  {Servant}},\ }\href {\doibase 10.1088/1475-7516/2011/07/024} {\bibfield
  {journal} {\bibinfo  {journal} {JCAP}\ }\textbf {\bibinfo {volume} {07}},\
  \bibinfo {pages} {024} (\bibinfo {year} {2011})},\ \Eprint
  {http://arxiv.org/abs/1104.4793} {arXiv:1104.4793 [hep-ph]} \BibitemShut
  {NoStop}%
\bibitem [{\citenamefont {Fujikura}\ \emph {et~al.}(2021)\citenamefont
  {Fujikura}, \citenamefont {Harigaya}, \citenamefont {Nakai},\ and\
  \citenamefont {Wang}}]{Fujikura:2021abj}%
  \BibitemOpen
  \bibfield  {author} {\bibinfo {author} {\bibfnamefont {K.}~\bibnamefont
  {Fujikura}}, \bibinfo {author} {\bibfnamefont {K.}~\bibnamefont {Harigaya}},
  \bibinfo {author} {\bibfnamefont {Y.}~\bibnamefont {Nakai}}, \ and\ \bibinfo
  {author} {\bibfnamefont {I.~R.}\ \bibnamefont {Wang}},\ }\href {\doibase
  10.1007/JHEP07(2021)224} {\bibfield  {journal} {\bibinfo  {journal} {JHEP}\
  }\textbf {\bibinfo {volume} {07}},\ \bibinfo {pages} {224} (\bibinfo {year}
  {2021})},\ \bibinfo {note} {[Erratum: JHEP 12, 192 (2021), Erratum: JHEP 1,
  156 (2022), Erratum: JHEP 01, 156 (2022)]},\ \Eprint
  {http://arxiv.org/abs/2103.05005} {arXiv:2103.05005 [hep-ph]} \BibitemShut
  {NoStop}%
\bibitem [{\citenamefont {Dasgupta}\ \emph {et~al.}(2022)\citenamefont
  {Dasgupta}, \citenamefont {Dev}, \citenamefont {Ghoshal},\ and\ \citenamefont
  {Mazumdar}}]{Dasgupta:2022isg}%
  \BibitemOpen
  \bibfield  {author} {\bibinfo {author} {\bibfnamefont {A.}~\bibnamefont
  {Dasgupta}}, \bibinfo {author} {\bibfnamefont {P.~S.~B.}\ \bibnamefont
  {Dev}}, \bibinfo {author} {\bibfnamefont {A.}~\bibnamefont {Ghoshal}}, \ and\
  \bibinfo {author} {\bibfnamefont {A.}~\bibnamefont {Mazumdar}},\ }\href
  {\doibase 10.1103/PhysRevD.106.075027} {\bibfield  {journal} {\bibinfo
  {journal} {Phys. Rev. D}\ }\textbf {\bibinfo {volume} {106}},\ \bibinfo
  {pages} {075027} (\bibinfo {year} {2022})},\ \Eprint
  {http://arxiv.org/abs/2206.07032} {arXiv:2206.07032 [hep-ph]} \BibitemShut
  {NoStop}%
\bibitem [{\citenamefont {Ellis}\ \emph {et~al.}(2023)\citenamefont {Ellis},
  \citenamefont {Lewicki}, \citenamefont {Merchand}, \citenamefont {No},\ and\
  \citenamefont {Zych}}]{Ellis:2022lft}%
  \BibitemOpen
  \bibfield  {author} {\bibinfo {author} {\bibfnamefont {J.}~\bibnamefont
  {Ellis}}, \bibinfo {author} {\bibfnamefont {M.}~\bibnamefont {Lewicki}},
  \bibinfo {author} {\bibfnamefont {M.}~\bibnamefont {Merchand}}, \bibinfo
  {author} {\bibfnamefont {J.~M.}\ \bibnamefont {No}}, \ and\ \bibinfo {author}
  {\bibfnamefont {M.}~\bibnamefont {Zych}},\ }\href {\doibase
  10.1007/JHEP01(2023)093} {\bibfield  {journal} {\bibinfo  {journal} {JHEP}\
  }\textbf {\bibinfo {volume} {01}},\ \bibinfo {pages} {093} (\bibinfo {year}
  {2023})},\ \Eprint {http://arxiv.org/abs/2210.16305} {arXiv:2210.16305
  [hep-ph]} \BibitemShut {NoStop}%
\bibitem [{\citenamefont {Chun}\ \emph {et~al.}(2023)\citenamefont {Chun},
  \citenamefont {Dutka}, \citenamefont {Jung}, \citenamefont {Nagels},\ and\
  \citenamefont {Vanvlasselaer}}]{Chun:2023ezg}%
  \BibitemOpen
  \bibfield  {author} {\bibinfo {author} {\bibfnamefont {E.~J.}\ \bibnamefont
  {Chun}}, \bibinfo {author} {\bibfnamefont {T.~P.}\ \bibnamefont {Dutka}},
  \bibinfo {author} {\bibfnamefont {T.~H.}\ \bibnamefont {Jung}}, \bibinfo
  {author} {\bibfnamefont {X.}~\bibnamefont {Nagels}}, \ and\ \bibinfo {author}
  {\bibfnamefont {M.}~\bibnamefont {Vanvlasselaer}},\ }\href {\doibase
  10.1007/JHEP09(2023)164} {\bibfield  {journal} {\bibinfo  {journal} {JHEP}\
  }\textbf {\bibinfo {volume} {09}},\ \bibinfo {pages} {164} (\bibinfo {year}
  {2023})},\ \Eprint {http://arxiv.org/abs/2305.10759} {arXiv:2305.10759
  [hep-ph]} \BibitemShut {NoStop}%
\bibitem [{\citenamefont {Cataldi}\ and\ \citenamefont
  {Shakya}(2024)}]{Cataldi:2024pgt}%
  \BibitemOpen
  \bibfield  {author} {\bibinfo {author} {\bibfnamefont {M.}~\bibnamefont
  {Cataldi}}\ and\ \bibinfo {author} {\bibfnamefont {B.}~\bibnamefont
  {Shakya}},\ }\href {\doibase 10.1088/1475-7516/2024/11/047} {\bibfield
  {journal} {\bibinfo  {journal} {JCAP}\ }\textbf {\bibinfo {volume} {11}},\
  \bibinfo {pages} {047} (\bibinfo {year} {2024})},\ \Eprint
  {http://arxiv.org/abs/2407.16747} {arXiv:2407.16747 [hep-ph]} \BibitemShut
  {NoStop}%
\bibitem [{\citenamefont {Fujikura}\ \emph {et~al.}(2024)\citenamefont
  {Fujikura}, \citenamefont {Girmohanta}, \citenamefont {Nakai},\ and\
  \citenamefont {Zhang}}]{Fujikura:2024jto}%
  \BibitemOpen
  \bibfield  {author} {\bibinfo {author} {\bibfnamefont {K.}~\bibnamefont
  {Fujikura}}, \bibinfo {author} {\bibfnamefont {S.}~\bibnamefont
  {Girmohanta}}, \bibinfo {author} {\bibfnamefont {Y.}~\bibnamefont {Nakai}}, \
  and\ \bibinfo {author} {\bibfnamefont {Z.}~\bibnamefont {Zhang}},\ }\href
  {\doibase 10.1016/j.physletb.2024.139045} {\bibfield  {journal} {\bibinfo
  {journal} {Phys. Lett. B}\ }\textbf {\bibinfo {volume} {858}},\ \bibinfo
  {pages} {139045} (\bibinfo {year} {2024})},\ \Eprint
  {http://arxiv.org/abs/2406.12956} {arXiv:2406.12956 [hep-ph]} \BibitemShut
  {NoStop}%
\bibitem [{\citenamefont {Agashe}\ \emph {et~al.}(2024)\citenamefont {Agashe},
  \citenamefont {Du}, \citenamefont {Ekhterachian}, \citenamefont {Fong},
  \citenamefont {Hong},\ and\ \citenamefont {Vecchi}}]{Agashe:2024uvp}%
  \BibitemOpen
  \bibfield  {author} {\bibinfo {author} {\bibfnamefont {K.}~\bibnamefont
  {Agashe}}, \bibinfo {author} {\bibfnamefont {P.}~\bibnamefont {Du}}, \bibinfo
  {author} {\bibfnamefont {M.}~\bibnamefont {Ekhterachian}}, \bibinfo {author}
  {\bibfnamefont {C.~S.}\ \bibnamefont {Fong}}, \bibinfo {author}
  {\bibfnamefont {S.}~\bibnamefont {Hong}}, \ and\ \bibinfo {author}
  {\bibfnamefont {L.}~\bibnamefont {Vecchi}},\ }\href@noop {} {\  (\bibinfo
  {year} {2024})},\ \Eprint {http://arxiv.org/abs/2410.00960} {arXiv:2410.00960
  [hep-ph]} \BibitemShut {NoStop}%
\bibitem [{\citenamefont {Witten}(1984)}]{Witten:1984rs}%
  \BibitemOpen
  \bibfield  {author} {\bibinfo {author} {\bibfnamefont {E.}~\bibnamefont
  {Witten}},\ }\href {\doibase 10.1103/PhysRevD.30.272} {\bibfield  {journal}
  {\bibinfo  {journal} {Phys. Rev. D}\ }\textbf {\bibinfo {volume} {30}},\
  \bibinfo {pages} {272} (\bibinfo {year} {1984})}\BibitemShut {NoStop}%
\bibitem [{\citenamefont {Chway}\ \emph {et~al.}(2020)\citenamefont {Chway},
  \citenamefont {Jung},\ and\ \citenamefont {Shin}}]{Chway:2019kft}%
  \BibitemOpen
  \bibfield  {author} {\bibinfo {author} {\bibfnamefont {D.}~\bibnamefont
  {Chway}}, \bibinfo {author} {\bibfnamefont {T.~H.}\ \bibnamefont {Jung}}, \
  and\ \bibinfo {author} {\bibfnamefont {C.~S.}\ \bibnamefont {Shin}},\ }\href
  {\doibase 10.1103/PhysRevD.101.095019} {\bibfield  {journal} {\bibinfo
  {journal} {Phys. Rev. D}\ }\textbf {\bibinfo {volume} {101}},\ \bibinfo
  {pages} {095019} (\bibinfo {year} {2020})},\ \Eprint
  {http://arxiv.org/abs/1912.04238} {arXiv:1912.04238 [hep-ph]} \BibitemShut
  {NoStop}%
\bibitem [{\citenamefont {Hong}\ \emph {et~al.}(2020)\citenamefont {Hong},
  \citenamefont {Jung},\ and\ \citenamefont {Xie}}]{Hong:2020est}%
  \BibitemOpen
  \bibfield  {author} {\bibinfo {author} {\bibfnamefont {J.-P.}\ \bibnamefont
  {Hong}}, \bibinfo {author} {\bibfnamefont {S.}~\bibnamefont {Jung}}, \ and\
  \bibinfo {author} {\bibfnamefont {K.-P.}\ \bibnamefont {Xie}},\ }\href
  {\doibase 10.1103/PhysRevD.102.075028} {\bibfield  {journal} {\bibinfo
  {journal} {Phys. Rev. D}\ }\textbf {\bibinfo {volume} {102}},\ \bibinfo
  {pages} {075028} (\bibinfo {year} {2020})},\ \Eprint
  {http://arxiv.org/abs/2008.04430} {arXiv:2008.04430 [hep-ph]} \BibitemShut
  {NoStop}%
\bibitem [{\citenamefont {Azatov}\ \emph {et~al.}(2021)\citenamefont {Azatov},
  \citenamefont {Vanvlasselaer},\ and\ \citenamefont {Yin}}]{Azatov:2021ifm}%
  \BibitemOpen
  \bibfield  {author} {\bibinfo {author} {\bibfnamefont {A.}~\bibnamefont
  {Azatov}}, \bibinfo {author} {\bibfnamefont {M.}~\bibnamefont
  {Vanvlasselaer}}, \ and\ \bibinfo {author} {\bibfnamefont {W.}~\bibnamefont
  {Yin}},\ }\href {\doibase 10.1007/JHEP03(2021)288} {\bibfield  {journal}
  {\bibinfo  {journal} {JHEP}\ }\textbf {\bibinfo {volume} {03}},\ \bibinfo
  {pages} {288} (\bibinfo {year} {2021})},\ \Eprint
  {http://arxiv.org/abs/2101.05721} {arXiv:2101.05721 [hep-ph]} \BibitemShut
  {NoStop}%
\bibitem [{\citenamefont {Kawana}\ and\ \citenamefont
  {Xie}(2022)}]{Kawana:2021tde}%
  \BibitemOpen
  \bibfield  {author} {\bibinfo {author} {\bibfnamefont {K.}~\bibnamefont
  {Kawana}}\ and\ \bibinfo {author} {\bibfnamefont {K.-P.}\ \bibnamefont
  {Xie}},\ }\href {\doibase 10.1016/j.physletb.2021.136791} {\bibfield
  {journal} {\bibinfo  {journal} {Phys. Lett. B}\ }\textbf {\bibinfo {volume}
  {824}},\ \bibinfo {pages} {136791} (\bibinfo {year} {2022})},\ \Eprint
  {http://arxiv.org/abs/2106.00111} {arXiv:2106.00111 [astro-ph.CO]}
  \BibitemShut {NoStop}%
\bibitem [{\citenamefont {Giudice}\ \emph {et~al.}(2024)\citenamefont
  {Giudice}, \citenamefont {Lee}, \citenamefont {Pomarol},\ and\ \citenamefont
  {Shakya}}]{Giudice:2024tcp}%
  \BibitemOpen
  \bibfield  {author} {\bibinfo {author} {\bibfnamefont {G.~F.}\ \bibnamefont
  {Giudice}}, \bibinfo {author} {\bibfnamefont {H.~M.}\ \bibnamefont {Lee}},
  \bibinfo {author} {\bibfnamefont {A.}~\bibnamefont {Pomarol}}, \ and\
  \bibinfo {author} {\bibfnamefont {B.}~\bibnamefont {Shakya}},\ }\href
  {\doibase 10.1007/JHEP12(2024)190} {\bibfield  {journal} {\bibinfo  {journal}
  {JHEP}\ }\textbf {\bibinfo {volume} {12}},\ \bibinfo {pages} {190} (\bibinfo
  {year} {2024})},\ \Eprint {http://arxiv.org/abs/2403.03252} {arXiv:2403.03252
  [hep-ph]} \BibitemShut {NoStop}%
\bibitem [{\citenamefont {Zhang}\ \emph {et~al.}(2024)\citenamefont {Zhang},
  \citenamefont {Zhang}, \citenamefont {Cai},\ and\ \citenamefont
  {Zhang}}]{Zhang:2024dgv}%
  \BibitemOpen
  \bibfield  {author} {\bibinfo {author} {\bibfnamefont {B.}~\bibnamefont
  {Zhang}}, \bibinfo {author} {\bibfnamefont {Z.}~\bibnamefont {Zhang}},
  \bibinfo {author} {\bibfnamefont {C.}~\bibnamefont {Cai}}, \ and\ \bibinfo
  {author} {\bibfnamefont {H.-H.}\ \bibnamefont {Zhang}},\ }\href {\doibase
  10.1103/PhysRevD.110.015023} {\bibfield  {journal} {\bibinfo  {journal}
  {Phys. Rev. D}\ }\textbf {\bibinfo {volume} {110}},\ \bibinfo {pages}
  {015023} (\bibinfo {year} {2024})},\ \Eprint
  {http://arxiv.org/abs/2404.05332} {arXiv:2404.05332 [hep-ph]} \BibitemShut
  {NoStop}%
\bibitem [{\citenamefont {Allahverdi}\ \emph {et~al.}(2024)\citenamefont
  {Allahverdi}, \citenamefont {Hauptmann},\ and\ \citenamefont
  {Huang}}]{Allahverdi:2024ofe}%
  \BibitemOpen
  \bibfield  {author} {\bibinfo {author} {\bibfnamefont {R.}~\bibnamefont
  {Allahverdi}}, \bibinfo {author} {\bibfnamefont {C.}~\bibnamefont
  {Hauptmann}}, \ and\ \bibinfo {author} {\bibfnamefont {P.}~\bibnamefont
  {Huang}},\ }\href {\doibase 10.1103/PhysRevD.110.115005} {\bibfield
  {journal} {\bibinfo  {journal} {Phys. Rev. D}\ }\textbf {\bibinfo {volume}
  {110}},\ \bibinfo {pages} {115005} (\bibinfo {year} {2024})},\ \Eprint
  {http://arxiv.org/abs/2409.02179} {arXiv:2409.02179 [hep-ph]} \BibitemShut
  {NoStop}%
\bibitem [{\citenamefont {Turner}\ and\ \citenamefont
  {Wilczek}(1990)}]{Turner:1990rc}%
  \BibitemOpen
  \bibfield  {author} {\bibinfo {author} {\bibfnamefont {M.~S.}\ \bibnamefont
  {Turner}}\ and\ \bibinfo {author} {\bibfnamefont {F.}~\bibnamefont
  {Wilczek}},\ }\href {\doibase 10.1103/PhysRevLett.65.3080} {\bibfield
  {journal} {\bibinfo  {journal} {Phys. Rev. Lett.}\ }\textbf {\bibinfo
  {volume} {65}},\ \bibinfo {pages} {3080} (\bibinfo {year}
  {1990})}\BibitemShut {NoStop}%
\bibitem [{\citenamefont {Turner}\ \emph {et~al.}(1992)\citenamefont {Turner},
  \citenamefont {Weinberg},\ and\ \citenamefont {Widrow}}]{Turner:1992tz}%
  \BibitemOpen
  \bibfield  {author} {\bibinfo {author} {\bibfnamefont {M.~S.}\ \bibnamefont
  {Turner}}, \bibinfo {author} {\bibfnamefont {E.~J.}\ \bibnamefont
  {Weinberg}}, \ and\ \bibinfo {author} {\bibfnamefont {L.~M.}\ \bibnamefont
  {Widrow}},\ }\href {\doibase 10.1103/PhysRevD.46.2384} {\bibfield  {journal}
  {\bibinfo  {journal} {Phys. Rev. D}\ }\textbf {\bibinfo {volume} {46}},\
  \bibinfo {pages} {2384} (\bibinfo {year} {1992})}\BibitemShut {NoStop}%
\bibitem [{\citenamefont {Kosowsky}\ \emph
  {et~al.}(1992{\natexlab{a}})\citenamefont {Kosowsky}, \citenamefont
  {Turner},\ and\ \citenamefont {Watkins}}]{Kosowsky:1992rz}%
  \BibitemOpen
  \bibfield  {author} {\bibinfo {author} {\bibfnamefont {A.}~\bibnamefont
  {Kosowsky}}, \bibinfo {author} {\bibfnamefont {M.~S.}\ \bibnamefont
  {Turner}}, \ and\ \bibinfo {author} {\bibfnamefont {R.}~\bibnamefont
  {Watkins}},\ }\href {\doibase 10.1103/PhysRevLett.69.2026} {\bibfield
  {journal} {\bibinfo  {journal} {Phys. Rev. Lett.}\ }\textbf {\bibinfo
  {volume} {69}},\ \bibinfo {pages} {2026} (\bibinfo {year}
  {1992}{\natexlab{a}})}\BibitemShut {NoStop}%
\bibitem [{\citenamefont {Kosowsky}\ and\ \citenamefont
  {Turner}(1993)}]{Kosowsky:1992vn}%
  \BibitemOpen
  \bibfield  {author} {\bibinfo {author} {\bibfnamefont {A.}~\bibnamefont
  {Kosowsky}}\ and\ \bibinfo {author} {\bibfnamefont {M.~S.}\ \bibnamefont
  {Turner}},\ }\href {\doibase 10.1103/PhysRevD.47.4372} {\bibfield  {journal}
  {\bibinfo  {journal} {Phys. Rev. D}\ }\textbf {\bibinfo {volume} {47}},\
  \bibinfo {pages} {4372} (\bibinfo {year} {1993})},\ \Eprint
  {http://arxiv.org/abs/astro-ph/9211004} {arXiv:astro-ph/9211004} \BibitemShut
  {NoStop}%
\bibitem [{\citenamefont {Hindmarsh}\ \emph {et~al.}(2014)\citenamefont
  {Hindmarsh}, \citenamefont {Huber}, \citenamefont {Rummukainen},\ and\
  \citenamefont {Weir}}]{Hindmarsh:2013xza}%
  \BibitemOpen
  \bibfield  {author} {\bibinfo {author} {\bibfnamefont {M.}~\bibnamefont
  {Hindmarsh}}, \bibinfo {author} {\bibfnamefont {S.~J.}\ \bibnamefont
  {Huber}}, \bibinfo {author} {\bibfnamefont {K.}~\bibnamefont {Rummukainen}},
  \ and\ \bibinfo {author} {\bibfnamefont {D.~J.}\ \bibnamefont {Weir}},\
  }\href {\doibase 10.1103/PhysRevLett.112.041301} {\bibfield  {journal}
  {\bibinfo  {journal} {Phys. Rev. Lett.}\ }\textbf {\bibinfo {volume} {112}},\
  \bibinfo {pages} {041301} (\bibinfo {year} {2014})},\ \Eprint
  {http://arxiv.org/abs/1304.2433} {arXiv:1304.2433 [hep-ph]} \BibitemShut
  {NoStop}%
\bibitem [{\citenamefont {Giblin}\ and\ \citenamefont
  {Mertens}(2014)}]{Giblin:2014qia}%
  \BibitemOpen
  \bibfield  {author} {\bibinfo {author} {\bibfnamefont {J.~T.}\ \bibnamefont
  {Giblin}}\ and\ \bibinfo {author} {\bibfnamefont {J.~B.}\ \bibnamefont
  {Mertens}},\ }\href {\doibase 10.1103/PhysRevD.90.023532} {\bibfield
  {journal} {\bibinfo  {journal} {Phys. Rev. D}\ }\textbf {\bibinfo {volume}
  {90}},\ \bibinfo {pages} {023532} (\bibinfo {year} {2014})},\ \Eprint
  {http://arxiv.org/abs/1405.4005} {arXiv:1405.4005 [astro-ph.CO]} \BibitemShut
  {NoStop}%
\bibitem [{\citenamefont {Hindmarsh}\ \emph {et~al.}(2015)\citenamefont
  {Hindmarsh}, \citenamefont {Huber}, \citenamefont {Rummukainen},\ and\
  \citenamefont {Weir}}]{Hindmarsh:2015qta}%
  \BibitemOpen
  \bibfield  {author} {\bibinfo {author} {\bibfnamefont {M.}~\bibnamefont
  {Hindmarsh}}, \bibinfo {author} {\bibfnamefont {S.~J.}\ \bibnamefont
  {Huber}}, \bibinfo {author} {\bibfnamefont {K.}~\bibnamefont {Rummukainen}},
  \ and\ \bibinfo {author} {\bibfnamefont {D.~J.}\ \bibnamefont {Weir}},\
  }\href {\doibase 10.1103/PhysRevD.92.123009} {\bibfield  {journal} {\bibinfo
  {journal} {Phys. Rev. D}\ }\textbf {\bibinfo {volume} {92}},\ \bibinfo
  {pages} {123009} (\bibinfo {year} {2015})},\ \Eprint
  {http://arxiv.org/abs/1504.03291} {arXiv:1504.03291 [astro-ph.CO]}
  \BibitemShut {NoStop}%
\bibitem [{\citenamefont {Hindmarsh}\ \emph {et~al.}(2017)\citenamefont
  {Hindmarsh}, \citenamefont {Huber}, \citenamefont {Rummukainen},\ and\
  \citenamefont {Weir}}]{Hindmarsh:2017gnf}%
  \BibitemOpen
  \bibfield  {author} {\bibinfo {author} {\bibfnamefont {M.}~\bibnamefont
  {Hindmarsh}}, \bibinfo {author} {\bibfnamefont {S.~J.}\ \bibnamefont
  {Huber}}, \bibinfo {author} {\bibfnamefont {K.}~\bibnamefont {Rummukainen}},
  \ and\ \bibinfo {author} {\bibfnamefont {D.~J.}\ \bibnamefont {Weir}},\
  }\href {\doibase 10.1103/PhysRevD.96.103520} {\bibfield  {journal} {\bibinfo
  {journal} {Phys. Rev. D}\ }\textbf {\bibinfo {volume} {96}},\ \bibinfo
  {pages} {103520} (\bibinfo {year} {2017})},\ \bibinfo {note} {[Erratum:
  Phys.Rev.D 101, 089902 (2020)]},\ \Eprint {http://arxiv.org/abs/1704.05871}
  {arXiv:1704.05871 [astro-ph.CO]} \BibitemShut {NoStop}%
\bibitem [{\citenamefont {Kamionkowski}\ \emph {et~al.}(1994)\citenamefont
  {Kamionkowski}, \citenamefont {Kosowsky},\ and\ \citenamefont
  {Turner}}]{Kamionkowski:1993fg}%
  \BibitemOpen
  \bibfield  {author} {\bibinfo {author} {\bibfnamefont {M.}~\bibnamefont
  {Kamionkowski}}, \bibinfo {author} {\bibfnamefont {A.}~\bibnamefont
  {Kosowsky}}, \ and\ \bibinfo {author} {\bibfnamefont {M.~S.}\ \bibnamefont
  {Turner}},\ }\href {\doibase 10.1103/PhysRevD.49.2837} {\bibfield  {journal}
  {\bibinfo  {journal} {Phys. Rev. D}\ }\textbf {\bibinfo {volume} {49}},\
  \bibinfo {pages} {2837} (\bibinfo {year} {1994})},\ \Eprint
  {http://arxiv.org/abs/astro-ph/9310044} {arXiv:astro-ph/9310044} \BibitemShut
  {NoStop}%
\bibitem [{\citenamefont {Kosowsky}\ \emph {et~al.}(2002)\citenamefont
  {Kosowsky}, \citenamefont {Mack},\ and\ \citenamefont
  {Kahniashvili}}]{Kosowsky:2001xp}%
  \BibitemOpen
  \bibfield  {author} {\bibinfo {author} {\bibfnamefont {A.}~\bibnamefont
  {Kosowsky}}, \bibinfo {author} {\bibfnamefont {A.}~\bibnamefont {Mack}}, \
  and\ \bibinfo {author} {\bibfnamefont {T.}~\bibnamefont {Kahniashvili}},\
  }\href {\doibase 10.1103/PhysRevD.66.024030} {\bibfield  {journal} {\bibinfo
  {journal} {Phys. Rev. D}\ }\textbf {\bibinfo {volume} {66}},\ \bibinfo
  {pages} {024030} (\bibinfo {year} {2002})},\ \Eprint
  {http://arxiv.org/abs/astro-ph/0111483} {arXiv:astro-ph/0111483} \BibitemShut
  {NoStop}%
\bibitem [{\citenamefont {Caprini}\ and\ \citenamefont
  {Durrer}(2006)}]{Caprini:2006jb}%
  \BibitemOpen
  \bibfield  {author} {\bibinfo {author} {\bibfnamefont {C.}~\bibnamefont
  {Caprini}}\ and\ \bibinfo {author} {\bibfnamefont {R.}~\bibnamefont
  {Durrer}},\ }\href {\doibase 10.1103/PhysRevD.74.063521} {\bibfield
  {journal} {\bibinfo  {journal} {Phys. Rev. D}\ }\textbf {\bibinfo {volume}
  {74}},\ \bibinfo {pages} {063521} (\bibinfo {year} {2006})},\ \Eprint
  {http://arxiv.org/abs/astro-ph/0603476} {arXiv:astro-ph/0603476} \BibitemShut
  {NoStop}%
\bibitem [{\citenamefont {Caprini}\ \emph {et~al.}(2009)\citenamefont
  {Caprini}, \citenamefont {Durrer},\ and\ \citenamefont
  {Servant}}]{Caprini:2009yp}%
  \BibitemOpen
  \bibfield  {author} {\bibinfo {author} {\bibfnamefont {C.}~\bibnamefont
  {Caprini}}, \bibinfo {author} {\bibfnamefont {R.}~\bibnamefont {Durrer}}, \
  and\ \bibinfo {author} {\bibfnamefont {G.}~\bibnamefont {Servant}},\ }\href
  {\doibase 10.1088/1475-7516/2009/12/024} {\bibfield  {journal} {\bibinfo
  {journal} {JCAP}\ }\textbf {\bibinfo {volume} {12}},\ \bibinfo {pages} {024}
  (\bibinfo {year} {2009})},\ \Eprint {http://arxiv.org/abs/0909.0622}
  {arXiv:0909.0622 [astro-ph.CO]} \BibitemShut {NoStop}%
\bibitem [{\citenamefont {Gogoberidze}\ \emph {et~al.}(2007)\citenamefont
  {Gogoberidze}, \citenamefont {Kahniashvili},\ and\ \citenamefont
  {Kosowsky}}]{Gogoberidze:2007an}%
  \BibitemOpen
  \bibfield  {author} {\bibinfo {author} {\bibfnamefont {G.}~\bibnamefont
  {Gogoberidze}}, \bibinfo {author} {\bibfnamefont {T.}~\bibnamefont
  {Kahniashvili}}, \ and\ \bibinfo {author} {\bibfnamefont {A.}~\bibnamefont
  {Kosowsky}},\ }\href {\doibase 10.1103/PhysRevD.76.083002} {\bibfield
  {journal} {\bibinfo  {journal} {Phys. Rev. D}\ }\textbf {\bibinfo {volume}
  {76}},\ \bibinfo {pages} {083002} (\bibinfo {year} {2007})},\ \Eprint
  {http://arxiv.org/abs/0705.1733} {arXiv:0705.1733 [astro-ph]} \BibitemShut
  {NoStop}%
\bibitem [{\citenamefont {Niksa}\ \emph {et~al.}(2018)\citenamefont {Niksa},
  \citenamefont {Schlederer},\ and\ \citenamefont {Sigl}}]{Niksa:2018ofa}%
  \BibitemOpen
  \bibfield  {author} {\bibinfo {author} {\bibfnamefont {P.}~\bibnamefont
  {Niksa}}, \bibinfo {author} {\bibfnamefont {M.}~\bibnamefont {Schlederer}}, \
  and\ \bibinfo {author} {\bibfnamefont {G.}~\bibnamefont {Sigl}},\ }\href
  {\doibase 10.1088/1361-6382/aac89c} {\bibfield  {journal} {\bibinfo
  {journal} {Class. Quant. Grav.}\ }\textbf {\bibinfo {volume} {35}},\ \bibinfo
  {pages} {144001} (\bibinfo {year} {2018})},\ \Eprint
  {http://arxiv.org/abs/1803.02271} {arXiv:1803.02271 [astro-ph.CO]}
  \BibitemShut {NoStop}%
\bibitem [{\citenamefont {Coleman}\ and\ \citenamefont
  {Weinberg}(1973)}]{Coleman:1973jx}%
  \BibitemOpen
  \bibfield  {author} {\bibinfo {author} {\bibfnamefont {S.~R.}\ \bibnamefont
  {Coleman}}\ and\ \bibinfo {author} {\bibfnamefont {E.~J.}\ \bibnamefont
  {Weinberg}},\ }\href {\doibase 10.1103/PhysRevD.7.1888} {\bibfield  {journal}
  {\bibinfo  {journal} {Phys. Rev. D}\ }\textbf {\bibinfo {volume} {7}},\
  \bibinfo {pages} {1888} (\bibinfo {year} {1973})}\BibitemShut {NoStop}%
\bibitem [{\citenamefont {Witten}(1981)}]{Witten:1980ez}%
  \BibitemOpen
  \bibfield  {author} {\bibinfo {author} {\bibfnamefont {E.}~\bibnamefont
  {Witten}},\ }\href {\doibase 10.1016/0550-3213(81)90182-6} {\bibfield
  {journal} {\bibinfo  {journal} {Nucl. Phys. B}\ }\textbf {\bibinfo {volume}
  {177}},\ \bibinfo {pages} {477} (\bibinfo {year} {1981})}\BibitemShut
  {NoStop}%
\bibitem [{\citenamefont {Iso}\ \emph {et~al.}(2017)\citenamefont {Iso},
  \citenamefont {Serpico},\ and\ \citenamefont {Shimada}}]{Iso:2017uuu}%
  \BibitemOpen
  \bibfield  {author} {\bibinfo {author} {\bibfnamefont {S.}~\bibnamefont
  {Iso}}, \bibinfo {author} {\bibfnamefont {P.~D.}\ \bibnamefont {Serpico}}, \
  and\ \bibinfo {author} {\bibfnamefont {K.}~\bibnamefont {Shimada}},\ }\href
  {\doibase 10.1103/PhysRevLett.119.141301} {\bibfield  {journal} {\bibinfo
  {journal} {Phys. Rev. Lett.}\ }\textbf {\bibinfo {volume} {119}},\ \bibinfo
  {pages} {141301} (\bibinfo {year} {2017})},\ \Eprint
  {http://arxiv.org/abs/1704.04955} {arXiv:1704.04955 [hep-ph]} \BibitemShut
  {NoStop}%
\bibitem [{\citenamefont {Jinno}\ and\ \citenamefont
  {Takimoto}(2017{\natexlab{a}})}]{Jinno:2016knw}%
  \BibitemOpen
  \bibfield  {author} {\bibinfo {author} {\bibfnamefont {R.}~\bibnamefont
  {Jinno}}\ and\ \bibinfo {author} {\bibfnamefont {M.}~\bibnamefont
  {Takimoto}},\ }\href {\doibase 10.1103/PhysRevD.95.015020} {\bibfield
  {journal} {\bibinfo  {journal} {Phys. Rev. D}\ }\textbf {\bibinfo {volume}
  {95}},\ \bibinfo {pages} {015020} (\bibinfo {year} {2017}{\natexlab{a}})},\
  \Eprint {http://arxiv.org/abs/1604.05035} {arXiv:1604.05035 [hep-ph]}
  \BibitemShut {NoStop}%
\bibitem [{\citenamefont {Hiramatsu}\ \emph {et~al.}(2015)\citenamefont
  {Hiramatsu}, \citenamefont {Miyamoto},\ and\ \citenamefont
  {Yokoyama}}]{Hiramatsu:2014uta}%
  \BibitemOpen
  \bibfield  {author} {\bibinfo {author} {\bibfnamefont {T.}~\bibnamefont
  {Hiramatsu}}, \bibinfo {author} {\bibfnamefont {Y.}~\bibnamefont {Miyamoto}},
  \ and\ \bibinfo {author} {\bibfnamefont {J.}~\bibnamefont {Yokoyama}},\
  }\href {\doibase 10.1088/1475-7516/2015/03/024} {\bibfield  {journal}
  {\bibinfo  {journal} {JCAP}\ }\textbf {\bibinfo {volume} {03}},\ \bibinfo
  {pages} {024} (\bibinfo {year} {2015})},\ \Eprint
  {http://arxiv.org/abs/1412.7814} {arXiv:1412.7814 [hep-ph]} \BibitemShut
  {NoStop}%
\bibitem [{\citenamefont {Coradeschi}\ \emph {et~al.}(2013)\citenamefont
  {Coradeschi}, \citenamefont {Lodone}, \citenamefont {Pappadopulo},
  \citenamefont {Rattazzi},\ and\ \citenamefont {Vitale}}]{Coradeschi:2013gda}%
  \BibitemOpen
  \bibfield  {author} {\bibinfo {author} {\bibfnamefont {F.}~\bibnamefont
  {Coradeschi}}, \bibinfo {author} {\bibfnamefont {P.}~\bibnamefont {Lodone}},
  \bibinfo {author} {\bibfnamefont {D.}~\bibnamefont {Pappadopulo}}, \bibinfo
  {author} {\bibfnamefont {R.}~\bibnamefont {Rattazzi}}, \ and\ \bibinfo
  {author} {\bibfnamefont {L.}~\bibnamefont {Vitale}},\ }\href {\doibase
  10.1007/JHEP11(2013)057} {\bibfield  {journal} {\bibinfo  {journal} {JHEP}\
  }\textbf {\bibinfo {volume} {11}},\ \bibinfo {pages} {057} (\bibinfo {year}
  {2013})},\ \Eprint {http://arxiv.org/abs/1306.4601} {arXiv:1306.4601
  [hep-th]} \BibitemShut {NoStop}%
\bibitem [{\citenamefont {Chacko}\ and\ \citenamefont
  {Mishra}(2013)}]{Chacko:2012sy}%
  \BibitemOpen
  \bibfield  {author} {\bibinfo {author} {\bibfnamefont {Z.}~\bibnamefont
  {Chacko}}\ and\ \bibinfo {author} {\bibfnamefont {R.~K.}\ \bibnamefont
  {Mishra}},\ }\href {\doibase 10.1103/PhysRevD.87.115006} {\bibfield
  {journal} {\bibinfo  {journal} {Phys. Rev. D}\ }\textbf {\bibinfo {volume}
  {87}},\ \bibinfo {pages} {115006} (\bibinfo {year} {2013})},\ \Eprint
  {http://arxiv.org/abs/1209.3022} {arXiv:1209.3022 [hep-ph]} \BibitemShut
  {NoStop}%
\bibitem [{\citenamefont {Maldacena}(1998)}]{Maldacena:1997re}%
  \BibitemOpen
  \bibfield  {author} {\bibinfo {author} {\bibfnamefont {J.~M.}\ \bibnamefont
  {Maldacena}},\ }\href {\doibase 10.4310/ATMP.1998.v2.n2.a1} {\bibfield
  {journal} {\bibinfo  {journal} {Adv. Theor. Math. Phys.}\ }\textbf {\bibinfo
  {volume} {2}},\ \bibinfo {pages} {231} (\bibinfo {year} {1998})},\ \Eprint
  {http://arxiv.org/abs/hep-th/9711200} {arXiv:hep-th/9711200} \BibitemShut
  {NoStop}%
\bibitem [{\citenamefont {Gubser}\ \emph {et~al.}(1998)\citenamefont {Gubser},
  \citenamefont {Klebanov},\ and\ \citenamefont {Polyakov}}]{Gubser:1998bc}%
  \BibitemOpen
  \bibfield  {author} {\bibinfo {author} {\bibfnamefont {S.~S.}\ \bibnamefont
  {Gubser}}, \bibinfo {author} {\bibfnamefont {I.~R.}\ \bibnamefont
  {Klebanov}}, \ and\ \bibinfo {author} {\bibfnamefont {A.~M.}\ \bibnamefont
  {Polyakov}},\ }\href {\doibase 10.1016/S0370-2693(98)00377-3} {\bibfield
  {journal} {\bibinfo  {journal} {Phys. Lett. B}\ }\textbf {\bibinfo {volume}
  {428}},\ \bibinfo {pages} {105} (\bibinfo {year} {1998})},\ \Eprint
  {http://arxiv.org/abs/hep-th/9802109} {arXiv:hep-th/9802109} \BibitemShut
  {NoStop}%
\bibitem [{\citenamefont {Witten}(1998)}]{Witten:1998qj}%
  \BibitemOpen
  \bibfield  {author} {\bibinfo {author} {\bibfnamefont {E.}~\bibnamefont
  {Witten}},\ }\href {\doibase 10.4310/ATMP.1998.v2.n2.a2} {\bibfield
  {journal} {\bibinfo  {journal} {Adv. Theor. Math. Phys.}\ }\textbf {\bibinfo
  {volume} {2}},\ \bibinfo {pages} {253} (\bibinfo {year} {1998})},\ \Eprint
  {http://arxiv.org/abs/hep-th/9802150} {arXiv:hep-th/9802150} \BibitemShut
  {NoStop}%
\bibitem [{\citenamefont {Arkani-Hamed}\ \emph {et~al.}(2001)\citenamefont
  {Arkani-Hamed}, \citenamefont {Porrati},\ and\ \citenamefont
  {Randall}}]{Arkani-Hamed:2000ijo}%
  \BibitemOpen
  \bibfield  {author} {\bibinfo {author} {\bibfnamefont {N.}~\bibnamefont
  {Arkani-Hamed}}, \bibinfo {author} {\bibfnamefont {M.}~\bibnamefont
  {Porrati}}, \ and\ \bibinfo {author} {\bibfnamefont {L.}~\bibnamefont
  {Randall}},\ }\href {\doibase 10.1088/1126-6708/2001/08/017} {\bibfield
  {journal} {\bibinfo  {journal} {JHEP}\ }\textbf {\bibinfo {volume} {08}},\
  \bibinfo {pages} {017} (\bibinfo {year} {2001})},\ \Eprint
  {http://arxiv.org/abs/hep-th/0012148} {arXiv:hep-th/0012148} \BibitemShut
  {NoStop}%
\bibitem [{\citenamefont {Rattazzi}\ and\ \citenamefont
  {Zaffaroni}(2001)}]{Rattazzi:2000hs}%
  \BibitemOpen
  \bibfield  {author} {\bibinfo {author} {\bibfnamefont {R.}~\bibnamefont
  {Rattazzi}}\ and\ \bibinfo {author} {\bibfnamefont {A.}~\bibnamefont
  {Zaffaroni}},\ }\href {\doibase 10.1088/1126-6708/2001/04/021} {\bibfield
  {journal} {\bibinfo  {journal} {JHEP}\ }\textbf {\bibinfo {volume} {04}},\
  \bibinfo {pages} {021} (\bibinfo {year} {2001})},\ \Eprint
  {http://arxiv.org/abs/hep-th/0012248} {arXiv:hep-th/0012248} \BibitemShut
  {NoStop}%
\bibitem [{\citenamefont {Randall}\ and\ \citenamefont
  {Sundrum}(1999)}]{Randall:1999ee}%
  \BibitemOpen
  \bibfield  {author} {\bibinfo {author} {\bibfnamefont {L.}~\bibnamefont
  {Randall}}\ and\ \bibinfo {author} {\bibfnamefont {R.}~\bibnamefont
  {Sundrum}},\ }\href {\doibase 10.1103/PhysRevLett.83.3370} {\bibfield
  {journal} {\bibinfo  {journal} {Phys. Rev. Lett.}\ }\textbf {\bibinfo
  {volume} {83}},\ \bibinfo {pages} {3370} (\bibinfo {year} {1999})},\ \Eprint
  {http://arxiv.org/abs/hep-ph/9905221} {arXiv:hep-ph/9905221} \BibitemShut
  {NoStop}%
\bibitem [{\citenamefont {Goldberger}\ and\ \citenamefont
  {Wise}(1999)}]{Goldberger:1999uk}%
  \BibitemOpen
  \bibfield  {author} {\bibinfo {author} {\bibfnamefont {W.~D.}\ \bibnamefont
  {Goldberger}}\ and\ \bibinfo {author} {\bibfnamefont {M.~B.}\ \bibnamefont
  {Wise}},\ }\href {\doibase 10.1103/PhysRevLett.83.4922} {\bibfield  {journal}
  {\bibinfo  {journal} {Phys. Rev. Lett.}\ }\textbf {\bibinfo {volume} {83}},\
  \bibinfo {pages} {4922} (\bibinfo {year} {1999})},\ \Eprint
  {http://arxiv.org/abs/hep-ph/9907447} {arXiv:hep-ph/9907447} \BibitemShut
  {NoStop}%
\bibitem [{\citenamefont {Garriga}\ \emph {et~al.}(2001)\citenamefont
  {Garriga}, \citenamefont {Pujolas},\ and\ \citenamefont
  {Tanaka}}]{Garriga:2000jb}%
  \BibitemOpen
  \bibfield  {author} {\bibinfo {author} {\bibfnamefont {J.}~\bibnamefont
  {Garriga}}, \bibinfo {author} {\bibfnamefont {O.}~\bibnamefont {Pujolas}}, \
  and\ \bibinfo {author} {\bibfnamefont {T.}~\bibnamefont {Tanaka}},\ }\href
  {\doibase 10.1016/S0550-3213(01)00144-4} {\bibfield  {journal} {\bibinfo
  {journal} {Nucl. Phys. B}\ }\textbf {\bibinfo {volume} {605}},\ \bibinfo
  {pages} {192} (\bibinfo {year} {2001})},\ \Eprint
  {http://arxiv.org/abs/hep-th/0004109} {arXiv:hep-th/0004109} \BibitemShut
  {NoStop}%
\bibitem [{\citenamefont {Goldberger}\ and\ \citenamefont
  {Rothstein}(2000)}]{Goldberger:2000dv}%
  \BibitemOpen
  \bibfield  {author} {\bibinfo {author} {\bibfnamefont {W.~D.}\ \bibnamefont
  {Goldberger}}\ and\ \bibinfo {author} {\bibfnamefont {I.~Z.}\ \bibnamefont
  {Rothstein}},\ }\href {\doibase 10.1016/S0370-2693(00)01047-9} {\bibfield
  {journal} {\bibinfo  {journal} {Phys. Lett. B}\ }\textbf {\bibinfo {volume}
  {491}},\ \bibinfo {pages} {339} (\bibinfo {year} {2000})},\ \Eprint
  {http://arxiv.org/abs/hep-th/0007065} {arXiv:hep-th/0007065} \BibitemShut
  {NoStop}%
\bibitem [{\citenamefont {Hofmann}\ \emph {et~al.}(2001)\citenamefont
  {Hofmann}, \citenamefont {Kanti},\ and\ \citenamefont
  {Pospelov}}]{Hofmann:2000cj}%
  \BibitemOpen
  \bibfield  {author} {\bibinfo {author} {\bibfnamefont {R.}~\bibnamefont
  {Hofmann}}, \bibinfo {author} {\bibfnamefont {P.}~\bibnamefont {Kanti}}, \
  and\ \bibinfo {author} {\bibfnamefont {M.}~\bibnamefont {Pospelov}},\ }\href
  {\doibase 10.1103/PhysRevD.63.124020} {\bibfield  {journal} {\bibinfo
  {journal} {Phys. Rev. D}\ }\textbf {\bibinfo {volume} {63}},\ \bibinfo
  {pages} {124020} (\bibinfo {year} {2001})},\ \Eprint
  {http://arxiv.org/abs/hep-ph/0012213} {arXiv:hep-ph/0012213} \BibitemShut
  {NoStop}%
\bibitem [{\citenamefont {Brevik}\ \emph {et~al.}(2001)\citenamefont {Brevik},
  \citenamefont {Milton}, \citenamefont {Nojiri},\ and\ \citenamefont
  {Odintsov}}]{Brevik:2000vt}%
  \BibitemOpen
  \bibfield  {author} {\bibinfo {author} {\bibfnamefont {I.~H.}\ \bibnamefont
  {Brevik}}, \bibinfo {author} {\bibfnamefont {K.~A.}\ \bibnamefont {Milton}},
  \bibinfo {author} {\bibfnamefont {S.}~\bibnamefont {Nojiri}}, \ and\ \bibinfo
  {author} {\bibfnamefont {S.~D.}\ \bibnamefont {Odintsov}},\ }\href {\doibase
  10.1016/S0550-3213(01)00026-8} {\bibfield  {journal} {\bibinfo  {journal}
  {Nucl. Phys. B}\ }\textbf {\bibinfo {volume} {599}},\ \bibinfo {pages} {305}
  (\bibinfo {year} {2001})},\ \Eprint {http://arxiv.org/abs/hep-th/0010205}
  {arXiv:hep-th/0010205} \BibitemShut {NoStop}%
\bibitem [{\citenamefont {Flachi}\ and\ \citenamefont
  {Toms}(2001)}]{Flachi:2001pq}%
  \BibitemOpen
  \bibfield  {author} {\bibinfo {author} {\bibfnamefont {A.}~\bibnamefont
  {Flachi}}\ and\ \bibinfo {author} {\bibfnamefont {D.~J.}\ \bibnamefont
  {Toms}},\ }\href {\doibase 10.1016/S0550-3213(01)00314-5} {\bibfield
  {journal} {\bibinfo  {journal} {Nucl. Phys. B}\ }\textbf {\bibinfo {volume}
  {610}},\ \bibinfo {pages} {144} (\bibinfo {year} {2001})},\ \Eprint
  {http://arxiv.org/abs/hep-th/0103077} {arXiv:hep-th/0103077} \BibitemShut
  {NoStop}%
\bibitem [{\citenamefont {Nojiri}\ \emph {et~al.}(2001)\citenamefont {Nojiri},
  \citenamefont {Odintsov},\ and\ \citenamefont {Ogushi}}]{Nojiri:2001ai}%
  \BibitemOpen
  \bibfield  {author} {\bibinfo {author} {\bibfnamefont {S.}~\bibnamefont
  {Nojiri}}, \bibinfo {author} {\bibfnamefont {S.~D.}\ \bibnamefont
  {Odintsov}}, \ and\ \bibinfo {author} {\bibfnamefont {S.}~\bibnamefont
  {Ogushi}},\ }\href {\doibase 10.1016/S0370-2693(01)00359-8} {\bibfield
  {journal} {\bibinfo  {journal} {Phys. Lett. B}\ }\textbf {\bibinfo {volume}
  {506}},\ \bibinfo {pages} {200} (\bibinfo {year} {2001})},\ \Eprint
  {http://arxiv.org/abs/hep-th/0102082} {arXiv:hep-th/0102082} \BibitemShut
  {NoStop}%
\bibitem [{\citenamefont {Garriga}\ and\ \citenamefont
  {Pomarol}(2003)}]{Garriga:2002vf}%
  \BibitemOpen
  \bibfield  {author} {\bibinfo {author} {\bibfnamefont {J.}~\bibnamefont
  {Garriga}}\ and\ \bibinfo {author} {\bibfnamefont {A.}~\bibnamefont
  {Pomarol}},\ }\href {\doibase 10.1016/S0370-2693(03)00301-0} {\bibfield
  {journal} {\bibinfo  {journal} {Phys. Lett. B}\ }\textbf {\bibinfo {volume}
  {560}},\ \bibinfo {pages} {91} (\bibinfo {year} {2003})},\ \Eprint
  {http://arxiv.org/abs/hep-th/0212227} {arXiv:hep-th/0212227} \BibitemShut
  {NoStop}%
\bibitem [{\citenamefont {Haba}\ and\ \citenamefont
  {Yamada}(2019)}]{Haba:2019zjc}%
  \BibitemOpen
  \bibfield  {author} {\bibinfo {author} {\bibfnamefont {N.}~\bibnamefont
  {Haba}}\ and\ \bibinfo {author} {\bibfnamefont {T.}~\bibnamefont {Yamada}},\
  }\href@noop {} {\  (\bibinfo {year} {2019})},\ \Eprint
  {http://arxiv.org/abs/1903.10160} {arXiv:1903.10160 [hep-ph]} \BibitemShut
  {NoStop}%
\bibitem [{\citenamefont {Fujikura}\ \emph {et~al.}(2020)\citenamefont
  {Fujikura}, \citenamefont {Nakai},\ and\ \citenamefont
  {Yamada}}]{Fujikura:2019oyi}%
  \BibitemOpen
  \bibfield  {author} {\bibinfo {author} {\bibfnamefont {K.}~\bibnamefont
  {Fujikura}}, \bibinfo {author} {\bibfnamefont {Y.}~\bibnamefont {Nakai}}, \
  and\ \bibinfo {author} {\bibfnamefont {M.}~\bibnamefont {Yamada}},\ }\href
  {\doibase 10.1007/JHEP02(2020)111} {\bibfield  {journal} {\bibinfo  {journal}
  {JHEP}\ }\textbf {\bibinfo {volume} {02}},\ \bibinfo {pages} {111} (\bibinfo
  {year} {2020})},\ \Eprint {http://arxiv.org/abs/1910.07546} {arXiv:1910.07546
  [hep-ph]} \BibitemShut {NoStop}%
\bibitem [{\citenamefont {Girmohanta}\ \emph
  {et~al.}(2024{\natexlab{a}})\citenamefont {Girmohanta}, \citenamefont
  {Nakai}, \citenamefont {Suzuki}, \citenamefont {Wang},\ and\ \citenamefont
  {Xu}}]{Girmohanta:2024kyx}%
  \BibitemOpen
  \bibfield  {author} {\bibinfo {author} {\bibfnamefont {S.}~\bibnamefont
  {Girmohanta}}, \bibinfo {author} {\bibfnamefont {Y.}~\bibnamefont {Nakai}},
  \bibinfo {author} {\bibfnamefont {M.}~\bibnamefont {Suzuki}}, \bibinfo
  {author} {\bibfnamefont {Y.}~\bibnamefont {Wang}}, \ and\ \bibinfo {author}
  {\bibfnamefont {J.}~\bibnamefont {Xu}},\ }\href {\doibase
  10.1007/JHEP08(2024)229} {\bibfield  {journal} {\bibinfo  {journal} {JHEP}\
  }\textbf {\bibinfo {volume} {08}},\ \bibinfo {pages} {229} (\bibinfo {year}
  {2024}{\natexlab{a}})},\ \Eprint {http://arxiv.org/abs/2404.05141}
  {arXiv:2404.05141 [hep-th]} \BibitemShut {NoStop}%
\bibitem [{\citenamefont {Girmohanta}\ \emph
  {et~al.}(2024{\natexlab{b}})\citenamefont {Girmohanta}, \citenamefont
  {Nakai}, \citenamefont {Qiu},\ and\ \citenamefont
  {Zhang}}]{Girmohanta:2024ywz}%
  \BibitemOpen
  \bibfield  {author} {\bibinfo {author} {\bibfnamefont {S.}~\bibnamefont
  {Girmohanta}}, \bibinfo {author} {\bibfnamefont {Y.}~\bibnamefont {Nakai}},
  \bibinfo {author} {\bibfnamefont {Y.-C.}\ \bibnamefont {Qiu}}, \ and\
  \bibinfo {author} {\bibfnamefont {Z.}~\bibnamefont {Zhang}},\ }\href@noop {}
  {\  (\bibinfo {year} {2024}{\natexlab{b}})},\ \Eprint
  {http://arxiv.org/abs/2411.16304} {arXiv:2411.16304 [hep-th]} \BibitemShut
  {NoStop}%
\bibitem [{\citenamefont {Creminelli}\ \emph {et~al.}(2002)\citenamefont
  {Creminelli}, \citenamefont {Nicolis},\ and\ \citenamefont
  {Rattazzi}}]{Creminelli:2001th}%
  \BibitemOpen
  \bibfield  {author} {\bibinfo {author} {\bibfnamefont {P.}~\bibnamefont
  {Creminelli}}, \bibinfo {author} {\bibfnamefont {A.}~\bibnamefont {Nicolis}},
  \ and\ \bibinfo {author} {\bibfnamefont {R.}~\bibnamefont {Rattazzi}},\
  }\href {\doibase 10.1088/1126-6708/2002/03/051} {\bibfield  {journal}
  {\bibinfo  {journal} {JHEP}\ }\textbf {\bibinfo {volume} {03}},\ \bibinfo
  {pages} {051} (\bibinfo {year} {2002})},\ \Eprint
  {http://arxiv.org/abs/hep-th/0107141} {arXiv:hep-th/0107141} \BibitemShut
  {NoStop}%
\bibitem [{\citenamefont {Miura}\ \emph {et~al.}(2019)\citenamefont {Miura},
  \citenamefont {Ohki}, \citenamefont {Otani},\ and\ \citenamefont
  {Yamawaki}}]{Miura:2018dsy}%
  \BibitemOpen
  \bibfield  {author} {\bibinfo {author} {\bibfnamefont {K.}~\bibnamefont
  {Miura}}, \bibinfo {author} {\bibfnamefont {H.}~\bibnamefont {Ohki}},
  \bibinfo {author} {\bibfnamefont {S.}~\bibnamefont {Otani}}, \ and\ \bibinfo
  {author} {\bibfnamefont {K.}~\bibnamefont {Yamawaki}},\ }\href {\doibase
  10.1007/JHEP10(2019)194} {\bibfield  {journal} {\bibinfo  {journal} {JHEP}\
  }\textbf {\bibinfo {volume} {10}},\ \bibinfo {pages} {194} (\bibinfo {year}
  {2019})},\ \Eprint {http://arxiv.org/abs/1811.05670} {arXiv:1811.05670
  [hep-ph]} \BibitemShut {NoStop}%
\bibitem [{\citenamefont {Azatov}\ and\ \citenamefont
  {Vanvlasselaer}(2020)}]{Azatov:2020nbe}%
  \BibitemOpen
  \bibfield  {author} {\bibinfo {author} {\bibfnamefont {A.}~\bibnamefont
  {Azatov}}\ and\ \bibinfo {author} {\bibfnamefont {M.}~\bibnamefont
  {Vanvlasselaer}},\ }\href {\doibase 10.1007/JHEP09(2020)085} {\bibfield
  {journal} {\bibinfo  {journal} {JHEP}\ }\textbf {\bibinfo {volume} {09}},\
  \bibinfo {pages} {085} (\bibinfo {year} {2020})},\ \Eprint
  {http://arxiv.org/abs/2003.10265} {arXiv:2003.10265 [hep-ph]} \BibitemShut
  {NoStop}%
\bibitem [{\citenamefont {Agazie}\ \emph {et~al.}(2023)\citenamefont {Agazie}
  \emph {et~al.}}]{NANOGrav:2023gor}%
  \BibitemOpen
  \bibfield  {author} {\bibinfo {author} {\bibfnamefont {G.}~\bibnamefont
  {Agazie}} \emph {et~al.} (\bibinfo {collaboration} {NANOGrav}),\ }\href
  {\doibase 10.3847/2041-8213/acdac6} {\bibfield  {journal} {\bibinfo
  {journal} {Astrophys. J. Lett.}\ }\textbf {\bibinfo {volume} {951}},\
  \bibinfo {pages} {L8} (\bibinfo {year} {2023})},\ \Eprint
  {http://arxiv.org/abs/2306.16213} {arXiv:2306.16213 [astro-ph.HE]}
  \BibitemShut {NoStop}%
\bibitem [{\citenamefont {Antoniadis}\ \emph {et~al.}(2023)\citenamefont
  {Antoniadis} \emph {et~al.}}]{EPTA:2023fyk}%
  \BibitemOpen
  \bibfield  {author} {\bibinfo {author} {\bibfnamefont {J.}~\bibnamefont
  {Antoniadis}} \emph {et~al.} (\bibinfo {collaboration} {EPTA, InPTA:}),\
  }\href {\doibase 10.1051/0004-6361/202346844} {\bibfield  {journal} {\bibinfo
   {journal} {Astron. Astrophys.}\ }\textbf {\bibinfo {volume} {678}},\
  \bibinfo {pages} {A50} (\bibinfo {year} {2023})},\ \Eprint
  {http://arxiv.org/abs/2306.16214} {arXiv:2306.16214 [astro-ph.HE]}
  \BibitemShut {NoStop}%
\bibitem [{\citenamefont {Xu}\ \emph {et~al.}(2023)\citenamefont {Xu} \emph
  {et~al.}}]{Xu:2023wog}%
  \BibitemOpen
  \bibfield  {author} {\bibinfo {author} {\bibfnamefont {H.}~\bibnamefont {Xu}}
  \emph {et~al.},\ }\href {\doibase 10.1088/1674-4527/acdfa5} {\bibfield
  {journal} {\bibinfo  {journal} {Res. Astron. Astrophys.}\ }\textbf {\bibinfo
  {volume} {23}},\ \bibinfo {pages} {075024} (\bibinfo {year} {2023})},\
  \Eprint {http://arxiv.org/abs/2306.16216} {arXiv:2306.16216 [astro-ph.HE]}
  \BibitemShut {NoStop}%
\bibitem [{\citenamefont {Reardon}\ \emph {et~al.}(2023)\citenamefont {Reardon}
  \emph {et~al.}}]{Reardon:2023gzh}%
  \BibitemOpen
  \bibfield  {author} {\bibinfo {author} {\bibfnamefont {D.~J.}\ \bibnamefont
  {Reardon}} \emph {et~al.},\ }\href {\doibase 10.3847/2041-8213/acdd02}
  {\bibfield  {journal} {\bibinfo  {journal} {Astrophys. J. Lett.}\ }\textbf
  {\bibinfo {volume} {951}},\ \bibinfo {pages} {L6} (\bibinfo {year} {2023})},\
  \Eprint {http://arxiv.org/abs/2306.16215} {arXiv:2306.16215 [astro-ph.HE]}
  \BibitemShut {NoStop}%
\bibitem [{\citenamefont {Intriligator}\ and\ \citenamefont
  {Seiberg}(2007)}]{Intriligator:2007cp}%
  \BibitemOpen
  \bibfield  {author} {\bibinfo {author} {\bibfnamefont {K.~A.}\ \bibnamefont
  {Intriligator}}\ and\ \bibinfo {author} {\bibfnamefont {N.}~\bibnamefont
  {Seiberg}},\ }\href {\doibase 10.1088/0264-9381/24/21/S02} {\bibfield
  {journal} {\bibinfo  {journal} {Class. Quant. Grav.}\ }\textbf {\bibinfo
  {volume} {24}},\ \bibinfo {pages} {S741} (\bibinfo {year} {2007})},\ \Eprint
  {http://arxiv.org/abs/hep-ph/0702069} {arXiv:hep-ph/0702069} \BibitemShut
  {NoStop}%
\bibitem [{\citenamefont {Peccei}\ and\ \citenamefont
  {Quinn}(1977)}]{Peccei:1977hh}%
  \BibitemOpen
  \bibfield  {author} {\bibinfo {author} {\bibfnamefont {R.~D.}\ \bibnamefont
  {Peccei}}\ and\ \bibinfo {author} {\bibfnamefont {H.~R.}\ \bibnamefont
  {Quinn}},\ }\href {\doibase 10.1103/PhysRevLett.38.1440} {\bibfield
  {journal} {\bibinfo  {journal} {Phys. Rev. Lett.}\ }\textbf {\bibinfo
  {volume} {38}},\ \bibinfo {pages} {1440} (\bibinfo {year}
  {1977})}\BibitemShut {NoStop}%
\bibitem [{\citenamefont {Nakai}\ and\ \citenamefont
  {Suzuki}(2021)}]{Nakai:2021nyf}%
  \BibitemOpen
  \bibfield  {author} {\bibinfo {author} {\bibfnamefont {Y.}~\bibnamefont
  {Nakai}}\ and\ \bibinfo {author} {\bibfnamefont {M.}~\bibnamefont {Suzuki}},\
  }\href {\doibase 10.1016/j.physletb.2021.136239} {\bibfield  {journal}
  {\bibinfo  {journal} {Phys. Lett. B}\ }\textbf {\bibinfo {volume} {816}},\
  \bibinfo {pages} {136239} (\bibinfo {year} {2021})},\ \Eprint
  {http://arxiv.org/abs/2102.01329} {arXiv:2102.01329 [hep-ph]} \BibitemShut
  {NoStop}%
\bibitem [{\citenamefont {Nakagawa}\ \emph
  {et~al.}(2024{\natexlab{a}})\citenamefont {Nakagawa}, \citenamefont {Nakai},
  \citenamefont {Yamada},\ and\ \citenamefont {Zhang}}]{Nakagawa:2023shi}%
  \BibitemOpen
  \bibfield  {author} {\bibinfo {author} {\bibfnamefont {S.}~\bibnamefont
  {Nakagawa}}, \bibinfo {author} {\bibfnamefont {Y.}~\bibnamefont {Nakai}},
  \bibinfo {author} {\bibfnamefont {M.}~\bibnamefont {Yamada}}, \ and\ \bibinfo
  {author} {\bibfnamefont {Y.}~\bibnamefont {Zhang}},\ }\href {\doibase
  10.1016/j.physletb.2024.138447} {\bibfield  {journal} {\bibinfo  {journal}
  {Phys. Lett. B}\ }\textbf {\bibinfo {volume} {849}},\ \bibinfo {pages}
  {138447} (\bibinfo {year} {2024}{\natexlab{a}})},\ \Eprint
  {http://arxiv.org/abs/2309.06964} {arXiv:2309.06964 [hep-ph]} \BibitemShut
  {NoStop}%
\bibitem [{\citenamefont {Nakagawa}\ \emph
  {et~al.}(2024{\natexlab{b}})\citenamefont {Nakagawa}, \citenamefont {Nakai},
  \citenamefont {Xu},\ and\ \citenamefont {Zhang}}]{Nakagawa:2024kcb}%
  \BibitemOpen
  \bibfield  {author} {\bibinfo {author} {\bibfnamefont {S.}~\bibnamefont
  {Nakagawa}}, \bibinfo {author} {\bibfnamefont {Y.}~\bibnamefont {Nakai}},
  \bibinfo {author} {\bibfnamefont {J.}~\bibnamefont {Xu}}, \ and\ \bibinfo
  {author} {\bibfnamefont {Y.}~\bibnamefont {Zhang}},\ }\href@noop {} {\
  (\bibinfo {year} {2024}{\natexlab{b}})},\ \Eprint
  {http://arxiv.org/abs/2412.08931} {arXiv:2412.08931 [hep-ph]} \BibitemShut
  {NoStop}%
\bibitem [{\citenamefont {Weinberg}(1974)}]{Weinberg:1974hy}%
  \BibitemOpen
  \bibfield  {author} {\bibinfo {author} {\bibfnamefont {S.}~\bibnamefont
  {Weinberg}},\ }\href {\doibase 10.1103/PhysRevD.9.3357} {\bibfield  {journal}
  {\bibinfo  {journal} {Phys. Rev. D}\ }\textbf {\bibinfo {volume} {9}},\
  \bibinfo {pages} {3357} (\bibinfo {year} {1974})}\BibitemShut {NoStop}%
\bibitem [{\citenamefont {Linde}(1980)}]{Linde:1980ts}%
  \BibitemOpen
  \bibfield  {author} {\bibinfo {author} {\bibfnamefont {A.~D.}\ \bibnamefont
  {Linde}},\ }\href {\doibase 10.1016/0370-2693(80)90769-8} {\bibfield
  {journal} {\bibinfo  {journal} {Phys. Lett. B}\ }\textbf {\bibinfo {volume}
  {96}},\ \bibinfo {pages} {289} (\bibinfo {year} {1980})}\BibitemShut
  {NoStop}%
\bibitem [{\citenamefont {Arnold}\ and\ \citenamefont
  {Espinosa}(1993)}]{Arnold:1992rz}%
  \BibitemOpen
  \bibfield  {author} {\bibinfo {author} {\bibfnamefont {P.~B.}\ \bibnamefont
  {Arnold}}\ and\ \bibinfo {author} {\bibfnamefont {O.}~\bibnamefont
  {Espinosa}},\ }\href {\doibase 10.1103/PhysRevD.47.3546} {\bibfield
  {journal} {\bibinfo  {journal} {Phys. Rev. D}\ }\textbf {\bibinfo {volume}
  {47}},\ \bibinfo {pages} {3546} (\bibinfo {year} {1993})},\ \bibinfo {note}
  {[Erratum: Phys.Rev.D 50, 6662 (1994)]},\ \Eprint
  {http://arxiv.org/abs/hep-ph/9212235} {arXiv:hep-ph/9212235} \BibitemShut
  {NoStop}%
\bibitem [{\citenamefont {Glioti}\ \emph {et~al.}(2019)\citenamefont {Glioti},
  \citenamefont {Rattazzi},\ and\ \citenamefont {Vecchi}}]{Glioti:2018roy}%
  \BibitemOpen
  \bibfield  {author} {\bibinfo {author} {\bibfnamefont {A.}~\bibnamefont
  {Glioti}}, \bibinfo {author} {\bibfnamefont {R.}~\bibnamefont {Rattazzi}}, \
  and\ \bibinfo {author} {\bibfnamefont {L.}~\bibnamefont {Vecchi}},\ }\href
  {\doibase 10.1007/JHEP04(2019)027} {\bibfield  {journal} {\bibinfo  {journal}
  {JHEP}\ }\textbf {\bibinfo {volume} {04}},\ \bibinfo {pages} {027} (\bibinfo
  {year} {2019})},\ \Eprint {http://arxiv.org/abs/1811.11740} {arXiv:1811.11740
  [hep-ph]} \BibitemShut {NoStop}%
\bibitem [{\citenamefont {Dolan}\ and\ \citenamefont
  {Jackiw}(1974)}]{Dolan:1973qd}%
  \BibitemOpen
  \bibfield  {author} {\bibinfo {author} {\bibfnamefont {L.}~\bibnamefont
  {Dolan}}\ and\ \bibinfo {author} {\bibfnamefont {R.}~\bibnamefont {Jackiw}},\
  }\href {\doibase 10.1103/PhysRevD.9.3320} {\bibfield  {journal} {\bibinfo
  {journal} {Phys. Rev. D}\ }\textbf {\bibinfo {volume} {9}},\ \bibinfo {pages}
  {3320} (\bibinfo {year} {1974})}\BibitemShut {NoStop}%
\bibitem [{\citenamefont {Quiros}(1999)}]{Quiros:1999jp}%
  \BibitemOpen
  \bibfield  {author} {\bibinfo {author} {\bibfnamefont {M.}~\bibnamefont
  {Quiros}},\ }in\ \href@noop {} {\emph {\bibinfo {booktitle} {{ICTP Summer
  School in High-Energy Physics and Cosmology}}}}\ (\bibinfo {year} {1999})\
  pp.\ \bibinfo {pages} {187--259},\ \Eprint
  {http://arxiv.org/abs/hep-ph/9901312} {arXiv:hep-ph/9901312} \BibitemShut
  {NoStop}%
\bibitem [{\citenamefont {Coleman}(1985)}]{Coleman:1985rnk}%
  \BibitemOpen
  \bibfield  {author} {\bibinfo {author} {\bibfnamefont {S.}~\bibnamefont
  {Coleman}},\ }\href {\doibase 10.1017/CBO9780511565045} {\emph {\bibinfo
  {title} {{Aspects of Symmetry}: {Selected Erice Lectures}}}}\ (\bibinfo
  {publisher} {Cambridge University Press},\ \bibinfo {address} {Cambridge,
  U.K.},\ \bibinfo {year} {1985})\BibitemShut {NoStop}%
\bibitem [{\citenamefont {Linde}(1983)}]{Linde:1981zj}%
  \BibitemOpen
  \bibfield  {author} {\bibinfo {author} {\bibfnamefont {A.~D.}\ \bibnamefont
  {Linde}},\ }\href {\doibase 10.1016/0550-3213(83)90072-X} {\bibfield
  {journal} {\bibinfo  {journal} {Nucl. Phys. B}\ }\textbf {\bibinfo {volume}
  {216}},\ \bibinfo {pages} {421} (\bibinfo {year} {1983})},\ \bibinfo {note}
  {[Erratum: Nucl.Phys.B 223, 544 (1983)]}\BibitemShut {NoStop}%
\bibitem [{\citenamefont {Coleman}\ \emph {et~al.}(1978)\citenamefont
  {Coleman}, \citenamefont {Glaser},\ and\ \citenamefont
  {Martin}}]{Coleman:1977th}%
  \BibitemOpen
  \bibfield  {author} {\bibinfo {author} {\bibfnamefont {S.~R.}\ \bibnamefont
  {Coleman}}, \bibinfo {author} {\bibfnamefont {V.}~\bibnamefont {Glaser}}, \
  and\ \bibinfo {author} {\bibfnamefont {A.}~\bibnamefont {Martin}},\ }\href
  {\doibase 10.1007/BF01609421} {\bibfield  {journal} {\bibinfo  {journal}
  {Commun. Math. Phys.}\ }\textbf {\bibinfo {volume} {58}},\ \bibinfo {pages}
  {211} (\bibinfo {year} {1978})}\BibitemShut {NoStop}%
\bibitem [{\citenamefont {Nardini}\ \emph {et~al.}(2007)\citenamefont
  {Nardini}, \citenamefont {Quiros},\ and\ \citenamefont
  {Wulzer}}]{Nardini:2007me}%
  \BibitemOpen
  \bibfield  {author} {\bibinfo {author} {\bibfnamefont {G.}~\bibnamefont
  {Nardini}}, \bibinfo {author} {\bibfnamefont {M.}~\bibnamefont {Quiros}}, \
  and\ \bibinfo {author} {\bibfnamefont {A.}~\bibnamefont {Wulzer}},\ }\href
  {\doibase 10.1088/1126-6708/2007/09/077} {\bibfield  {journal} {\bibinfo
  {journal} {JHEP}\ }\textbf {\bibinfo {volume} {09}},\ \bibinfo {pages} {077}
  (\bibinfo {year} {2007})},\ \Eprint {http://arxiv.org/abs/0706.3388}
  {arXiv:0706.3388 [hep-ph]} \BibitemShut {NoStop}%
\bibitem [{\citenamefont {Caprini}\ \emph {et~al.}(2016)\citenamefont {Caprini}
  \emph {et~al.}}]{Caprini:2015zlo}%
  \BibitemOpen
  \bibfield  {author} {\bibinfo {author} {\bibfnamefont {C.}~\bibnamefont
  {Caprini}} \emph {et~al.},\ }\href {\doibase 10.1088/1475-7516/2016/04/001}
  {\bibfield  {journal} {\bibinfo  {journal} {JCAP}\ }\textbf {\bibinfo
  {volume} {04}},\ \bibinfo {pages} {001} (\bibinfo {year} {2016})},\ \Eprint
  {http://arxiv.org/abs/1512.06239} {arXiv:1512.06239 [astro-ph.CO]}
  \BibitemShut {NoStop}%
\bibitem [{\citenamefont {Kosowsky}\ \emph
  {et~al.}(1992{\natexlab{b}})\citenamefont {Kosowsky}, \citenamefont
  {Turner},\ and\ \citenamefont {Watkins}}]{Kosowsky:1991ua}%
  \BibitemOpen
  \bibfield  {author} {\bibinfo {author} {\bibfnamefont {A.}~\bibnamefont
  {Kosowsky}}, \bibinfo {author} {\bibfnamefont {M.~S.}\ \bibnamefont
  {Turner}}, \ and\ \bibinfo {author} {\bibfnamefont {R.}~\bibnamefont
  {Watkins}},\ }\href {\doibase 10.1103/PhysRevD.45.4514} {\bibfield  {journal}
  {\bibinfo  {journal} {Phys. Rev. D}\ }\textbf {\bibinfo {volume} {45}},\
  \bibinfo {pages} {4514} (\bibinfo {year} {1992}{\natexlab{b}})}\BibitemShut
  {NoStop}%
\bibitem [{\citenamefont {Jinno}\ and\ \citenamefont
  {Takimoto}(2017{\natexlab{b}})}]{Jinno:2016vai}%
  \BibitemOpen
  \bibfield  {author} {\bibinfo {author} {\bibfnamefont {R.}~\bibnamefont
  {Jinno}}\ and\ \bibinfo {author} {\bibfnamefont {M.}~\bibnamefont
  {Takimoto}},\ }\href {\doibase 10.1103/PhysRevD.95.024009} {\bibfield
  {journal} {\bibinfo  {journal} {Phys. Rev. D}\ }\textbf {\bibinfo {volume}
  {95}},\ \bibinfo {pages} {024009} (\bibinfo {year} {2017}{\natexlab{b}})},\
  \Eprint {http://arxiv.org/abs/1605.01403} {arXiv:1605.01403 [astro-ph.CO]}
  \BibitemShut {NoStop}%
\bibitem [{\citenamefont {Jinno}\ and\ \citenamefont
  {Takimoto}(2019)}]{Jinno:2017fby}%
  \BibitemOpen
  \bibfield  {author} {\bibinfo {author} {\bibfnamefont {R.}~\bibnamefont
  {Jinno}}\ and\ \bibinfo {author} {\bibfnamefont {M.}~\bibnamefont
  {Takimoto}},\ }\href {\doibase 10.1088/1475-7516/2019/01/060} {\bibfield
  {journal} {\bibinfo  {journal} {JCAP}\ }\textbf {\bibinfo {volume} {01}},\
  \bibinfo {pages} {060} (\bibinfo {year} {2019})},\ \Eprint
  {http://arxiv.org/abs/1707.03111} {arXiv:1707.03111 [hep-ph]} \BibitemShut
  {NoStop}%
\bibitem [{\citenamefont {Espinosa}\ \emph {et~al.}(2010)\citenamefont
  {Espinosa}, \citenamefont {Konstandin}, \citenamefont {No},\ and\
  \citenamefont {Servant}}]{Espinosa:2010hh}%
  \BibitemOpen
  \bibfield  {author} {\bibinfo {author} {\bibfnamefont {J.~R.}\ \bibnamefont
  {Espinosa}}, \bibinfo {author} {\bibfnamefont {T.}~\bibnamefont
  {Konstandin}}, \bibinfo {author} {\bibfnamefont {J.~M.}\ \bibnamefont {No}},
  \ and\ \bibinfo {author} {\bibfnamefont {G.}~\bibnamefont {Servant}},\ }\href
  {\doibase 10.1088/1475-7516/2010/06/028} {\bibfield  {journal} {\bibinfo
  {journal} {JCAP}\ }\textbf {\bibinfo {volume} {06}},\ \bibinfo {pages} {028}
  (\bibinfo {year} {2010})},\ \Eprint {http://arxiv.org/abs/1004.4187}
  {arXiv:1004.4187 [hep-ph]} \BibitemShut {NoStop}%
\bibitem [{\citenamefont {Bodeker}\ and\ \citenamefont
  {Moore}(2009)}]{Bodeker:2009qy}%
  \BibitemOpen
  \bibfield  {author} {\bibinfo {author} {\bibfnamefont {D.}~\bibnamefont
  {Bodeker}}\ and\ \bibinfo {author} {\bibfnamefont {G.~D.}\ \bibnamefont
  {Moore}},\ }\href {\doibase 10.1088/1475-7516/2009/05/009} {\bibfield
  {journal} {\bibinfo  {journal} {JCAP}\ }\textbf {\bibinfo {volume} {05}},\
  \bibinfo {pages} {009} (\bibinfo {year} {2009})},\ \Eprint
  {http://arxiv.org/abs/0903.4099} {arXiv:0903.4099 [hep-ph]} \BibitemShut
  {NoStop}%
\bibitem [{\citenamefont {Bodeker}\ and\ \citenamefont
  {Moore}(2017)}]{Bodeker:2017cim}%
  \BibitemOpen
  \bibfield  {author} {\bibinfo {author} {\bibfnamefont {D.}~\bibnamefont
  {Bodeker}}\ and\ \bibinfo {author} {\bibfnamefont {G.~D.}\ \bibnamefont
  {Moore}},\ }\href {\doibase 10.1088/1475-7516/2017/05/025} {\bibfield
  {journal} {\bibinfo  {journal} {JCAP}\ }\textbf {\bibinfo {volume} {05}},\
  \bibinfo {pages} {025} (\bibinfo {year} {2017})},\ \Eprint
  {http://arxiv.org/abs/1703.08215} {arXiv:1703.08215 [hep-ph]} \BibitemShut
  {NoStop}%
\bibitem [{\citenamefont {Dorsch}\ \emph {et~al.}(2018)\citenamefont {Dorsch},
  \citenamefont {Huber},\ and\ \citenamefont {Konstandin}}]{Dorsch:2018pat}%
  \BibitemOpen
  \bibfield  {author} {\bibinfo {author} {\bibfnamefont {G.~C.}\ \bibnamefont
  {Dorsch}}, \bibinfo {author} {\bibfnamefont {S.~J.}\ \bibnamefont {Huber}}, \
  and\ \bibinfo {author} {\bibfnamefont {T.}~\bibnamefont {Konstandin}},\
  }\href {\doibase 10.1088/1475-7516/2018/12/034} {\bibfield  {journal}
  {\bibinfo  {journal} {JCAP}\ }\textbf {\bibinfo {volume} {12}},\ \bibinfo
  {pages} {034} (\bibinfo {year} {2018})},\ \Eprint
  {http://arxiv.org/abs/1809.04907} {arXiv:1809.04907 [hep-ph]} \BibitemShut
  {NoStop}%
\bibitem [{\citenamefont {Barroso~Mancha}\ \emph {et~al.}(2021)\citenamefont
  {Barroso~Mancha}, \citenamefont {Prokopec},\ and\ \citenamefont
  {Swiezewska}}]{BarrosoMancha:2020fay}%
  \BibitemOpen
  \bibfield  {author} {\bibinfo {author} {\bibfnamefont {M.}~\bibnamefont
  {Barroso~Mancha}}, \bibinfo {author} {\bibfnamefont {T.}~\bibnamefont
  {Prokopec}}, \ and\ \bibinfo {author} {\bibfnamefont {B.}~\bibnamefont
  {Swiezewska}},\ }\href {\doibase 10.1007/JHEP01(2021)070} {\bibfield
  {journal} {\bibinfo  {journal} {JHEP}\ }\textbf {\bibinfo {volume} {01}},\
  \bibinfo {pages} {070} (\bibinfo {year} {2021})},\ \Eprint
  {http://arxiv.org/abs/2005.10875} {arXiv:2005.10875 [hep-th]} \BibitemShut
  {NoStop}%
\bibitem [{\citenamefont {Laurent}\ and\ \citenamefont
  {Cline}(2020)}]{Laurent:2020gpg}%
  \BibitemOpen
  \bibfield  {author} {\bibinfo {author} {\bibfnamefont {B.}~\bibnamefont
  {Laurent}}\ and\ \bibinfo {author} {\bibfnamefont {J.~M.}\ \bibnamefont
  {Cline}},\ }\href {\doibase 10.1103/PhysRevD.102.063516} {\bibfield
  {journal} {\bibinfo  {journal} {Phys. Rev. D}\ }\textbf {\bibinfo {volume}
  {102}},\ \bibinfo {pages} {063516} (\bibinfo {year} {2020})},\ \Eprint
  {http://arxiv.org/abs/2007.10935} {arXiv:2007.10935 [hep-ph]} \BibitemShut
  {NoStop}%
\bibitem [{\citenamefont {Azatov}\ and\ \citenamefont
  {Vanvlasselaer}(2021)}]{Azatov:2020ufh}%
  \BibitemOpen
  \bibfield  {author} {\bibinfo {author} {\bibfnamefont {A.}~\bibnamefont
  {Azatov}}\ and\ \bibinfo {author} {\bibfnamefont {M.}~\bibnamefont
  {Vanvlasselaer}},\ }\href {\doibase 10.1088/1475-7516/2021/01/058} {\bibfield
   {journal} {\bibinfo  {journal} {JCAP}\ }\textbf {\bibinfo {volume} {01}},\
  \bibinfo {pages} {058} (\bibinfo {year} {2021})},\ \Eprint
  {http://arxiv.org/abs/2010.02590} {arXiv:2010.02590 [hep-ph]} \BibitemShut
  {NoStop}%
\bibitem [{\citenamefont {Wang}\ \emph {et~al.}(2020)\citenamefont {Wang},
  \citenamefont {Huang},\ and\ \citenamefont {Zhang}}]{Wang:2020zlf}%
  \BibitemOpen
  \bibfield  {author} {\bibinfo {author} {\bibfnamefont {X.}~\bibnamefont
  {Wang}}, \bibinfo {author} {\bibfnamefont {F.~P.}\ \bibnamefont {Huang}}, \
  and\ \bibinfo {author} {\bibfnamefont {X.}~\bibnamefont {Zhang}},\
  }\href@noop {} {\  (\bibinfo {year} {2020})},\ \Eprint
  {http://arxiv.org/abs/2011.12903} {arXiv:2011.12903 [hep-ph]} \BibitemShut
  {NoStop}%
\bibitem [{\citenamefont {Gouttenoire}\ \emph {et~al.}(2022)\citenamefont
  {Gouttenoire}, \citenamefont {Jinno},\ and\ \citenamefont
  {Sala}}]{Gouttenoire:2021kjv}%
  \BibitemOpen
  \bibfield  {author} {\bibinfo {author} {\bibfnamefont {Y.}~\bibnamefont
  {Gouttenoire}}, \bibinfo {author} {\bibfnamefont {R.}~\bibnamefont {Jinno}},
  \ and\ \bibinfo {author} {\bibfnamefont {F.}~\bibnamefont {Sala}},\ }\href
  {\doibase 10.1007/JHEP05(2022)004} {\bibfield  {journal} {\bibinfo  {journal}
  {JHEP}\ }\textbf {\bibinfo {volume} {05}},\ \bibinfo {pages} {004} (\bibinfo
  {year} {2022})},\ \Eprint {http://arxiv.org/abs/2112.07686} {arXiv:2112.07686
  [hep-ph]} \BibitemShut {NoStop}%
\bibitem [{\citenamefont {Ai}\ \emph {et~al.}(2023)\citenamefont {Ai},
  \citenamefont {Laurent},\ and\ \citenamefont {van~de Vis}}]{Ai:2023see}%
  \BibitemOpen
  \bibfield  {author} {\bibinfo {author} {\bibfnamefont {W.-Y.}\ \bibnamefont
  {Ai}}, \bibinfo {author} {\bibfnamefont {B.}~\bibnamefont {Laurent}}, \ and\
  \bibinfo {author} {\bibfnamefont {J.}~\bibnamefont {van~de Vis}},\ }\href
  {\doibase 10.1088/1475-7516/2023/07/002} {\bibfield  {journal} {\bibinfo
  {journal} {JCAP}\ }\textbf {\bibinfo {volume} {07}},\ \bibinfo {pages} {002}
  (\bibinfo {year} {2023})},\ \Eprint {http://arxiv.org/abs/2303.10171}
  {arXiv:2303.10171 [astro-ph.CO]} \BibitemShut {NoStop}%
\bibitem [{\citenamefont {Krajewski}\ \emph {et~al.}(2023)\citenamefont
  {Krajewski}, \citenamefont {Lewicki},\ and\ \citenamefont
  {Zych}}]{Krajewski:2023clt}%
  \BibitemOpen
  \bibfield  {author} {\bibinfo {author} {\bibfnamefont {T.}~\bibnamefont
  {Krajewski}}, \bibinfo {author} {\bibfnamefont {M.}~\bibnamefont {Lewicki}},
  \ and\ \bibinfo {author} {\bibfnamefont {M.}~\bibnamefont {Zych}},\ }\href
  {\doibase 10.1103/PhysRevD.108.103523} {\bibfield  {journal} {\bibinfo
  {journal} {Phys. Rev. D}\ }\textbf {\bibinfo {volume} {108}},\ \bibinfo
  {pages} {103523} (\bibinfo {year} {2023})},\ \Eprint
  {http://arxiv.org/abs/2303.18216} {arXiv:2303.18216 [astro-ph.CO]}
  \BibitemShut {NoStop}%
\bibitem [{\citenamefont {Li}\ \emph {et~al.}(2023)\citenamefont {Li},
  \citenamefont {Wang},\ and\ \citenamefont {Yuwen}}]{Li:2023xto}%
  \BibitemOpen
  \bibfield  {author} {\bibinfo {author} {\bibfnamefont {L.}~\bibnamefont
  {Li}}, \bibinfo {author} {\bibfnamefont {S.-J.}\ \bibnamefont {Wang}}, \ and\
  \bibinfo {author} {\bibfnamefont {Z.-Y.}\ \bibnamefont {Yuwen}},\ }\href
  {\doibase 10.1103/PhysRevD.108.096033} {\bibfield  {journal} {\bibinfo
  {journal} {Phys. Rev. D}\ }\textbf {\bibinfo {volume} {108}},\ \bibinfo
  {pages} {096033} (\bibinfo {year} {2023})},\ \Eprint
  {http://arxiv.org/abs/2302.10042} {arXiv:2302.10042 [hep-th]} \BibitemShut
  {NoStop}%
\bibitem [{\citenamefont {Abbott}\ \emph {et~al.}(2021)\citenamefont {Abbott}
  \emph {et~al.}}]{KAGRA:2021kbb}%
  \BibitemOpen
  \bibfield  {author} {\bibinfo {author} {\bibfnamefont {R.}~\bibnamefont
  {Abbott}} \emph {et~al.} (\bibinfo {collaboration} {KAGRA, Virgo, LIGO
  Scientific}),\ }\href {\doibase 10.1103/PhysRevD.104.022004} {\bibfield
  {journal} {\bibinfo  {journal} {Phys. Rev. D}\ }\textbf {\bibinfo {volume}
  {104}},\ \bibinfo {pages} {022004} (\bibinfo {year} {2021})},\ \Eprint
  {http://arxiv.org/abs/2101.12130} {arXiv:2101.12130 [gr-qc]} \BibitemShut
  {NoStop}%
\bibitem [{\citenamefont {Aghanim}\ \emph {et~al.}(2020)\citenamefont {Aghanim}
  \emph {et~al.}}]{Planck:2018vyg}%
  \BibitemOpen
  \bibfield  {author} {\bibinfo {author} {\bibfnamefont {N.}~\bibnamefont
  {Aghanim}} \emph {et~al.} (\bibinfo {collaboration} {Planck}),\ }\href
  {\doibase 10.1051/0004-6361/201833910} {\bibfield  {journal} {\bibinfo
  {journal} {Astron. Astrophys.}\ }\textbf {\bibinfo {volume} {641}},\ \bibinfo
  {pages} {A6} (\bibinfo {year} {2020})},\ \bibinfo {note} {[Erratum:
  Astron.Astrophys. 652, C4 (2021)]},\ \Eprint
  {http://arxiv.org/abs/1807.06209} {arXiv:1807.06209 [astro-ph.CO]}
  \BibitemShut {NoStop}%
\bibitem [{\citenamefont {Sehgal}\ \emph {et~al.}(2019)\citenamefont {Sehgal}
  \emph {et~al.}}]{Sehgal:2019ewc}%
  \BibitemOpen
  \bibfield  {author} {\bibinfo {author} {\bibfnamefont {N.}~\bibnamefont
  {Sehgal}} \emph {et~al.},\ }\href@noop {} {\  (\bibinfo {year} {2019})},\
  \Eprint {http://arxiv.org/abs/1906.10134} {arXiv:1906.10134 [astro-ph.CO]}
  \BibitemShut {NoStop}%
\bibitem [{\citenamefont {Aiola}\ \emph {et~al.}(2022)\citenamefont {Aiola}
  \emph {et~al.}}]{CMB-HD:2022bsz}%
  \BibitemOpen
  \bibfield  {author} {\bibinfo {author} {\bibfnamefont {S.}~\bibnamefont
  {Aiola}} \emph {et~al.} (\bibinfo {collaboration} {CMB-HD}),\ }\href@noop {}
  {\  (\bibinfo {year} {2022})},\ \Eprint {http://arxiv.org/abs/2203.05728}
  {arXiv:2203.05728 [astro-ph.CO]} \BibitemShut {NoStop}%
\bibitem [{\citenamefont {Nakai}\ \emph {et~al.}(2021)\citenamefont {Nakai},
  \citenamefont {Suzuki}, \citenamefont {Takahashi},\ and\ \citenamefont
  {Yamada}}]{Nakai:2020oit}%
  \BibitemOpen
  \bibfield  {author} {\bibinfo {author} {\bibfnamefont {Y.}~\bibnamefont
  {Nakai}}, \bibinfo {author} {\bibfnamefont {M.}~\bibnamefont {Suzuki}},
  \bibinfo {author} {\bibfnamefont {F.}~\bibnamefont {Takahashi}}, \ and\
  \bibinfo {author} {\bibfnamefont {M.}~\bibnamefont {Yamada}},\ }\href
  {\doibase 10.1016/j.physletb.2021.136238} {\bibfield  {journal} {\bibinfo
  {journal} {Phys. Lett. B}\ }\textbf {\bibinfo {volume} {816}},\ \bibinfo
  {pages} {136238} (\bibinfo {year} {2021})},\ \Eprint
  {http://arxiv.org/abs/2009.09754} {arXiv:2009.09754 [astro-ph.CO]}
  \BibitemShut {NoStop}%
\bibitem [{\citenamefont {Fujikura}\ \emph {et~al.}(2023)\citenamefont
  {Fujikura}, \citenamefont {Girmohanta}, \citenamefont {Nakai},\ and\
  \citenamefont {Suzuki}}]{Fujikura:2023lkn}%
  \BibitemOpen
  \bibfield  {author} {\bibinfo {author} {\bibfnamefont {K.}~\bibnamefont
  {Fujikura}}, \bibinfo {author} {\bibfnamefont {S.}~\bibnamefont
  {Girmohanta}}, \bibinfo {author} {\bibfnamefont {Y.}~\bibnamefont {Nakai}}, \
  and\ \bibinfo {author} {\bibfnamefont {M.}~\bibnamefont {Suzuki}},\ }\href
  {\doibase 10.1016/j.physletb.2023.138203} {\bibfield  {journal} {\bibinfo
  {journal} {Phys. Lett. B}\ }\textbf {\bibinfo {volume} {846}},\ \bibinfo
  {pages} {138203} (\bibinfo {year} {2023})},\ \Eprint
  {http://arxiv.org/abs/2306.17086} {arXiv:2306.17086 [hep-ph]} \BibitemShut
  {NoStop}%
\bibitem [{\citenamefont {Bringmann}\ \emph {et~al.}(2023)\citenamefont
  {Bringmann}, \citenamefont {Depta}, \citenamefont {Konstandin}, \citenamefont
  {Schmidt-Hoberg},\ and\ \citenamefont {Tasillo}}]{Bringmann:2023opz}%
  \BibitemOpen
  \bibfield  {author} {\bibinfo {author} {\bibfnamefont {T.}~\bibnamefont
  {Bringmann}}, \bibinfo {author} {\bibfnamefont {P.~F.}\ \bibnamefont
  {Depta}}, \bibinfo {author} {\bibfnamefont {T.}~\bibnamefont {Konstandin}},
  \bibinfo {author} {\bibfnamefont {K.}~\bibnamefont {Schmidt-Hoberg}}, \ and\
  \bibinfo {author} {\bibfnamefont {C.}~\bibnamefont {Tasillo}},\ }\href
  {\doibase 10.1088/1475-7516/2023/11/053} {\bibfield  {journal} {\bibinfo
  {journal} {JCAP}\ }\textbf {\bibinfo {volume} {11}},\ \bibinfo {pages} {053}
  (\bibinfo {year} {2023})},\ \Eprint {http://arxiv.org/abs/2306.09411}
  {arXiv:2306.09411 [astro-ph.CO]} \BibitemShut {NoStop}%
\bibitem [{\citenamefont {Moore}\ \emph {et~al.}(2017)\citenamefont {Moore},
  \citenamefont {Mihaylov}, \citenamefont {Lasenby},\ and\ \citenamefont
  {Gilmore}}]{Moore:2017ity}%
  \BibitemOpen
  \bibfield  {author} {\bibinfo {author} {\bibfnamefont {C.~J.}\ \bibnamefont
  {Moore}}, \bibinfo {author} {\bibfnamefont {D.~P.}\ \bibnamefont {Mihaylov}},
  \bibinfo {author} {\bibfnamefont {A.}~\bibnamefont {Lasenby}}, \ and\
  \bibinfo {author} {\bibfnamefont {G.}~\bibnamefont {Gilmore}},\ }\href
  {\doibase 10.1103/PhysRevLett.119.261102} {\bibfield  {journal} {\bibinfo
  {journal} {Phys. Rev. Lett.}\ }\textbf {\bibinfo {volume} {119}},\ \bibinfo
  {pages} {261102} (\bibinfo {year} {2017})},\ \Eprint
  {http://arxiv.org/abs/1707.06239} {arXiv:1707.06239 [astro-ph.IM]}
  \BibitemShut {NoStop}%
\bibitem [{\citenamefont {Boehm}\ \emph {et~al.}(2017)\citenamefont {Boehm}
  \emph {et~al.}}]{Theia:2017xtk}%
  \BibitemOpen
  \bibfield  {author} {\bibinfo {author} {\bibfnamefont {C.}~\bibnamefont
  {Boehm}} \emph {et~al.} (\bibinfo {collaboration} {Theia}),\ }\href@noop {}
  {\  (\bibinfo {year} {2017})},\ \Eprint {http://arxiv.org/abs/1707.01348}
  {arXiv:1707.01348 [astro-ph.IM]} \BibitemShut {NoStop}%
\bibitem [{\citenamefont {Garcia-Bellido}\ \emph {et~al.}(2021)\citenamefont
  {Garcia-Bellido}, \citenamefont {Murayama},\ and\ \citenamefont
  {White}}]{Garcia-Bellido:2021zgu}%
  \BibitemOpen
  \bibfield  {author} {\bibinfo {author} {\bibfnamefont {J.}~\bibnamefont
  {Garcia-Bellido}}, \bibinfo {author} {\bibfnamefont {H.}~\bibnamefont
  {Murayama}}, \ and\ \bibinfo {author} {\bibfnamefont {G.}~\bibnamefont
  {White}},\ }\href {\doibase 10.1088/1475-7516/2021/12/023} {\bibfield
  {journal} {\bibinfo  {journal} {JCAP}\ }\textbf {\bibinfo {volume} {12}},\
  \bibinfo {pages} {023} (\bibinfo {year} {2021})},\ \Eprint
  {http://arxiv.org/abs/2104.04778} {arXiv:2104.04778 [hep-ph]} \BibitemShut
  {NoStop}%
\bibitem [{\citenamefont {Breitbach}\ \emph {et~al.}(2019)\citenamefont
  {Breitbach}, \citenamefont {Kopp}, \citenamefont {Madge}, \citenamefont
  {Opferkuch},\ and\ \citenamefont {Schwaller}}]{Breitbach:2018ddu}%
  \BibitemOpen
  \bibfield  {author} {\bibinfo {author} {\bibfnamefont {M.}~\bibnamefont
  {Breitbach}}, \bibinfo {author} {\bibfnamefont {J.}~\bibnamefont {Kopp}},
  \bibinfo {author} {\bibfnamefont {E.}~\bibnamefont {Madge}}, \bibinfo
  {author} {\bibfnamefont {T.}~\bibnamefont {Opferkuch}}, \ and\ \bibinfo
  {author} {\bibfnamefont {P.}~\bibnamefont {Schwaller}},\ }\href {\doibase
  10.1088/1475-7516/2019/07/007} {\bibfield  {journal} {\bibinfo  {journal}
  {JCAP}\ }\textbf {\bibinfo {volume} {07}},\ \bibinfo {pages} {007} (\bibinfo
  {year} {2019})},\ \Eprint {http://arxiv.org/abs/1811.11175} {arXiv:1811.11175
  [hep-ph]} \BibitemShut {NoStop}%
\bibitem [{\citenamefont {Sesana}\ \emph {et~al.}(2021)\citenamefont {Sesana}
  \emph {et~al.}}]{Sesana:2019vho}%
  \BibitemOpen
  \bibfield  {author} {\bibinfo {author} {\bibfnamefont {A.}~\bibnamefont
  {Sesana}} \emph {et~al.},\ }\href {\doibase 10.1007/s10686-021-09709-9}
  {\bibfield  {journal} {\bibinfo  {journal} {Exper. Astron.}\ }\textbf
  {\bibinfo {volume} {51}},\ \bibinfo {pages} {1333} (\bibinfo {year}
  {2021})},\ \Eprint {http://arxiv.org/abs/1908.11391} {arXiv:1908.11391
  [astro-ph.IM]} \BibitemShut {NoStop}%
\bibitem [{\citenamefont {Cornish}\ and\ \citenamefont
  {Robson}(2017)}]{Cornish:2017vip}%
  \BibitemOpen
  \bibfield  {author} {\bibinfo {author} {\bibfnamefont {N.}~\bibnamefont
  {Cornish}}\ and\ \bibinfo {author} {\bibfnamefont {T.}~\bibnamefont
  {Robson}},\ }\href {\doibase 10.1088/1742-6596/840/1/012024} {\bibfield
  {journal} {\bibinfo  {journal} {J. Phys. Conf. Ser.}\ }\textbf {\bibinfo
  {volume} {840}},\ \bibinfo {pages} {012024} (\bibinfo {year} {2017})},\
  \Eprint {http://arxiv.org/abs/1703.09858} {arXiv:1703.09858 [astro-ph.IM]}
  \BibitemShut {NoStop}%
\bibitem [{\citenamefont {Robson}\ \emph {et~al.}(2019)\citenamefont {Robson},
  \citenamefont {Cornish},\ and\ \citenamefont {Liu}}]{Robson:2018ifk}%
  \BibitemOpen
  \bibfield  {author} {\bibinfo {author} {\bibfnamefont {T.}~\bibnamefont
  {Robson}}, \bibinfo {author} {\bibfnamefont {N.~J.}\ \bibnamefont {Cornish}},
  \ and\ \bibinfo {author} {\bibfnamefont {C.}~\bibnamefont {Liu}},\ }\href
  {\doibase 10.1088/1361-6382/ab1101} {\bibfield  {journal} {\bibinfo
  {journal} {Class. Quant. Grav.}\ }\textbf {\bibinfo {volume} {36}},\ \bibinfo
  {pages} {105011} (\bibinfo {year} {2019})},\ \Eprint
  {http://arxiv.org/abs/1803.01944} {arXiv:1803.01944 [astro-ph.HE]}
  \BibitemShut {NoStop}%
\bibitem [{\citenamefont {Yagi}\ and\ \citenamefont
  {Seto}(2011)}]{Yagi:2011wg}%
  \BibitemOpen
  \bibfield  {author} {\bibinfo {author} {\bibfnamefont {K.}~\bibnamefont
  {Yagi}}\ and\ \bibinfo {author} {\bibfnamefont {N.}~\bibnamefont {Seto}},\
  }\href {\doibase 10.1103/PhysRevD.83.044011} {\bibfield  {journal} {\bibinfo
  {journal} {Phys. Rev. D}\ }\textbf {\bibinfo {volume} {83}},\ \bibinfo
  {pages} {044011} (\bibinfo {year} {2011})},\ \bibinfo {note} {[Erratum:
  Phys.Rev.D 95, 109901 (2017)]},\ \Eprint {http://arxiv.org/abs/1101.3940}
  {arXiv:1101.3940 [astro-ph.CO]} \BibitemShut {NoStop}%
\bibitem [{\citenamefont {Kawamura}\ \emph {et~al.}(2021)\citenamefont
  {Kawamura} \emph {et~al.}}]{Kawamura:2020pcg}%
  \BibitemOpen
  \bibfield  {author} {\bibinfo {author} {\bibfnamefont {S.}~\bibnamefont
  {Kawamura}} \emph {et~al.},\ }\href {\doibase 10.1093/ptep/ptab019}
  {\bibfield  {journal} {\bibinfo  {journal} {PTEP}\ }\textbf {\bibinfo
  {volume} {2021}},\ \bibinfo {pages} {05A105} (\bibinfo {year} {2021})},\
  \Eprint {http://arxiv.org/abs/2006.13545} {arXiv:2006.13545 [gr-qc]}
  \BibitemShut {NoStop}%
\bibitem [{\citenamefont {El-Neaj}\ \emph {et~al.}(2020)\citenamefont {El-Neaj}
  \emph {et~al.}}]{AEDGE:2019nxb}%
  \BibitemOpen
  \bibfield  {author} {\bibinfo {author} {\bibfnamefont {Y.~A.}\ \bibnamefont
  {El-Neaj}} \emph {et~al.} (\bibinfo {collaboration} {AEDGE}),\ }\href
  {\doibase 10.1140/epjqt/s40507-020-0080-0} {\bibfield  {journal} {\bibinfo
  {journal} {EPJ Quant. Technol.}\ }\textbf {\bibinfo {volume} {7}},\ \bibinfo
  {pages} {6} (\bibinfo {year} {2020})},\ \Eprint
  {http://arxiv.org/abs/1908.00802} {arXiv:1908.00802 [gr-qc]} \BibitemShut
  {NoStop}%
\bibitem [{\citenamefont {Hild}\ \emph {et~al.}(2011)\citenamefont {Hild} \emph
  {et~al.}}]{Hild:2010id}%
  \BibitemOpen
  \bibfield  {author} {\bibinfo {author} {\bibfnamefont {S.}~\bibnamefont
  {Hild}} \emph {et~al.},\ }\href {\doibase 10.1088/0264-9381/28/9/094013}
  {\bibfield  {journal} {\bibinfo  {journal} {Class. Quant. Grav.}\ }\textbf
  {\bibinfo {volume} {28}},\ \bibinfo {pages} {094013} (\bibinfo {year}
  {2011})},\ \Eprint {http://arxiv.org/abs/1012.0908} {arXiv:1012.0908 [gr-qc]}
  \BibitemShut {NoStop}%
\bibitem [{\citenamefont {Reitze}\ \emph {et~al.}(2019)\citenamefont {Reitze}
  \emph {et~al.}}]{Reitze:2019iox}%
  \BibitemOpen
  \bibfield  {author} {\bibinfo {author} {\bibfnamefont {D.}~\bibnamefont
  {Reitze}} \emph {et~al.},\ }\href@noop {} {\bibfield  {journal} {\bibinfo
  {journal} {Bull. Am. Astron. Soc.}\ }\textbf {\bibinfo {volume} {51}},\
  \bibinfo {pages} {035} (\bibinfo {year} {2019})},\ \Eprint
  {http://arxiv.org/abs/1907.04833} {arXiv:1907.04833 [astro-ph.IM]}
  \BibitemShut {NoStop}%
\bibitem [{\citenamefont {Aasi}\ \emph {et~al.}(2015)\citenamefont {Aasi} \emph
  {et~al.}}]{LIGOScientific:2014pky}%
  \BibitemOpen
  \bibfield  {author} {\bibinfo {author} {\bibfnamefont {J.}~\bibnamefont
  {Aasi}} \emph {et~al.} (\bibinfo {collaboration} {LIGO Scientific}),\ }\href
  {\doibase 10.1088/0264-9381/32/7/074001} {\bibfield  {journal} {\bibinfo
  {journal} {Class. Quant. Grav.}\ }\textbf {\bibinfo {volume} {32}},\ \bibinfo
  {pages} {074001} (\bibinfo {year} {2015})},\ \Eprint
  {http://arxiv.org/abs/1411.4547} {arXiv:1411.4547 [gr-qc]} \BibitemShut
  {NoStop}%
\bibitem [{\citenamefont {Thrane}\ and\ \citenamefont
  {Romano}(2013)}]{Thrane:2013oya}%
  \BibitemOpen
  \bibfield  {author} {\bibinfo {author} {\bibfnamefont {E.}~\bibnamefont
  {Thrane}}\ and\ \bibinfo {author} {\bibfnamefont {J.~D.}\ \bibnamefont
  {Romano}},\ }\href {\doibase 10.1103/PhysRevD.88.124032} {\bibfield
  {journal} {\bibinfo  {journal} {Phys. Rev. D}\ }\textbf {\bibinfo {volume}
  {88}},\ \bibinfo {pages} {124032} (\bibinfo {year} {2013})},\ \Eprint
  {http://arxiv.org/abs/1310.5300} {arXiv:1310.5300 [astro-ph.IM]} \BibitemShut
  {NoStop}%
\bibitem [{\citenamefont {Allen}\ and\ \citenamefont
  {Romano}(1999)}]{Allen:1997ad}%
  \BibitemOpen
  \bibfield  {author} {\bibinfo {author} {\bibfnamefont {B.}~\bibnamefont
  {Allen}}\ and\ \bibinfo {author} {\bibfnamefont {J.~D.}\ \bibnamefont
  {Romano}},\ }\href {\doibase 10.1103/PhysRevD.59.102001} {\bibfield
  {journal} {\bibinfo  {journal} {Phys. Rev. D}\ }\textbf {\bibinfo {volume}
  {59}},\ \bibinfo {pages} {102001} (\bibinfo {year} {1999})},\ \Eprint
  {http://arxiv.org/abs/gr-qc/9710117} {arXiv:gr-qc/9710117} \BibitemShut
  {NoStop}%
\bibitem [{\citenamefont {Kudoh}\ \emph {et~al.}(2006)\citenamefont {Kudoh},
  \citenamefont {Taruya}, \citenamefont {Hiramatsu},\ and\ \citenamefont
  {Himemoto}}]{Kudoh:2005as}%
  \BibitemOpen
  \bibfield  {author} {\bibinfo {author} {\bibfnamefont {H.}~\bibnamefont
  {Kudoh}}, \bibinfo {author} {\bibfnamefont {A.}~\bibnamefont {Taruya}},
  \bibinfo {author} {\bibfnamefont {T.}~\bibnamefont {Hiramatsu}}, \ and\
  \bibinfo {author} {\bibfnamefont {Y.}~\bibnamefont {Himemoto}},\ }\href
  {\doibase 10.1103/PhysRevD.73.064006} {\bibfield  {journal} {\bibinfo
  {journal} {Phys. Rev. D}\ }\textbf {\bibinfo {volume} {73}},\ \bibinfo
  {pages} {064006} (\bibinfo {year} {2006})},\ \Eprint
  {http://arxiv.org/abs/gr-qc/0511145} {arXiv:gr-qc/0511145} \BibitemShut
  {NoStop}%
\bibitem [{\citenamefont {Caprini}\ \emph {et~al.}(2019)\citenamefont
  {Caprini}, \citenamefont {Figueroa}, \citenamefont {Flauger}, \citenamefont
  {Nardini}, \citenamefont {Peloso}, \citenamefont {Pieroni}, \citenamefont
  {Ricciardone},\ and\ \citenamefont {Tasinato}}]{Caprini:2019pxz}%
  \BibitemOpen
  \bibfield  {author} {\bibinfo {author} {\bibfnamefont {C.}~\bibnamefont
  {Caprini}}, \bibinfo {author} {\bibfnamefont {D.~G.}\ \bibnamefont
  {Figueroa}}, \bibinfo {author} {\bibfnamefont {R.}~\bibnamefont {Flauger}},
  \bibinfo {author} {\bibfnamefont {G.}~\bibnamefont {Nardini}}, \bibinfo
  {author} {\bibfnamefont {M.}~\bibnamefont {Peloso}}, \bibinfo {author}
  {\bibfnamefont {M.}~\bibnamefont {Pieroni}}, \bibinfo {author} {\bibfnamefont
  {A.}~\bibnamefont {Ricciardone}}, \ and\ \bibinfo {author} {\bibfnamefont
  {G.}~\bibnamefont {Tasinato}},\ }\href {\doibase
  10.1088/1475-7516/2019/11/017} {\bibfield  {journal} {\bibinfo  {journal}
  {JCAP}\ }\textbf {\bibinfo {volume} {11}},\ \bibinfo {pages} {017} (\bibinfo
  {year} {2019})},\ \Eprint {http://arxiv.org/abs/1906.09244} {arXiv:1906.09244
  [astro-ph.CO]} \BibitemShut {NoStop}%
\bibitem [{\citenamefont {Abazajian}\ \emph {et~al.}(2022)\citenamefont
  {Abazajian} \emph {et~al.}}]{CMB-S4:2022ght}%
  \BibitemOpen
  \bibfield  {author} {\bibinfo {author} {\bibfnamefont {K.}~\bibnamefont
  {Abazajian}} \emph {et~al.} (\bibinfo {collaboration} {CMB-S4}),\ }\href@noop
  {} {\  (\bibinfo {year} {2022})},\ \Eprint {http://arxiv.org/abs/2203.08024}
  {arXiv:2203.08024 [astro-ph.CO]} \BibitemShut {NoStop}%
\end{thebibliography}%

\end{document}